\documentclass[11pt]{article}
\usepackage{amsmath}
\usepackage{amsfonts}
\usepackage{amssymb}
\usepackage[utf8]{inputenc}
\usepackage{amsmath}
\usepackage{amsfonts}
\usepackage{graphicx,color}
\usepackage{hyperref}
\usepackage{booktabs}
\usepackage[super]{nth}
\hypersetup{
  colorlinks=TRUE,
  linkcolor=blue,
  urlcolor=blue,
  citecolor=blue
}
\usepackage{authblk}
\usepackage{epsfig}
\usepackage{amssymb}
\usepackage{fancyhdr}
\usepackage{amsmath}
\usepackage{multirow}
\usepackage{algorithm}
\usepackage[round,authoryear,sort&compress]{natbib}
\usepackage{geometry}
\usepackage{xcolor}

\usepackage{algorithm}
\usepackage{algorithmic}
\usepackage{float}
\usepackage{rotating}

\title{Variational Bayesian Sparse Negative Binomial Regression}
\author[1]{Mitra Kharabati\thanks{kharabatimitra@gmail.com}}
\author[1]{Morteza Amini\thanks{Corresponding author, e-mail: morteza.amini@ut.ac.ir}}
\author[2]{Mohammad Arashi\thanks{arashi@um.ac.ir}}
\affil[1]{Department of Statistics, School of Mathematics, Statistics, and Computer Science, College of Science, University of Tehran, Tehran, Iran}
\affil[2]{Department of Statistics, Faculty of Mathematical Sciences, Ferdowsi University of Mashhad, Mashhad 9177948974, Razavi Khorasan, Iran}
\begin{document}
\maketitle

\begin{abstract}
Count data with overdispersion and high-dimensional predictors pose significant challenges in modern applications. While negative binomial regression offers a flexible modeling framework, existing Bayesian approaches rely on computationally expensive MCMC methods that become impractical in high-dimensional settings. This paper develops a variational Bayesian framework for sparse negative binomial regression using horseshoe and continuous spike-and-slab priors. Our proposed methods achieve estimation accuracy and variable selection performance comparable to MCMC benchmarks while offering substantial computational savings over MCMC. Extensive simulations demonstrate that the negative binomial specification is essential for overdispersed data, as Poisson-based approaches exhibit substantial performance degradation under overdispersion. Conversely, our methods remain robust when the data are Poisson, making them a safer default choice. Applications to real benchmark datasets further confirm the practical utility of our approach. 
\end{abstract}

\section{Introduction}

Count data arise frequently in numerous scientific disciplines, including epidemiology, ecology, finance, and social sciences, where the response variable of interest takes non-negative integer values. The Poisson regression model \citep{gr03} serves as the foundational approach for count data, assuming that the conditional mean equals the conditional variance. However, this equidispersion assumption is often violated in practice, as real count data frequently exhibit overdispersion where the variance exceeds the mean \citep{cox09}. The negative binomial regression model, which introduces an additional dispersion parameter to accommodate overdispersion, has emerged as a flexible and widely-used alternative \citep{hilbe11}. 

The Bayesian paradigm offers a principled framework for count regression modeling, enabling coherent uncertainty quantification. Bayesian Poisson regression has been extensively studied \citep{el73, ch00, ch97}, while Bayesian negative binomial regression has received increasing attention due to its ability to handle overdispersion \citep{fu15, kim13}. However, in modern high-dimensional applications where the number of predictors $p$ may exceed the sample size $n$, these classical Bayesian approaches face significant challenges due to the curse of dimensionality and the need for variable selection.

Sparse regression has become an essential tool for high-dimensional data analysis, with penalized likelihood methods such as the LASSO \citep{t96} and SCAD \citep{fan01} being widely adopted in the frequentist framework. For count data, penalized Poisson regression has been developed \citep{a15, sea11, jia19}, and the elastic-net regularized negative binomial regression has been proposed for high-dimensional settings \citep{zhang22}. In the Bayesian context, sparsity is typically induced through carefully constructed prior distributions, including spike-and-slab priors \citep{mit88, geo93, ish05}, the horseshoe prior \citep{car9}, and the Bayesian bridge \citep{pol14}. For count regression, Bayesian sparse methods have been developed for Poisson models \citep{datta16} and negative binomial models \citep{fu16}. These approaches provide probabilistic variable selection and uncertainty quantification, which are valuable advantages over frequentist penalized methods.

Despite these advances, Bayesian inference for sparse count regression models often relies on Markov chain Monte Carlo (MCMC) methods, which can be computationally prohibitive in high-dimensional settings with large sample sizes. Variational Bayes (VB) has emerged as a computationally efficient alternative to MCMC, providing fast approximate posterior inference through optimization \citep{b17}. VB methods have been successfully applied to various Bayesian models, including sparse regression \citep{orm17, ray22}. For count data, variational inference has been developed for Poisson regression \citep{luts15, kharabati26} and semiparametric count response regression \citep{murru25}. However, to the best of our knowledge, a comprehensive variational Bayesian framework for sparse negative binomial regression that simultaneously handles overdispersion and high-dimensional variable selection has not been fully developed.

The existing literature on variational inference for count data primarily focuses on Poisson models, which may fail in the presence of overdispersion. While \citet{luts15} developed variational inference for count response semiparametric regression, their approach does not consider the variable selection problem. The negative binomial model presents additional challenges for variational inference due to the non-conjugate relationship between the dispersion parameter and the regression coefficients. Recent work by \citet{murru25} addresses variational inference for count response semiparametric regression with convex solutions, yet their framework does not specifically target the sparse high-dimensional negative binomial setting with the mean-field approximation.

This paper fills this gap by developing a variational Bayesian framework for sparse negative binomial regression. We propose efficient mean-field variational inference algorithms for two sparsity-inducing priors: the horseshoe prior and the continuous spike-and-slab prior. Our approach handles the non-conjugate structure of the negative binomial model through the non-conjugate variational Bayes framework \citep{wand11}, providing closed-form updates for the variational parameters. We demonstrate through extensive simulation studies that our proposed VB methods achieve comparable performance to MCMC benchmarks in terms of estimation accuracy, variable selection, and uncertainty quantification, while offering substantial computational advantages. Additionally, we evaluate the predictive performance of our methods on several real benchmark datasets, showing their practical utility in diverse applications.

The remainder of this paper is organized as follows. Section 2 reviews the non-conjugate mean-field variational Bayes framework, which forms the theoretical foundation for our proposed methodology. Section 3 presents the variational Bayesian sparse negative binomial regression model, including the prior specifications for the horseshoe (HS) and continuous spike and slab (CS) priors, the derivation of the variational update equations, the sparsity-enforcing thresholding procedure, and the approximation of the posterior predictive mass function. Section 4 evaluates the performance of our proposed methods through extensive simulation studies under both low-dimensional and high-dimensional scenarios, considering data generated from both negative binomial and Poisson models. Section 5 applies the proposed methods to several benchmark real datasets, demonstrating their practical utility in diverse applications. Section 6 concludes the paper with a discussion of the main findings and potential directions for future research. Additional results, including further simulation studies and derivations, are provided in the supplementary material.

\section{Non-conjugate mean-field variational Bayes}

Let $\mathbf{x}$ denote a vector of observed data and $\boldsymbol\theta$ a vector of parameters with joint distribution $p(\mathbf{x}, \boldsymbol\theta)$. In the Bayesian inference framework, inference about $\boldsymbol\theta$ is based on the posterior distribution $p(\boldsymbol\theta \mid \mathbf{x}) = p(\mathbf{x}, \boldsymbol\theta) / p(\mathbf{x})$, where $p(\mathbf{x}) = \int p(\mathbf{x}, \boldsymbol\theta) d\boldsymbol\theta$ is the marginal likelihood.

Variational Bayes (VB) is a method for obtaining an approximate distribution $q(\boldsymbol\theta)$ to the posterior distribution $p(\boldsymbol\theta \mid \mathbf{x})$ by minimizing the Kullback-Leibler (KL) divergence 
\begin{equation}
\text{KL}\left[q(\boldsymbol\theta) \mid\mid p(\boldsymbol\theta \mid \mathbf{x})\right] = \int q(\boldsymbol\theta) \log \frac{q(\boldsymbol\theta)}{p(\boldsymbol\theta \mid \mathbf{x})} d\boldsymbol\theta = \log p(\mathbf{x}) - \int q(\boldsymbol\theta) \log \frac{p(\mathbf{x}, \boldsymbol\theta)}{q(\boldsymbol\theta)} d\boldsymbol\theta,\label{kl}
\end{equation}
 as a measure of discrepancy. In the mean-field variational Bayes framework, we assume that the parameter vector $\boldsymbol\theta$ is partitioned into $M$ blocks $\{\boldsymbol\theta_1, \ldots, \boldsymbol\theta_M\}$, and we approximate $p(\boldsymbol\theta \mid \mathbf{x})$ by
\begin{equation}
q(\boldsymbol\theta) = \prod_{j=1}^{M} q(\boldsymbol\theta_j),
\end{equation}
i.e., the mean-field approximation assumes that $\boldsymbol\theta_1, \ldots, \boldsymbol\theta_M$ are mutually independent.

The optimal variational Bayes approximation, denoted by $q^{\ast}(\cdot)$, is obtained as
\begin{equation*}
q^{\ast}(\boldsymbol\theta) = \arg\min_q \text{KL}\left[q(\boldsymbol\theta) \mid\mid p(\boldsymbol\theta \mid \mathbf{x})\right],
\end{equation*}
which leads to the well-known variational Bayes update equation
\begin{equation}
\log q(\boldsymbol\theta_j) = \mathbb{E}_{(-\boldsymbol\theta_j)}\left[\log p(\mathbf{x}, \boldsymbol\theta)\right] + \text{Const.}, \qquad j = 1, \ldots, M,\label{eq4}
\end{equation}
where the expectation $\mathbb{E}_{(-\boldsymbol\theta_j)}$ is taken with respect to $q(-\boldsymbol\theta_j) = \prod_{i \neq j} q(\boldsymbol\theta_i)$.

From \eqref{kl}, it is evident that minimizing the KL divergence is equivalent to maximizing the evidence lower bound (ELBO), defined as
\begin{equation}
\text{ELBO} = \log p(\mathbf{y};q) = \int q(\boldsymbol\theta) \log \frac{p(\mathbf{x}, \boldsymbol\theta)}{q(\boldsymbol\theta)} d\boldsymbol\theta.\label{eq5}
\end{equation}
The ELBO equals $\log p(\mathbf{x})$ when the KL divergence is zero, i.e., when the approximation is exact. When the approximation is not exact, we have
\begin{equation*}
\text{ELBO} < \log p(\mathbf{x}).
\end{equation*}

In many statistical models, particularly when the prior is not conjugate to the likelihood, equation \eqref{eq4} does not yield a closed-form or recognizable distribution for $q(\boldsymbol\theta_j)$. This situation is known as the \emph{non-conjugate case} in variational Bayes. Under such circumstances, additional approximation techniques are required to  update the hyper-parameters of the approximate distribution. In such settings, one efficient strategy is to assume an exponential family distribution for the non-conjugate parameter block. Without loss of generality, we take the non-conjugate block to be $\boldsymbol\theta_1$. This approach, developed by \citet{wand11}, considers the following approximate distribution for the parameter $\boldsymbol\theta_1$
\begin{equation}
q(\boldsymbol\theta_1 \mid \boldsymbol\eta_1) = a(\boldsymbol\eta_1) b(\boldsymbol\theta_1) \exp\left(\boldsymbol\eta_1^\top T(\boldsymbol\theta_1)\right), 
\end{equation}
so that the full approximation factorizes as
\begin{equation}
q(\boldsymbol\theta) = q(\boldsymbol\theta_1 \mid \boldsymbol\eta_1) \prod_{j=2}^M q(\boldsymbol\theta_j).
\end{equation}

The goal is to find the natural parameters $\boldsymbol\eta_1$ that minimize the KL divergence between $q(\boldsymbol\theta)$ and $p(\boldsymbol\theta \mid \mathbf{x})$. It has been shown \citep{wand11} that the optimality condition for the natural parameters is
\begin{equation}
\boldsymbol\eta_1 = \text{Var}_{q(\boldsymbol\theta_1)}(\boldsymbol\theta_1)^{-1} \frac{\partial}{\partial \boldsymbol\eta_1} \mathbb{E}_{q(\boldsymbol\theta)}\left[\log p(\mathbf{x}, \boldsymbol\theta)\right].
\end{equation}

In the special case where the approximate distribution for $\boldsymbol\theta_1$ is assumed to be multivariate normal with mean $\boldsymbol\mu_1$ and covariance matrix $\boldsymbol\Sigma_1$, \citet{wand11} showed that by defining the function $\mathcal{G}(\boldsymbol\mu_1, \boldsymbol\Sigma_1) = \mathbb{E}_{q(\boldsymbol\theta)}[\log p(\mathbf{x}, \boldsymbol\theta)]$, the fixed-point updates for the covariance matrix and mean are given by
\begin{equation}
\boldsymbol\Sigma_1^{\text{new}} \longleftarrow \left[ -2 \frac{\partial \mathcal{G}}{\partial \boldsymbol\Sigma_1} \right]^{-1},
\end{equation}
\begin{equation}
\boldsymbol\mu_1^{\text{new}} \longleftarrow \boldsymbol\mu_1^{\text{old}} + \boldsymbol\Sigma_1 \left[ \frac{\partial \mathcal{G}}{\partial \boldsymbol\mu_1} \right]^\top,
\end{equation}
where $\frac{\partial \mathcal{G}}{\partial \boldsymbol\Sigma_1}$ and $\frac{\partial \mathcal{G}}{\partial \boldsymbol\mu_1}$ denote the derivatives of the function $\mathcal{G}$ with respect to the covariance matrix and mean vector, respectively.

\section{Variational Bayesian sparse negitive binomial regression}

Suppose that $y_i \mathop{\sim}\limits^{\mathrm{ind}} \text{NB} (\kappa, \mu_i = \exp(\mathbf{X}_i { \boldsymbol\beta})), \quad i=1, \ldots , n $, where $\mathbf{X}_i = (1,X_{i1},\ldots,X_{i(p-1)})$ and ${ \boldsymbol\beta = (\beta_0,\beta_1,\ldots,\beta_{p-1})^\top}$. As mentioned by \cite{luts15}, $y_i \mathop{\sim}\limits^{\mathrm{ind}} \text{NB} (\kappa, \mu_i = \exp(\mathbf{X}_i { \boldsymbol\beta}))$, can be replaced by 
$$y_i|g_i \mathop{\sim}\limits^{\mathrm{ind}}  \;\text{Poiss} (g_i), \quad
g_i|{ \boldsymbol\beta}\mathop{\sim}\limits^{\mathrm{ind}}  \;\text{Gamma}(\kappa, \kappa \exp(-\mathbf{X}_i { \boldsymbol\beta}))).$$

One of the most famous {sparsity-enforcing} priors is the spike and slab prior \citep{mit88,geo97}, which is a mixture of a point mass $\delta(\cdot)$ and a continuous prior, which is usually considered to be Gaussian. Let {$\mathbf{Z} = (Z_1,\ldots,Z_{p-1})^\top$} be a vector of the latent binary {variables}. Then, the spike and slab prior {for} the vector of regression coefficient ${ \boldsymbol\beta}$ {is} defined as
\begin{equation}\label{ssp12}
p({ \boldsymbol\beta} ) = \prod_{j=1}^{p-1} \left[ Z_j {\rm N} (\beta_j;0, { \sigma^2} ) + ( 1 -Z_j ) \delta(\beta_j) \right],
\end{equation}
where $\delta(\beta_j) = 1,$ if $\beta_j = 0$ and $=0$, otherwise. 

Continuous relaxations of \eqref{ssp12}, with $\delta(\cdot)$ replaced by a peaked
continuous density, {are} considered by many authors \citep[see e.g.,][among others]{geo93,ir03,ish05}. One of the most famous variations of the continuous spike and slab prior is as follows 
\begin{equation}\label{ssp13}
p({ \boldsymbol\beta} ) = \prod_{j=1}^{p-1} \left[ Z_j {\rm N} ({ \boldsymbol\beta}_j;0, { \sigma^2} ) +
 ( 1 -Z_j ) {\rm N} ({ \boldsymbol\beta}_j; 0 , c { \sigma^2} ) \right],
\end{equation}
where the constant $c$ is chosen sufficiently small to induce sparsity on $\boldsymbol\beta_j$. Based on a sensitivity analysis, we set $c = 0.001$, as the results were found to be insensitive to small perturbations of this value.

Using prior \eqref{ssp13}, the CS-VB-NB model is defined to be as follows
\begin{align}
y_{i}| g_i \mathop{\sim}\limits^{\mathrm{ind}} &  \;\text{Poiss} ( g_i ) , \quad  
g_i \vert \kappa , { \boldsymbol\beta} \mathop{\sim}  \text{Gamma} (\kappa , \kappa \exp(- \mathbf{X}_i { \boldsymbol\beta})),\; i=1, \ldots , n,\nonumber\\
\mathbf{X}_i = & \;(1,X_{i1},\ldots,X_{i(p-1)}), \; { \boldsymbol\beta}  = (\beta_0,\beta_1,\ldots,\beta_{p-1})^\top, \nonumber\\
\kappa \vert b \mathop{\sim} & \text{Gamma} (\frac{1}{2} , \frac{1}{b}), \quad
b \mathop{\sim}  \text{Inv-Gammma} (1/2 , 1/\mathcal{M}_{\kappa}), \nonumber\\
p( { \boldsymbol\beta}  \vert \mathbf{Z}, { \sigma^2}) = & \; \prod_{j=0}^{p-1} \left[ Z_j {\rm N} ({ \boldsymbol\beta}_j;0, { \sigma^2} ) +
 ( 1 -Z_j ) {\rm N} ({ \boldsymbol\beta}_j; 0 , c { \sigma^2} ) \right] , \nonumber\\
Z_0 = 1, & \;\;{\rm w.p.1} \nonumber\\
Z_j|\zeta_j  \mathop{\sim}\limits^{\mathrm{ind}} & \;\text{Ber} ( \zeta_j ) ,  \;\; 
\zeta_j \mathop{\sim}\limits^{\mathrm{iid}}  \text{Beta}  ( {\alpha_1 , \alpha_2}  )  \quad { j = 1 , \ldots , p-1} ,\nonumber\\
{ \sigma^2}|a \sim &\; \text{Inv-Gamma} \left( \dfrac{1}{2} , \dfrac{1}{a} \right), \;\;
a \sim  \;\text{Inv-Gamma} \left( \dfrac{1}{2} , \dfrac{1}{\mathcal{M}_{\sigma^2}} \right). \label{model22}
\end{align}

The hierarchical hyper-priors for $ \sigma^2|a$ and $a$ in \eqref{model22} {result} in the marginal hyper-prior $\sqrt{\sigma^2} \sim {\rm HalfCauchy}(\sqrt{\mathcal{M}_{\sigma^2}})$, which is suggested by \cite{luts15}.

The second prior considered in this study for promoting sparsity in the regression coefficients is the Horseshoe prior. The Horseshoe prior introduces adaptive shrinkage through coefficient--specific local scales and a common global scale, allowing strong shrinkage of negligible effects while preserving significant signals. Specifically, each regression coefficient is modeled as
\begin{align*}
\beta_j \mid \Lambda , \Xi \sim N(0,\Lambda_j \Xi),\; j = 0,1, \ldots, p-1, 
\end{align*}
where $\Lambda_j$ and $\Xi$ denote the local and global shrinkage parameters, respectively. The Horseshoe prior places Half-Cauchy distributions on these scales,
\begin{align*}
\Lambda_j \sim C^+(0,\mathcal{M}_{\lambda}), \; j = 1, \ldots, p-1, 
\qquad
\Xi \sim C^+(0,\mathcal{M}_{\xi}),
\end{align*}
which yields heavy tails and an infinite spike at zero, making the prior particularly suitable for sparse Bayesian regression models.

Again, we employ the auxiliary inverse-gamma representation of the Half-Cauchy distribution, for computational tractability within the variational Bayesian framework. In particular, the Half-Cauchy distribution admits the representation
\begin{align*}
\Lambda_j|\lambda &\sim \text{Inv-Gamma}\!\left(\frac12,\frac1{\lambda}\right),
\qquad
\lambda \sim \text{Inv-Gamma}\!\left(\frac12,\frac{1}{\mathcal{M}_{\lambda}}\right), \\
\Xi|\xi &\sim \text{Inv-Gamma}\!\left(\frac12,\frac1{\xi}\right),
\qquad
\xi \sim \text{Inv-Gamma}\!\left(\frac12,\frac{1}{\mathcal{M}_{\xi}}\right).
\end{align*}

Let $\boldsymbol\Lambda = (\Lambda_0,\lambda_1,\ldots,\Lambda_{p-1})^\top$.  
Using the above representation, the HS-VB-NB model is defined as follows
\begin{align}
y_i|g_i \mathop{\sim}\limits^{\mathrm{ind}} & \text{Poiss}(g_i), \quad g_i| \kappa , {\boldsymbol\beta}\mathop{\sim}\limits^{\mathrm{ind}} \text{Gamma}\big(\kappa,\kappa\exp(-\mathbf \mathbf{X}_i\boldsymbol\beta)\big), \quad i=1,\ldots,n, \nonumber\\
\kappa|b &\sim \text{Gamma}(1/2,1/b),  \quad 
b \sim \text{Inv-Gamma}(1/2,1/\mathcal{M}_\kappa),\nonumber\\
\beta_0|\tau_0 &\sim  \;{\rm N} (0 , \tau_0), \;\; 
\;\tau_0|a \sim \text{Inv-Gamma} \left(\frac{1}{2},\frac{1}{a}\right), \;\;
a \sim \text{Inv-Gamma}\left(\frac{1}{2}, \frac{1}{\mathcal{M}_{\tau_0}}\right),\nonumber\\
\beta_j|\Lambda_j,\Xi &\sim N(0,\Xi\Lambda_j),\quad j=1,\ldots,p-1,\nonumber\\
\Lambda_j|\lambda &\sim \text{Inv-Gamma}\!\left(\frac12,\frac1{\lambda}\right),
\qquad
\lambda \sim \text{Inv-Gamma}\!\left(\frac12,\frac{1}{\mathcal{M}_{\lambda}}\right),  \nonumber\\
\Xi|\xi &\sim \text{Inv-Gamma}\!\left(\frac12,\frac1{\xi}\right),
\qquad
\xi \sim \text{Inv-Gamma}\!\left(\frac12,\frac{1}{\mathcal{M}_{\lambda}}\right).
\label{model32}
\end{align}

The mean-field VB approximation density $q$  is assumed to be 
$$q({\boldsymbol\beta},\textbf{g},{ \boldsymbol\tau},{ \eta},a,b, \kappa) = q({\boldsymbol\beta})q(\textbf{g})q({ \boldsymbol\tau})q({ \eta})q(a)q(b)q(\kappa),$$
where ${ \boldsymbol\tau} = (\tau_0,\tau_1,\ldots,\tau_{p-1})^\top$, and $\textbf{g} = (g_1,\ldots,g_n)^\top$.

To provide conjugacy in the VB component $q({ \boldsymbol\beta})$, we assume for $M = C, H$, representing the CS and HS priors, respectively, that 
$$q({ \boldsymbol\beta}) = {\rm N}({ \boldsymbol\beta}; { \boldsymbol\mu}_{{ \boldsymbol\beta}(M)}, { \boldsymbol\Sigma}_{{ \boldsymbol\beta}(M)}),$$
where ${\rm N}(x; { \boldsymbol\mu}, { \boldsymbol\Sigma})$ stands for the multivariate normal density function, and 
$\left\{{ \boldsymbol\mu}_{{ \boldsymbol\beta}(M)}, { \boldsymbol\Sigma}_{{ \boldsymbol\beta}(M)}\right\}$, is the set of hyper-parameters of the variational approximations, associated with the model $M$, for $M = C,H$. Also, let $D^{(M)}_{{ \boldsymbol\beta}} = { \boldsymbol\mu}_{{ \boldsymbol\beta}(M)}{ \boldsymbol\mu}_{{ \boldsymbol\beta}(M)}^\top + { \boldsymbol\Sigma}_{{ \boldsymbol\beta}(M)}$, for $M = C, H$. 

Applying the formulation presented by \cite{luts15}, we have, for $M = C, H$, 
\begin{equation}\label{smub}
{ \boldsymbol\Sigma}_{{ \boldsymbol\beta}(M)} =\left( -2 \text{d}_{{\boldsymbol\Sigma}_{{ \boldsymbol\beta}(M)}}{\rm E}_{q} \right)^{-1}
  , \quad 
{ \boldsymbol\mu}_{{ \boldsymbol\beta}(M)}^{new} = { \boldsymbol\mu}_{{ \boldsymbol\beta}(M)}^{old}  + {\boldsymbol\Sigma}_{{ \boldsymbol\beta}(M)} \left( \text{d}_{{ \boldsymbol\mu}_{{ \boldsymbol\beta}(M)}} {\rm E}_q \right)^{\top}
\end{equation}
where $ \text{d}_{{\boldsymbol\Sigma}_{{ \boldsymbol\beta}(M)}}{\rm E}_{q}$ and $ \text{d}_{{ \boldsymbol\mu}_{{ \boldsymbol\beta}(M)}}{\rm E}_q$ denote the derivaties of 
${\rm E}_q = {\rm E}_{q(\theta)}(\log P(y,\theta))$ with respect to $ { \boldsymbol\mu}_{{ \boldsymbol\beta}(M)}$ and $ { \boldsymbol\Sigma}_{{ \boldsymbol\beta}(M)} $, respectively, in which  $\theta = ({\boldsymbol\beta},\textbf{g},{ \boldsymbol\tau},a,\kappa,b, \textbf{Z} , \zeta)^\top$, for CS-VB-NB model, and $\theta = (\mathbf{g} , {\boldsymbol\beta} , \kappa , b , {\Xi} , \xi , {\boldsymbol\Lambda} , \lambda)^\top$, for HS-VB-NB model.  

As demonstrated in Appendix A, 
\begin{align*}
d_{{ \boldsymbol\mu }_{{ \boldsymbol\beta}(C)}} {\rm E}_q  = & -\mu_{\kappa}\mathbf{X}^\top (\mathbf{1}_n - \mu_{\boldsymbol g} \odot \boldsymbol\omega^{(C)}) -  {\rm E}(\tau^{-2}) ((1-1/c)P^{(C)}+1/c\mathbf{1}_p)\odot \mu_{\beta(C)},\\
d_{{ \boldsymbol\Sigma}_{{ \boldsymbol\beta}(C)}} {\rm E}_q = &-\frac{1}{2} \mu_{\kappa}\mathbf{X}^\top \text{diag}(\mu_{\boldsymbol g} \odot \boldsymbol\omega^{(C)}) \mathbf{X}- \frac{1}{2} {\rm E}(\tau^{-2}) \text{diag}((1-1/c)P^{(C)}+1/c\mathbf{1}_p), 
\end{align*}
and 
\begin{align}
d_{{ \boldsymbol\mu }_{{ \boldsymbol\beta}(H)}} {\rm E}_q  = & -\mu_{\kappa}\mathbf{X}^\top (\mathbf{1}_n - \mu_{\boldsymbol g} \odot \boldsymbol\omega^{(H)}) - {\rm E}({\Xi^{-1} })  \text{diag} \left( {\rm E} ({\boldsymbol\Lambda^{-1}}) \right) { \boldsymbol\mu }_{{ \boldsymbol\beta}(H)},\nonumber\\
d_{{ \boldsymbol\Sigma}_{{ \boldsymbol\beta}(H)}} {\rm E}_q = &-\frac{1}{2} \mu_{\kappa}\mathbf{X}^\top \text{diag}(\mu_{\boldsymbol g} \odot \boldsymbol\omega^{(H)}) \mathbf{X}-\frac{1}{2} {\rm E }({\Xi^{-1} }) \text{diag} \left( {\rm E} ({\boldsymbol\Lambda^{-1}}) \right),\label{dH}
\end{align}
where, for $M = C, H$, $\omega_i^{(M)} = \exp(-\mathbf{X}_i  { \boldsymbol\mu}_{{ \boldsymbol\beta}(M)} + \frac{1}{2} \mathbf{X}_i^{\top} { \boldsymbol\Sigma}_{{ \boldsymbol\beta}(M)}  \mathbf{X}_i ) $, and $\boldsymbol\omega^{(M)} = (\omega_1^{(M)},\ldots,\omega_n^{(M)})^\top$.

Furthermore, from Appendix A, for $M = C, H$, we obtain 
 $$q(g_i) = \text{Gamma} (g_i ; y_i + \mu_{\kappa} , 1+ \mu_{\kappa}\omega_i^{(M)}),\; i=1,\ldots,n, $$
and, the optimal variational distribution for $ \kappa $ is obtained as 
\begin{equation}\label{qka}
q(\kappa) = \dfrac{\exp \left\lbrace n ( \kappa \log \kappa - \log \Gamma (\kappa)) - C^{(M)} \kappa \right\rbrace }{ \kappa^{\frac{1}{2}} \mathcal{H} (-\frac{1}{2} , 0,1, n , C^{(M)} )}
\end{equation}
where
$$\mathcal{H} (p,q,\kappa, s,t) = \int_{0}^{\infty} x^p \log (1 +\kappa x )^q  \left( \dfrac{x^x}{\Gamma (x)}  \right) ^s \exp (-tx)  dx, $$
$$ C^{(M)} = 1^T X \mu_{\boldsymbol\beta}^{(M)} - 1^T E (\log\boldsymbol g) + \mu_{\boldsymbol g}^T \boldsymbol\omega^{(M)} + E(b^{-1}).$$

Thus, $\mu_{\kappa} = {\rm E}(\kappa) = \dfrac{\mathcal{H} (\frac{1}{2}, 0, 1, n , C^{(M)} )}{\mathcal{H}  (-\frac{1}{2} , 0, 1, n , C^{(M)} )}$.

Also, we see that
\begin{equation}\label{qb}
q(b) = \text{Inv-Gamma} (b;1,\mathcal{M}_{\kappa}+ \mu_{\kappa}),
\end{equation}
with $E(b^{-1} ) =  ({\mathcal{M}_{\kappa} + \mu_{\kappa}})^{-1}$. For the CS-VB-NB model,  we observe that 
$$q({ \sigma^2}) = \text{Inv-Gamma}\left({ \sigma^2}; \alpha_{\sigma^2}, \beta_{ \sigma^2}\right),
$$
where $\alpha_{\sigma^2} = \dfrac{p-1}{2} $ and 
$$ \beta_{ \sigma^2}= \frac{1}{2} \text{tr(diag}({ P^{(C)}}) D_{{ \boldsymbol\beta}}^{(C)}) + \dfrac{1}{2c}\text{tr(diag}(1-{ P^{(C)}}) D_{{ \boldsymbol\beta}}^{(C)}) + {\rm E}(a^{-1}),$$
and thus, 
${\rm E}({\sigma^{-2}}) = \frac{\alpha_{\sigma^2}}{ \beta_{ \sigma^2}}.$ Furthermore, we obtain 
$$q(a) = \text{Inv-Gamma}\left(a ; 1/2 , {\rm E}({\sigma^{-2}})+ \mathcal{M}_{\sigma^2}^{-1}\right),
$$
hence, 
${\rm E}(a^{-1}) = 0.5({{\rm E}({\sigma^{-2}})+ \mathcal{M}_{\sigma^2}^{-1}})^{-1}.$ The update for $q(Z_j)$ is derived as 
$$q(Z_j)  =\text{Ber} (Z_j; { P_j^{(C)}}),\quad { j=1,\ldots,p-1},
$$
where 
\begin{equation}\label{pj}
{ P_j^{(C)}} = \sigma \left({\rm E}(\log \zeta_j) - {\rm E}(\log (1-\zeta_j)) - \frac{1}{2} {\rm E}({\tau^{-2}}) { d_{jj}^{(C)}} (1-\frac{1}{c})\right), 
\end{equation}
and $ { P^{(C)}} = (1,{ P_1^{(C)}},\ldots,{ P_{p-1}^{(C)}}), $
{ in which $\sigma(v) = (1+\exp(-v))^{-1}$ is the sigmoid function,} and ${ d_{jj}^{(C)}} = (D_{{ \boldsymbol\beta}}^{(C)})_{jj}$. Also, 
$$q(\zeta_j) = \text{Beta} (\zeta_j;{ P_j^{(C)}} +{\alpha_1} , {\alpha_2} - { P_j^{(C)}} +1),
$$
and therefore 
${\rm E}(\log \zeta_j) = \psi({\alpha_1} + { P_j^{(C)}}) - \psi({\alpha_1 + \alpha_2} +1)$
and
$ {\rm E}(\log (1-\zeta_j))= \psi({\alpha_2} - { P_j^{(C)}} + 1) - \psi({\alpha_1 + \alpha_2} +1)$.

For the HS-VB-NB model, and for $j=1,\ldots,p-1$, we see in Appendix A that 
\[
q(\Lambda_j) = \text{InvGamma} \left(\Lambda_j; 1, \frac{1}{2}{\rm E}(\Xi^{-1})d_{jj}^{(H)}+{\rm E}(\lambda^{-1})\right).
\]
Also, 
\[
q(\Xi) = \text{InvGamma} \left(\Xi; \frac{p+1}{2}, \frac{1}{2}\sum_{j=1}^{p-1}{\rm E}(\Lambda_j^{-1})d_{jj}^{(H)}+{\rm E}(\xi^{-1})\right),
\]
\[
q(\lambda) = \text{InvGamma} \left(\lambda; \frac{1}{2}, \sum_{j=1}^{p-1}{\rm E}(\Lambda_j^{-1})+\frac{1}{\mathcal{M}_{\lambda}}\right),
\]
and
\[
q(\xi) = \text{InvGamma} \left(\xi; \frac{1}{2}, {\rm E}(\Xi^{-1})+\frac{1}{\mathcal{M}_{\xi}}\right).
\]
Furthermore,  
$$q(\tau_0) = \text{Inv-Gamma} (\tau_0 ; 1 , \frac{1}{2} d_{00}^{(H)} + {\rm E}(a^{-1})),
$$
and  
$$q(a) =  \text{Inv-Gamma} (a ; 1 , {\rm E}(\tau_0^{-1}  )+ \mathcal{M}_{\tau_0}^{-1}).
$$

For the CS-VB-NB model, the ELBO is also obtained as follows {(see Appendix A)}
\begin{align}
\log p(\mathbf{y};q) = \;&
{\rm E}_q (\log p (\mathbf{y}\vert { \boldsymbol\beta}, g )) +
{\rm E}_q (\log p (g \vert  { \boldsymbol\beta} , \kappa)) +
{\rm E}_q (\log p (\kappa \vert b)) +
{\rm E}_q (\log p (b)) \nonumber\allowdisplaybreaks\\
&+
{\rm E}_q(\log p({ \boldsymbol\beta} \vert \mathbf{Z},\tau^2)) +
{\rm E}_q(\log p({ \mathbf{Z} \vert \zeta})) +
{\rm E}_q (\log p(\zeta)) +
{\rm E}_q(\log p(\tau^2 \vert a)) \nonumber\allowdisplaybreaks\\
&+
{\rm E}_q (\log p(a)) -
{\rm E}_q (\log q (g)) -
{\rm E}_q (\log q(\kappa)) -
{\rm E}_q(\log q(b)) -
{\rm E}_q(\log q({ \boldsymbol\beta})) \nonumber\allowdisplaybreaks\\
&-
{\rm E}_q (\log q ({ \mathbf{Z}})) -
{\rm E}_q(\log q (\zeta) ) -
{\rm E}_q(\log q (\tau^2) ) -
{\rm E}_q(\log q (a)) \nonumber\allowdisplaybreaks\\
=&  -\frac12{\rm E}(\sigma^{-2}) \left[ (1-\tfrac1c) \sum_{j=1}^{p-1} P_j^{(C)}d_{jj}^{(C)}  +\tfrac1c\sum_{j=1}^{p-1} d_{jj}^{(C)}  \right]  \nonumber\allowdisplaybreaks\\ 
 & +{\rm E} (\sigma^{-2}) {\rm E}(a^{-1})  + \sum_{i=1}^{n} \log \Gamma(y_i + \mu_{\kappa}) \nonumber\allowdisplaybreaks\\
 & -(\mu_{\kappa}-1)1^\top {\rm E}(\log\boldsymbol g) + \mu_{\kappa} \mathbf{1}_n^\top (\mu_{\boldsymbol g}\odot\boldsymbol\omega^{(C)}) + \mathcal{H} (-\frac{1}{2} , 0, 1, n, C^{(C)}) \nonumber\allowdisplaybreaks\\
 & -\log (\mathcal{M}_{\kappa} + \mu_{\kappa}) + \frac{1}{2} \log \vert { \boldsymbol\Sigma}_{{ \boldsymbol\beta}(C)} \vert \nonumber\allowdisplaybreaks\\
 &  - {\sum_{j=1}^{p-1}} \left[ { P_j^{(C)}} \log ({ P_j^{(C)}}) + (1-{ P_j^{(C)}}) \log(1-{ P_j^{(C)}}) \right] \nonumber\allowdisplaybreaks\\
& + \sum_{j=1}^{p-1} \log \Gamma ({ P_j^{(C)}} + \alpha_1)  +\sum_{j=1}^{p-1} \log \Gamma (\alpha_2 - { P_j^{(C)}} + 1) \nonumber\allowdisplaybreaks\\
& - \alpha_{\sigma^2} \log \beta_{\sigma^2} + \log \Gamma (\alpha_{\sigma^2}) \nonumber\allowdisplaybreaks\\
& + \log \Big(\mathcal{M}_{\sigma^2}^{-1} + {\rm E}({ \sigma^{-2}}) \Big) + \text{Const.} \label{elbo22}
\end{align}

The ELBO for the HS-VB-NB model is also obtained as follows {(see Appendix A)}
\begin{align}
\log p(\mathbf{y};q) = \;&
{\rm E}_q (\log p (\mathbf{y}\vert { \boldsymbol\beta}, g )) +
{\rm E}_q (\log p (g \vert  { \boldsymbol\beta} , \kappa)) +
{\rm E}_q (\log p (\kappa \vert b)) +
{\rm E}_q (\log p (b)) \nonumber\allowdisplaybreaks\\
&+
{\rm E}_q(\log p({ \boldsymbol\beta} \vert \mathbf{\Lambda},\Xi)) +
{\rm E}_q(\log p({ \mathbf{\Lambda} \vert \lambda})) +
{\rm E}_q (\log p(\lambda)) +
{\rm E}_q(\log p(\Xi \vert \xi)) \nonumber\allowdisplaybreaks\\
&+
{\rm E}_q (\log p(\xi)) -
{\rm E}_q (\log q (g)) -
{\rm E}_q (\log q(\kappa)) -
{\rm E}_q(\log q(b)) -
{\rm E}_q(\log q({ \boldsymbol\beta})) \nonumber\allowdisplaybreaks\\
&-
{\rm E}_q (\log q ({ \mathbf{\Lambda}})) -
{\rm E}_q(\log q (\lambda) ) -
{\rm E}_q(\log q (\Xi) ) -
{\rm E}_q(\log q (\xi)) \nonumber\allowdisplaybreaks\\
=& -(\mu_{\kappa}-1)1^\top {\rm E}(\log\boldsymbol g) + \mu_{\kappa}\mathbf{1}_n^\top (\mu_{\boldsymbol g}\odot\boldsymbol\omega^{(H)})\nonumber\allowdisplaybreaks \\
&+\mathcal{H} (-\frac{1}{2} , 0, 1, n, C^{(H)}) +
{\rm E} (b^{-1}) \mu_{\kappa} -\frac{1}{2}\log (\mathcal{M}_{\kappa} + \mu_{\kappa})\nonumber\allowdisplaybreaks \\
& + \frac{1}{2} \log \vert { \boldsymbol\Sigma}_{{ \boldsymbol\beta}(H)} \vert + \frac{1}{2} {\rm E}(\Xi^{-1}) \sum_{j=1}^{p-1} {\rm E}(\Lambda_j^{-1})d_{jj}^{(H)}\nonumber\allowdisplaybreaks\\
&+{\rm E} (a^{-1}) {\rm E} (\tau_0^{-1}) -\frac{1}{2}\log ({\rm E} (\tau_0^{-1}) + A^{-1})\nonumber\allowdisplaybreaks \\
 &+ {\rm E}(\lambda^{-1})\sum_{j=1}^{p-1}{\rm E}(\Lambda_j^{-1})+ {\rm E}(\Xi^{-1}){\rm E}(\xi^{-1})+\text{Const.} \label{elbo32}
\end{align}

Algorithms \ref{al22} and \ref{al32} presents the algorithm for the CS-VB-NB, and HS-VB-NB methods, respectively. 

\begin{algorithm}
\small
    \caption{CS-VB-NB method for sparse Negative Binomial regression model.}\label{al22}
    \begin{algorithmic}
        \STATE $\bullet$ Set proper initial values for the hyper-parameters of $q(\cdot)$ functions,
	\STATE $\bullet$ Set $\epsilon$ equal to an arbitrary small value,
            \WHILE{The absolute relative change in ELBO is greater than $\epsilon$}
          \STATE $\bullet$ update  $d_{{ \boldsymbol\Sigma}_{{ \boldsymbol\beta}(C)}} {\rm E}_q \leftarrow   -\frac{1}{2}\mu_{\kappa}\mathbf{X}^\top \text{diag}(\mu_{\boldsymbol g} \odot \boldsymbol\omega^{(C)}) \mathbf{X} - \frac{1}{2} {\rm E}(\tau^{-2}) \text{diag}((1-1/c)P^{(C)}+1/c\mathbf{1}_p)$,
                \STATE $\bullet$ update 
${ \boldsymbol\Sigma}_{{ \boldsymbol\beta}(C)} \leftarrow  (-2  d_{{ \boldsymbol\Sigma}_{{ \boldsymbol\beta}(C)}} {\rm E}_q)^{-1}$,
  	\STATE $\bullet$ update  $d_{{ \boldsymbol\mu }_{{ \boldsymbol\beta}(C)}} {\rm E}_q  \leftarrow  -\mu_{\kappa}\mathbf{X}^\top (\mathbf{1}_n - \mu_{\boldsymbol g} \odot \boldsymbol\omega^{(C)})
-  {\rm E}(\tau^{-2}) ((1-1/c)P^{(C)}+1/c\mathbf{1}_p)\odot \mu_{\beta(C)}$,
                \STATE $\bullet$ update 
    ${ \boldsymbol\mu}_{{ \boldsymbol\beta}(C)}^{new} \leftarrow  { \boldsymbol\mu}_{{ \boldsymbol\beta}(C)}^{old} + { \boldsymbol\Sigma}_{{ \boldsymbol\beta}(C)} (d_{{ \boldsymbol\mu}_{{ \boldsymbol\beta}(C)}} {\rm E}_q)^{\top}$
        \STATE $\bullet$ update   $ D_{\boldsymbol\beta}^{(C)} \leftarrow { \boldsymbol\mu}_{{ \boldsymbol\beta}(C)}{ \boldsymbol\mu}_{{ \boldsymbol\beta}(C)}^{\top} + { \boldsymbol\Sigma}_{{ \boldsymbol\beta}(C)}$,
         \STATE $\bullet$  update   $\mu_{\kappa} \leftarrow \frac{\mathcal{H} (\frac{1}{2} , 0, n , C^{(C)} )}{\mathcal{H}  (-\frac{1}{2} , 0, n , C^{(C)})}$ 
               \STATE $\bullet$ update  $\mu_{g_j} \leftarrow \frac{y_i + \mu_{\kappa}}{ 1 + \mu_{\kappa}\omega_i^{(C)} } $, for $i=1, \cdots, n$, 
               \STATE $\bullet$ update  ${\rm E}({ \sigma^{-2}}) \leftarrow  \frac{\alpha_{\sigma^2}}{\beta_{\sigma^2}}$,
               \STATE $\bullet$ update ${\rm E}(a^{-1}) \leftarrow 0.5 \left({\rm E}({ \sigma^{-2}} )+\mathcal{M}_{\sigma^2}^{-1}\right)^{-1}$,
               \STATE $\bullet$ update   ${\rm E}(\log \zeta_j) \leftarrow \psi({ \alpha}_1 + { P_j^{(C)}}) - \psi({ \alpha}_1 + { \alpha}_2 +1)$, for $j=1, \cdots , p-1$,
 \STATE $\bullet$ update  $ {\rm E}(\log (1-\zeta_j))= \psi({ \alpha}_2 - { P_j^{(C)}} + 1) - \psi({ \alpha}_1 + { \alpha}_2 +1)$, for $j=1, \cdots , p-1$,
               \STATE $\bullet$ update  ${ P_j^{(C)}} \leftarrow \sigma \left({\rm E}(\log\zeta_j) - {\rm E}(\log (1-\zeta_j))- \frac{1}{2} {\rm E}({ \tau^{-2}}) { d_{jj}^{(C)}} (1-\frac{1}{c})\right) $,
               \STATE  $\bullet$ update  $ {\rm E}(b^{-1}) \leftarrow (\mathcal{M}_{\kappa} + \mu_{\kappa})^{-1}$, 
		\STATE $\bullet$ calculate ELBO from \eqref{elbo22}
            \ENDWHILE
    \end{algorithmic}
\end{algorithm}

\begin{algorithm}
\small
    \caption{HS-VB-NB method for sparse Negative Binomial regression model.}\label{al32}
    \begin{algorithmic}
        \STATE $\bullet$ Set proper initial values for the hyper-parameters of $q(\cdot)$ functions,
	\STATE $\bullet$ Set $\epsilon$ equal to an arbitrary small value,
            \WHILE{The absolute relative change in ELBO is greater than $\epsilon$}
          \STATE $\bullet$ update  $d_{{ \boldsymbol\Sigma}_{{ \boldsymbol\beta}(H)}} {\rm E}_q \leftarrow   -\frac{1}{2}\mu_{\kappa}\mathbf{X}^\top \text{diag}(\mu_{\boldsymbol g} \odot \boldsymbol\omega^{(H)}) \mathbf{X} -\frac{1}{2} {\rm E }({\Xi^{-1} }) \text{diag} \left( {\rm E} ({\boldsymbol\Lambda^{-1}}) \right),$
               \STATE $\bullet$ update 
${ \boldsymbol\Sigma}_{{ \boldsymbol\beta}(H)} \leftarrow  (-2  d_{{ \boldsymbol\Sigma}_{{ \boldsymbol\beta}(H)}} {\rm E}_q)^{-1}$,
  	\STATE $\bullet$ update  $d_{{ \boldsymbol\mu }_{{ \boldsymbol\beta}(H)}} {\rm E}_q  \leftarrow  -\mu_{\kappa}\mathbf{X}^\top (\mathbf{1}_n - \mu_{\boldsymbol g} \odot \boldsymbol\omega^{(H)})
- {\rm E}({\Xi^{-1} })  \text{diag} \left( {\rm E} ({\boldsymbol\Lambda^{-1}}) \right) { \boldsymbol\mu }_{{ \boldsymbol\beta}(H)}$,
            \STATE $\bullet$ update 
    ${ \boldsymbol\mu}_{{ \boldsymbol\beta}(H)}^{new} \leftarrow  { \boldsymbol\mu}_{{ \boldsymbol\beta}(H)}^{old} + { \boldsymbol\Sigma}_{{ \boldsymbol\beta}(H)} (d_{{ \boldsymbol\mu}_{{ \boldsymbol\beta}(H)}} {\rm E}_q)^{\top}$, 
       \STATE $\bullet$ update   $ D_{\boldsymbol\beta}^{(H)} \leftarrow { \boldsymbol\mu}_{{ \boldsymbol\beta}(H)}{ \boldsymbol\mu}_{{ \boldsymbol\beta}(H)}^{\top} + { \boldsymbol\Sigma}_{{ \boldsymbol\beta}(H)}$ , 
        \STATE $\bullet$  update   $\mu_{\kappa} \leftarrow \frac{\mathcal{H} (\frac{1}{2} , 0, n , C^{(H)} )}{\mathcal{H}  (-\frac{1}{2} , 0, n ,C^{(H)} )}$ 
               \STATE $\bullet$ update  $\mu_{g_j} \leftarrow \frac{y_i + \mu_{\kappa}}{ 1 + \mu_{\kappa}\omega_i^{(H)} } $, for $i=1, \cdots, n$, 
               \STATE $\bullet$ update  ${\rm E}(\Lambda_j^{-1}) \leftarrow  \left(\frac{1}{2}{\rm E}(\Xi^{-1})d_{jj}^{(H)}+{\rm E}(\lambda^{-1})\right)^{-1}$, for $j=0,\ldots,p-1$, 
               \STATE $\bullet$ update  ${\rm E}(\Xi^{-1}) \leftarrow  (p+1)/2\left(\frac{1}{2}\sum_{j=0}^{p}{\rm E}(\Lambda_j^{-1})d_{jj}^{(H)}+{\rm E}(\xi^{-1})\right)^{-1}$ 
               \STATE $\bullet$ update  ${\rm E}(\lambda^{-1}) \leftarrow  \left(2\sum_{j=0}^{p-1}{\rm E}(\Lambda_j^{-1})+\frac{2}{\mathcal{M}_{\lambda}}\right)^{-1}$ 
       \STATE $\bullet$ update  ${\rm E}(\xi^{-1}) \leftarrow  \left(2{\rm E}(\Xi^{-1})+\frac{2}{\mathcal{M}_{\xi}}\right)^{-1}$ 
                    \STATE  $\bullet$ update  $ {\rm E}(b^{-1}) \leftarrow 
(\mathcal{M}_{\kappa}^{-1} + \mu_{\kappa})^{-1}$, 
		\STATE $\bullet$ calculate ELBO from \eqref{elbo32}
            \ENDWHILE
    \end{algorithmic}
\end{algorithm}

We aim to compare the CS-VB-NB and HS-VB-NB models with their corresponding VB models in the Poisson regression models, say, CS-VB-POISS and HS-VB-POISS models. The CS-VB-POISS model is introduced by \cite{kharabati26} (called CS-VB model by the authors). We have also introduced HS-VB-POISS model in the supplementary material. 

\subsection{Sparsity-enforcing thresholds}

Due to the structure of the VB methods, the posterior means of the coefficients (${ \boldsymbol\mu}_{ \boldsymbol\beta(M)}$, $H, C$) are not sparse estimators. To obtain a sparse estimator, we propose sparsity-enforcing thresholds.

We use the hard threshold (subset selection) strategy \citep[see e.g.,][]{dj94}, that is 
$$\widehat{\beta}_j^M = \left\{
\begin{array}{lr}
{ \boldsymbol\mu}_{{ \boldsymbol\beta}(M)j}, & |{ \boldsymbol\mu}_{{ \boldsymbol\beta}(M)j}|>\hat{\kappa},\\
0, & {\rm o.w.,} 
\end{array}
\right.$$
for $M = H, C$, where $\hat{\kappa}$ is optimally selected over a grid of values by minimizing Akaike Information Criterion 
$${\rm AIC} = -\log \widehat{L}({ \boldsymbol\beta}|x) + 2 {\rm df},$$
where $\widehat{L}({ \boldsymbol\beta}|x)$ is the estimated likelihood of the model and ${\rm df}$ is the {degrees of freedom} of the model, that is, the number of nonzero coefficients in the regression model. 

The AIC serves as an efficient proxy for leave-one-out cross-validation, with its primary advantage lying in predictive accuracy. We opted for AIC over BIC because the latter tends to favor overly simplified models in practical settings, particularly when the data-generating process is intricate and not well-represented by any candidate model. This inclination toward simpler structures often results in excessive bias and diminished forecasting ability. In the context of high-dimensional variable screening, exploring every possible subset is infeasible. The AIC offered a fast, deterministic alternative to computationally expensive resampling techniques, which would have been impractical given the scale of our problem. Within our methodological framework, we implement a crisp decision rule for variable retention rather than relying on continuous penalties or probabilistic inclusion. While smooth shrinkage approaches effectively reduce dimensionality, they inevitably compress active effects, distorting their magnitudes. Meanwhile, the variational Bayes approximations we employ tend to produce overly confident posterior distributions, compromising the reliability of inclusion probabilities as a basis for selection. Consequently, imposing a rigid cutoff on the estimated coefficients emerges as the most natural and transparent approach for deriving a final predictive model, delivering unambiguous variable choices while avoiding the attenuation that accompanies alternative regularization strategies.

\subsection{Posterior predictive mass function}

The posterior predictive mass function (ppmf) of the response variable associated with the covariates of a new sample $x_0$ given the training data set $(X,y)$ is obtained as follows
\begin{align*}
p(y_0|x_0,X,y) & = \int p(y_0|x_0,{ \boldsymbol\theta}) p({ \boldsymbol\theta} | X,y) d{ \boldsymbol\theta} \\
& \approx \int {p}(y_0|x_0,{ \boldsymbol\theta}) q({ \boldsymbol\theta}) d{ \boldsymbol\theta},
\end{align*}
where $p(y_0|x_0,{ \boldsymbol\theta})$ is the negative binomial likelihood, and ${ \boldsymbol\theta} = ({ \boldsymbol\beta}, \kappa)^\top$. 

Thus, we have 
\begin{align*}
p(y_0|x_0,X,y) & \approx \int  \int \frac{\Gamma(y_0+\kappa)\kappa^\kappa \exp\{y_0 x_0 { \boldsymbol\beta}\}}{\Gamma(\kappa)\Gamma(y_0+1)(\kappa +  \exp\{x_0 { \boldsymbol\beta}\})^{y_0+\kappa}}
q({ \boldsymbol\beta}) q(\kappa) d{ \boldsymbol\beta} d\kappa,
\end{align*}

To compute the integral with respect to $\kappa$, we apply a second-order Taylor expansion of the integrand around the mean of the variational distribution $q(\kappa)$, denoted by $\mu_\kappa$. Define the function
\begin{align*}
g(\kappa| y_0, z_0 = e^{x_0 \boldsymbol\beta}) = \frac{\Gamma(y_0+\kappa)\kappa^\kappa}{\Gamma(\kappa)(\kappa + z_0)^{y_0+\kappa}}.
\end{align*}

Expanding $g(\kappa| y_0, z_0)$ around $\mu_\kappa$ up to second order, and substituting this expansion into the integral with respect to $q(\kappa)$ and using the properties of the variational distribution, we obtain
\begin{align*}
\int g(\kappa| y_0, z_0) q(\kappa) d\kappa 
& \approx g(\mu_\kappa| y_0, z_0) \int q(\kappa) d\kappa 
+ g'(\mu_\kappa| y_0, z_0) \int (\kappa - \mu_\kappa) q(\kappa) d\kappa \\
& \qquad + \frac{1}{2} g''(\mu_\kappa| y_0, z_0) \int (\kappa - \mu_\kappa)^2 q(\kappa) d\kappa\\
& =  g(\mu_\kappa| y_0, z_0) + \frac{1}{2} g''(\mu_\kappa| y_0, z_0) \sigma^2_\kappa,
\end{align*}
where $\sigma^2_\kappa = \int (\kappa - \mu_\kappa)^2 q(\kappa) d\kappa = \dfrac{\mathcal{H} (\frac{3}{2}, 0, 1, n , C^{(M)} )}{\mathcal{H}  (-\frac{1}{2} , 0, 1, n , C^{(M)} )} - \mu_{\kappa}^2 $, for $M = H, C$. The second derivative of $g(\kappa)$ with respect to $\kappa$, evaluated at $\mu_\kappa$, is given by
\begin{align*}
g''(\mu_\kappa| y_0, z_0) = g(\mu_\kappa| y_0, z_0) \Bigg\{ & \left[ \psi(y_0+\mu_\kappa) - \psi(\mu_\kappa) + \log \mu_\kappa + 1 - \log(\mu_\kappa +  z_0) - \frac{y_0+\mu_\kappa}{\mu_\kappa +  z_0} \right]^2 \\
& + \psi_1(y_0+\mu_\kappa) - \psi_1(\mu_\kappa) + \frac{1}{\mu_\kappa} - \frac{\mu_\kappa + 2z_0 - y_0}{(\mu_\kappa + z_0)^2} \Bigg\},
\end{align*}
where $\psi(\cdot)$, $\psi_1(\cdot)$ denote the digamma and trigamma functions, respectively. 

Consequently, using re-parametrization $z_0 = e^{x_0{ \boldsymbol\beta}}$, we see that, for $M = H, C$, the posterior predictive mass function can be approximated as
\begin{align}
p(y_0|x_0,X,y) & \approx \frac{1}{y_0 !} \int z_0^{y_0} \left(g(\mu_\kappa| y_0, z_0) + \frac{1}{2} g''(\mu_\kappa| y_0, z_0) \sigma^2_\kappa\right)\nonumber\\
&\qquad \qquad \qquad {\rm LN}(z_0; x_0{ \boldsymbol\mu}_{{ \boldsymbol\beta}(M)}, x_0{ \boldsymbol\Sigma}_{{ \boldsymbol\beta}(M)}x_0^\top) \;dz_0,\label{Lpd}
\end{align}
where ${\rm LN}(z_0; \mu_0, \sigma^2_0)$ stands for the pdf of the log-normal distribution. The point predictor of $y_0$ is computed as 
$$\hat{y_0} = \arg\max_{y_0} p(y_0|x_0,X,y).$$

\section{Simulation study}

{ To examine} the performance of the proposed models, we have conducted a simulation study as follows. In each of $N=1000$ replications, we have generated samples of size $n$, {from} a sparse negative binomial regression model ($y_i\sim {\rm NB}(r,r/(r+\exp\{\mathbf{X}_i  { \boldsymbol\beta}  \}), \; i=1,\ldots,n$), with $r=3$, and vector of coefficients ${ \boldsymbol\beta}_{p\times 1}$, whose elements are first generated from $N(\mu_0,\sigma_0)$ and then {the vector is multiplied by a vector $\mathbf{z}$ of length $p$ with 0, 1, and -1 elements, to set some elements of ${ \boldsymbol\beta}$ to zero, while determining the sign of the non-zero elements.} The rows of the covariate matrix $X$ are generated independently from a multivariate normal distribution with {mean $\mu_X$} and a variance-covariance matrix with elements ${\Sigma}_{ij} = \sigma^2_X \cdot 0.3^{|i-j|}$, to model a slight multicollinearity. We consider the cases $n=30,100$, and $p=10,200$, 
to successfully cover the low- and high-dimensional scenarios. We have also considered a scenario in which the data is generated from the Poisson regression model ($y_i\sim {\rm Poiss}(\exp\{\mathbf{X}_i  { \boldsymbol\beta}  \}), \; i=1,\ldots,n$), which is disscussed in the supplementary material. For the cases with $p=10$, we let $\mu_0 = 1.5, \sigma_0 = 0.3, \mu_X = 0.1, \sigma^2_X = 1$, and we set $\mathbf{z} = (1, 0,1,0,0,0,-1,0,1,0)^\top$, and for the cases with $p = 200$, we let $\mu_0 = 0.1, \sigma_0 = 0.6, \mu_X = 0.1, \sigma^2_X =0.05$, and we randomly choose 60 (including the first one) out of 200 values of $\mathbf{z}$ to be equal 1, and the remaining are set to 0. The case $n=100, \; p = 10$, called the low-dimensional scenario, and the case $n=30, \; p=200$, called the high-dimensional scenario, are studied in the following subsections, while the two remaining cases ($n=30,\;p=10$ and $n=100,\;p=200$) are given in the supplementary material for the sake of brevity.

Two proposed VB models (HS, and CS) for the competitor Negative-Binomial (NB) and Poisson models, are fitted to each generated dataset. Furthermore, two frequentist alternatives, LASSO and SCAD-penalized NB and Poisson regression models, are considered for comparison. The sparsity parameters of both LASSO and SCAD models are optimized based on the Corrected AIC 
$${\rm AICc} = -\log \widehat{L}({ \boldsymbol\beta}|x) + 2 {\rm df} + 2{\rm df}({\rm df}+1)/(n-{\rm df}-1),$$
where $\widehat{L}({ \boldsymbol\beta}|x)$ is the estimated likelihood of the model and ${\rm df}$ is the {degrees of freedom} of the model. 

To compare the performance of the proposed models with each other and with the other competitor models, the regression coefficient relative errors of methods $M = H, C$ are computed as follows
\begin{equation}
\text{CRE} = \frac{\frac{1}{N}\sum_{t = 1}^N (\widehat{\boldsymbol\beta}^M - \boldsymbol\beta)^\top (\widehat{\boldsymbol\beta}^M - \boldsymbol\beta)}{\frac{1}{N}\sum_{t = 1}^N \boldsymbol\beta^\top \boldsymbol\beta}.
\end{equation}

Furthermore, the train-set and test-set relative errors are obtained, respectively, as
\begin{equation}
\text{TRRE} = \frac{\frac{1}{N}\sum_{t = 1}^N (\widehat{y}_{t}^{\text{train}} - y_{t}^{\text{train}})^\top (\widehat{y}_{t}^{\text{train}} - y_{t}^{\text{train}})}{\frac{1}{N}\sum_{t = 1}^N (y_{t}^{\text{train}} - \bar{y}_{t}^{\text{train}})^\top (y_{t}^{\text{train}} - \bar{y}_{t}^{\text{train}})},
\end{equation}
and
\begin{equation}
\text{TSRE} = \frac{\frac{1}{N}\sum_{t = 1}^N (\widehat{y}_{t}^{\text{test}} - y_{t}^{\text{test}})^\top (\widehat{y}_{t}^{\text{test}} - y_{t}^{\text{test}})}{\frac{1}{N}\sum_{t = 1}^N (y_{t}^{\text{test}} - \bar{y}_{t}^{\text{test}})^\top (y_{t}^{\text{test}} - \bar{y}_{t}^{\text{test}})},
\end{equation}
where $y_{t}^{\text{train}} = (y_{1t},\ldots,y_{nt})^\top$ and $y_{t}^{\text{test}} = (y'_{1t},\ldots,y'_{n't})^\top$ are the train-set (80\% of the full sample) and test-set (20\% of the full sample) response samples, respectively, generated during the $t$th iteration, with $\bar{y}_{t}^{\text{train}} = \frac{1}{n}\sum_{i=1}^n y_{it}$ and $\bar{y}_{t}^{\text{test}} = \frac{1}{n'}\sum_{i=1}^{n'} y'_{it}$.

To evaluate the full predictive performance, we employ the Continuous Ranked Probability Score (CRPS), which is a proper scoring rule that measures the discrepancy between the predicted cumulative distribution function and the observed value. For a discrete response variable following a negative binomial distribution, the CRPS is defined as
\begin{equation}
\text{CRPS}(F_{y_0}, y_0^{\text{obs}}) = \sum_{k=0}^{\infty} \left( F_{y_0}(k) - \mathbf{1}\{y_0^{\text{obs}} \leq k\} \right)^2,
\end{equation}
where $F_{y_0}(k) = p(y_0 \leq k \mid x_0, X, y)$ is the predictive cumulative distribution function evaluated at $k$, and $\mathbf{1}\{\cdot\}$ is the indicator function. The predictive distribution is obtained from the posterior predictive mass function (ppmf) derived in Section 3.3. The cumulative distribution function is then obtained by summing the ppmf over $k = 0, 1, \ldots, K$ for a sufficiently large $K$. 

For the MCMC benchmarks, in the low-dimensional scenarios, the CRPS is computed using the equivalent representation
\begin{equation}
\text{CRPS}(F, y_0^{\text{obs}}) = \mathbb{E}_{Y \sim F}|Y - y_0^{\text{obs}}| - \frac{1}{2}\mathbb{E}_{Y, Y' \sim F}|Y - Y'|,
\end{equation}
where $Y$ and $Y'$ are independent draws from the predictive distribution $F$. This formulation allows for straightforward computation from MCMC samples: for each test observation, we draw $S$ samples from the posterior predictive distribution using the MCMC output, and compute
\begin{equation}
\widehat{\text{CRPS}} = \frac{1}{S}\sum_{s=1}^S |Y_s - y_0^{\text{obs}}| - \frac{1}{2S^2}\sum_{s=1}^S\sum_{s'=1}^S |Y_s - Y_{s'}|.
\end{equation}
This estimator is unbiased and consistent for the true CRPS, providing a direct comparison between the VB approximations and the MCMC benchmarks.

The false negative rate and the false positive rate are also defined as follows
\begin{equation}
\text{FNR} = \frac{\#\{j: 1 \leq j \leq p,\; \beta_j \neq 0,\; \hat{\beta}^M_j = 0\}}{\#\{j: 1 \leq j \leq p,\; \beta_j \neq 0\}},
\end{equation}
\begin{equation}
\text{FPR} = \frac{\#\{j: 1 \leq j \leq p,\; \beta_j = 0,\; \hat{\beta}^M_j \neq 0\}}{\#\{j: 1 \leq j \leq p,\; \beta_j = 0\}},
\end{equation}
for $M = H, C$, where $\#A$ stands for the cardinality of the set $A$. These criteria are used to examine the sparsity performance of the competitive methods. For the VB methods, the hard-thresholding procedure described in Section 3.1 is applied to obtain point estimates $\hat{\beta}^M_j$, while for the MCMC benchmarks, the posterior means are used.
The computation time of all competitor models are also computed in each iteration. 

The hyper-parameters ${\alpha}_1$ and ${\alpha}_2$ for the CS-VB-NB method are both set to 1. The hyperparameters $\mathcal{M}_{\eta}$ is set to 0.01 for $\eta = \kappa, \sigma^2, \lambda, \xi$. 

{ \subsection{Low dimensional scenario}

Figures \ref{box1} -- \ref{box3} show the box plots of all aforementioned criteria for all competitive methods, for the low-dimensional scenario, with {$p=10$, $n=100$, and $\mathbf{z} = (1,0,1,0,1)^\top$}, comparing the proposed variational Bayes methods (HS-VB-NB and CS-VB-NB) with their Poisson counterparts (HS-VB-POISS and CS-VB-POISS), frequentist penalized methods (LASSO-NB, SCAD-NB, LASSO-POISS, SCAD-POISS), and MCMC benchmarks (HS-MCMC-NB and CS-MCMC-NB). In implementing the MCMC methods, we utilized the R package rjags within R (version 4.5.2), in conjunction with JAGS (version 4.3.1). Each chain comprised 10,000 iterations, with a burn-in of 5,000 and a thinning interval of 10. The Gelman-Rubin diagnostics are computed for each parameter and reported in the supplementary material. These values confirm the convergence of the MCMC algorithm for most parameters.

Figure \ref{box1} displays the coefficient relative error (CRE), train relative error (TRRE), and test relative error (TSRE) across all methods. The negative binomial VB methods (HS-VB-NB and CS-VB-NB) achieve CRE values that are comparable to their MCMC counterparts, with median CRE around 0.01, demonstrating that the variational approximation does not substantially degrade estimation accuracy. In contrast, the Poisson-based VB methods exhibit considerably larger CRE values, with medians ranging from 0.02 to 0.04, reflecting the impact of model misspecification when the data are generated from a negative binomial distribution. The frequentist LASSO-NB and SCAD-NB methods show competitive performance, though with slightly higher variability than the VB approaches. 

The train relative error (TRRE) and test relative error (TSRE) plots indicate that the Poisson models tend to overfit the training data. Nevertheless, the TSRE values are comparable across all models, suggesting that they perform similarly in predicting test set responses, at least in the low-dimensional setting.

Note that for the computation of TSRE values for the LASSO and SCAD methods, the prediction of the response is computed as 
$$\hat{y}_i = \exp(X_{\rm test}\hat{{ \boldsymbol\beta}}_{\rm sparse}),$$
in which $\hat{{ \boldsymbol\beta}}_{\rm sparse}$ is the sparse estimator obtained by the LASSO and SCAD methods, while the TSRE values for the VB methods are computed using the ppmf functions \eqref{Lpd}. The mode of \eqref{Lpd} is considered as the point predictor of the response value. 

Figure \ref{box2} presents the false positive rate (FPR), false negative rate (FNR), and the Continuous Ranked Probability Score (CRPS). The VB methods tend to exhibit slightly higher false positive rates (FPR) compared to their MCMC counterparts, while all models achieve low false negative rates (FNR). This suggests that the variational approximations are somewhat more liberal in selecting predictors, potentially including a few extra noise variables, but they successfully recover the true active set without missing important signals. The low FNR across all methods indicates that the sparsity-inducing priors effectively identify the relevant predictors, while the moderate FPR values reflect the inherent trade-off between false discoveries and true positives in high-dimensional variable selection.

The CRPS, which evaluates the full predictive distribution, is substantially lower for the HS-VB-NB method compared to other methods, while almost all methods achieve similar CRPS values. This confirms that the HS-VB-NB method provides better-calibrated predictive distributions, though the differences among most methods are not substantial.

Figure \ref{box3} shows the relative computation time of each method compared to CS-MCMC-NB. 
The variational Bayes methods are dramatically faster than MCMC, with HS-VB-NB and CS-VB-NB requiring less than 0.1\% of the computation time of CS-MCMC-NB. These results highlight the primary advantage of the VB approach: it achieves accuracy comparable to MCMC at a fraction of the computational cost, making it feasible for large-scale applications where MCMC would be prohibitively expensive. Note that this comparison is not yet fair, since the MCMC methods use {C++} programming for accelerating the computation, while we have done all computations in the R software. So, it seems that the VB methods are very fast, as expected. All computations were performed using R version 4.5.2 on a machine with a Core i5-10210U CPU.

Table \ref{cover-nb} presents the average (standard deviation) of the coverage probabilities of the highest posterior density (HPD) confidence intervals for the regression coefficients at level 0.95. The VB methods under the negative binomial model (HS-VB-NB and CS-VB-NB) achieve coverage probabilities close to the nominal level, ranging from 0.79 to 0.92, and closely replicate the MCMC benchmarks (HS-MCMC-NB and CS-MCMC-NB) which attain coverage near or above 0.95. In stark contrast, the Poisson-based VB methods (HS-VB-POISS and CS-VB-POISS) exhibit severely poor coverage probabilities between 0.14 and 0.29, with substantially larger standard deviations, indicating that the Poisson model is misspecified for these overdispersed data. 

\begin{table}[htbp]
\centering
\caption{Average (standard deviation) of coverage probabilities of the HPD intervals for the regression coefficients at level 0.95.}
\label{cover-nb}
\begin{tabular}{lcccccc}
\toprule
 & $\beta_0$ & $\beta_1$ & $\beta_2$ & $\beta_3$ & $\beta_4$ \\
\hline\hline
HS-VB-NB & 0.79 (0.41) & 0.92 (0.27) & 0.85 (0.36) & 0.86 (0.35) & 0.88 (0.32) \\
CS-VB-NB & 0.80 (0.40) & 0.91 (0.29) & 0.84 (0.37) & 0.85 (0.36) & 0.84 (0.37) \\
HS-VB-POISS & 0.24 (0.43) & 0.26 (0.44) & 0.17 (0.38) & 0.20 (0.40) & 0.28 (0.45) \\
CS-VB-POISS & 0.29 (0.45) & 0.25 (0.43) & 0.21 (0.41) & 0.17 (0.38) & 0.23 (0.42) \\
HS-MCMC-NB & 0.96 (0.20) & 0.99 (0.10) & 0.97 (0.17) & 0.95 (0.22) & 0.96 (0.20) \\
CS-MCMC-NB & 0.98 (0.14) & 1.00 (0.00) & 1.00 (0.00) & 0.98 (0.14) & 0.96 (0.20) \\
\hline\hline
\end{tabular}
\vspace{0.5cm}
\begin{tabular}{lcccccc}
 & $\beta_5$ & $\beta_6$ & $\beta_7$ & $\beta_8$ & $\beta_9$ \\
\hline\hline
HS-VB-NB & 0.89 (0.31) & 0.82 (0.38) & 0.88 (0.32) & 0.89 (0.31) & 0.91 (0.29) \\
CS-VB-NB & 0.86 (0.35) & 0.82 (0.38) & 0.85 (0.36) & 0.88 (0.32) & 0.88 (0.32) \\
HS-VB-POISS & 0.23 (0.42) & 0.14 (0.35) & 0.24 (0.43) & 0.20 (0.40) & 0.29 (0.45) \\
CS-VB-POISS & 0.22 (0.41) & 0.15 (0.36) & 0.19 (0.39) & 0.25 (0.43) & 0.22 (0.41) \\
HS-MCMC-NB & 1.00 (0.00) & 0.97 (0.17) & 0.95 (0.22) & 0.96 (0.20) & 0.99 (0.10) \\
CS-MCMC-NB & 1.00 (0.00) & 0.96 (0.20) & 0.98 (0.14) & 0.94 (0.24) & 0.96 (0.20) \\
\hline\hline
\end{tabular}
\end{table}

\begin{table}[htbp]
\centering
\caption{Average (standard error) of accuracy values for the parameters against an MCMC benchmark for different priors.}
\label{tab-acc-nb}
\begin{tabular}{lcccccc}
\toprule
 & Intercept & $\beta_1$ & $\beta_2$ & $\beta_3$ & $\beta_4$ \\
\hline\hline
HS-VB-NB & 76.41 (4.05) & 88.18 (5.05) & 82.43 (4.13) & 88.06 (5.39) & 87.60 (4.95) \\
CS-VB-NB & 76.96 (4.03) & 84.53 (9.84) & 81.81 (4.75) & 81.92 (10.33) & 82.56 (9.41) \\
\hline
 & $\beta_5$ & $\beta_6$ & $\beta_7$ & $\beta_8$ & $\beta_9$ \\
\hline
HS-VB-NB & 87.95 (4.88) & 82.70 (3.33) & 88.00 (5.30) & 81.12 (5.17) & 87.21 (4.12) \\
CS-VB-NB & 82.98 (9.26) & 81.76 (4.86) & 83.72 (9.22) & 80.58 (4.60) & 82.22 (7.48) \\
\hline\hline
\end{tabular}
\begin{tabular}{lcccccc}
 & $\sigma$ & ${Z}_1$ & ${Z}_2$ & ${Z}_3$ & ${Z}_4$ \\
\hline\hline
CS-VB-NB & 79.23 (6.69) & 14.16 (4.27) & 97.60 (3.07) & 14.20 (4.31) & 17.20 (7.64) \\
\hline
 & ${Z}_5$ & ${Z}_6$ & ${Z}_7$ & ${Z}_8$ & ${Z}_9$ \\
\hline
CS-VB-NB & 15.47 (4.55) & 98.48 (3.32) & 15.35 (7.60) & 98.27 (2.98) & 14.90 (4.52) \\
\hline\hline
\end{tabular}
\begin{tabular}{lcccccc}
 & & $\pi_1$ & $\pi_2$ & $\pi_3$ & $\pi_4$ \\
\hline\hline
CS-VB-NB & & 60.61 (2.50) & 94.64 (1.00) & 59.75 (2.38) & 60.55 (3.39) \\
\hline
 & $\pi_5$ & $\pi_6$ & $\pi_7$ & $\pi_8$ & $\pi_9$ \\
\hline
CS-VB-NB & 60.54 (1.94) & 94.62 (0.92) & 60.76 (3.59) & 94.82 (1.02) & 60.81 (2.81) \\
\hline\hline
\end{tabular}
\begin{tabular}{lcccccc}
 & & $\lambda_1$ & $\lambda_2$ & $\lambda_3$ & $\lambda_4$ \\
\hline\hline
CS-VB-NB & & 65.52 (0.96) & 65.18 (0.80) & 65.29 (0.79) & 65.43 (0.86) \\
\hline
 & $\lambda_5$ & $\lambda_6$ & $\lambda_7$ & $\lambda_8$ & $\lambda_9$ \\
\hline
CS-VB-NB & 65.42 (0.80) & 65.11 (0.78) & 65.48 (0.99) & 65.20 (0.85) & 65.54 (0.90) \\
\hline\hline
\end{tabular}
\end{table}

\begin{sidewaysfigure}
\centerline{\includegraphics[scale=0.6]{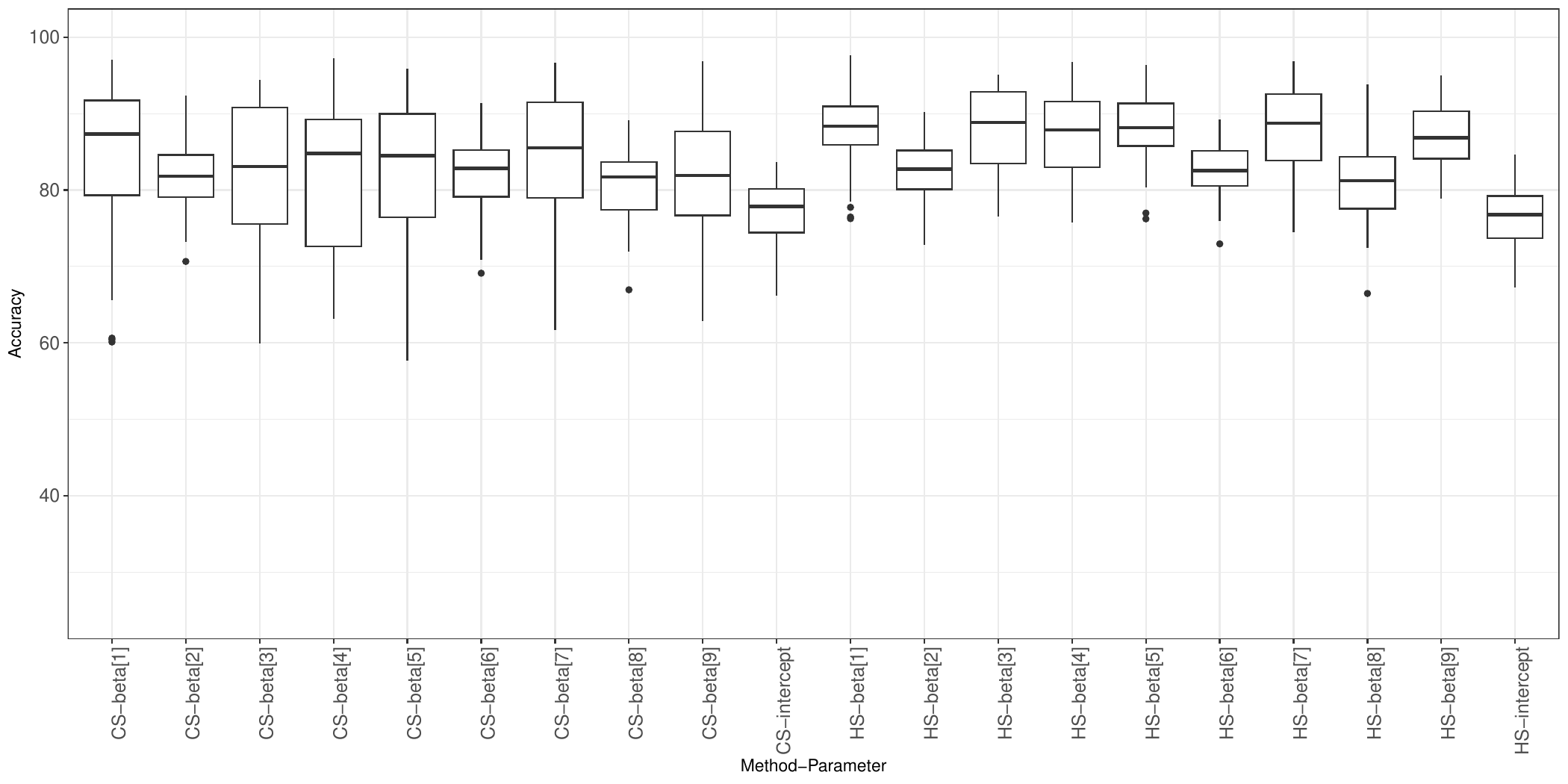}}
\caption{The boxplots of the accuracies for the {\color{black} regression} coefficients for different VB-methods.}\label{boxac1}
\end{sidewaysfigure}

\begin{figure}
\centerline{\includegraphics[scale=0.4]{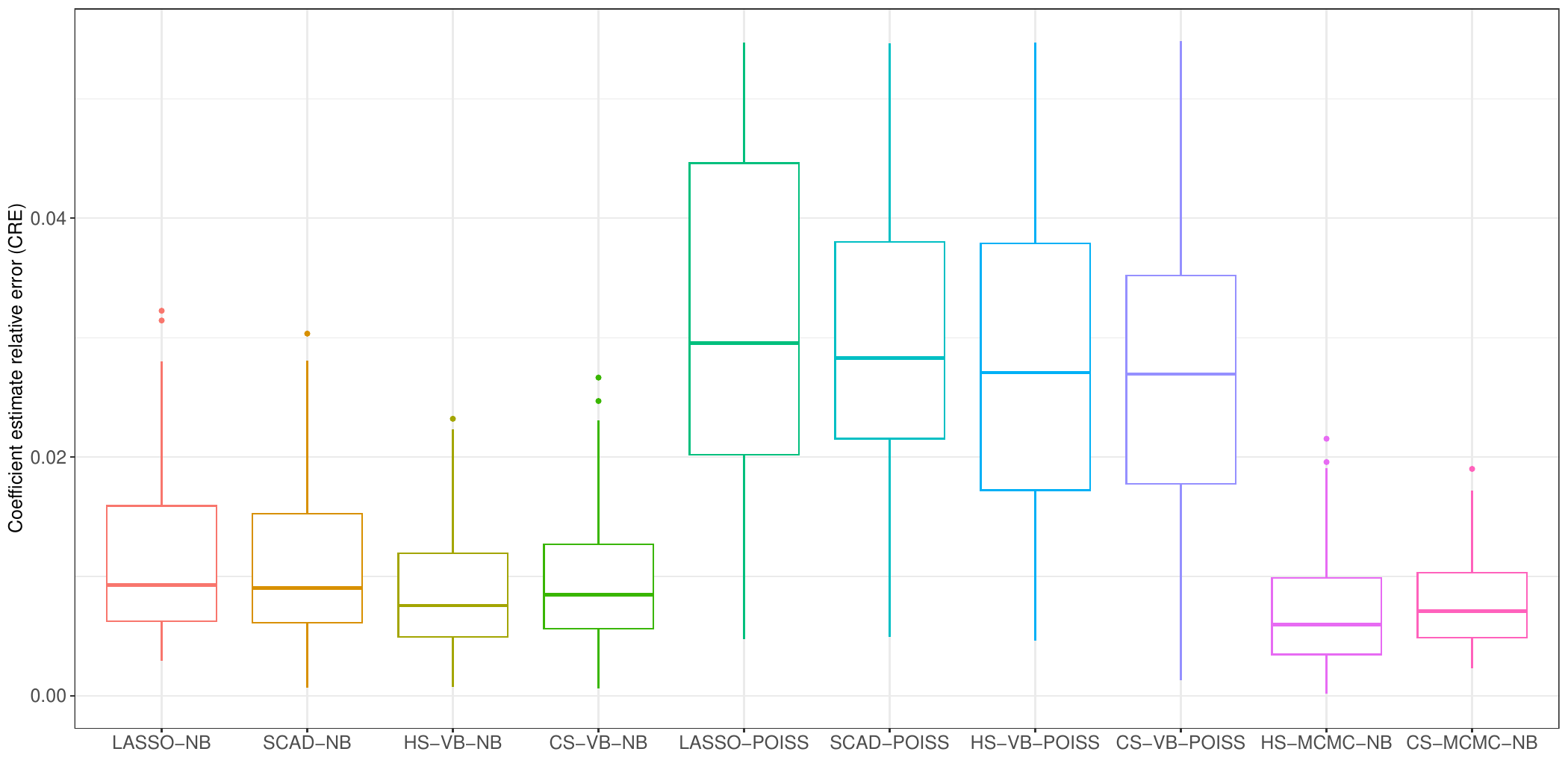}}
\centerline{\includegraphics[scale=0.4]{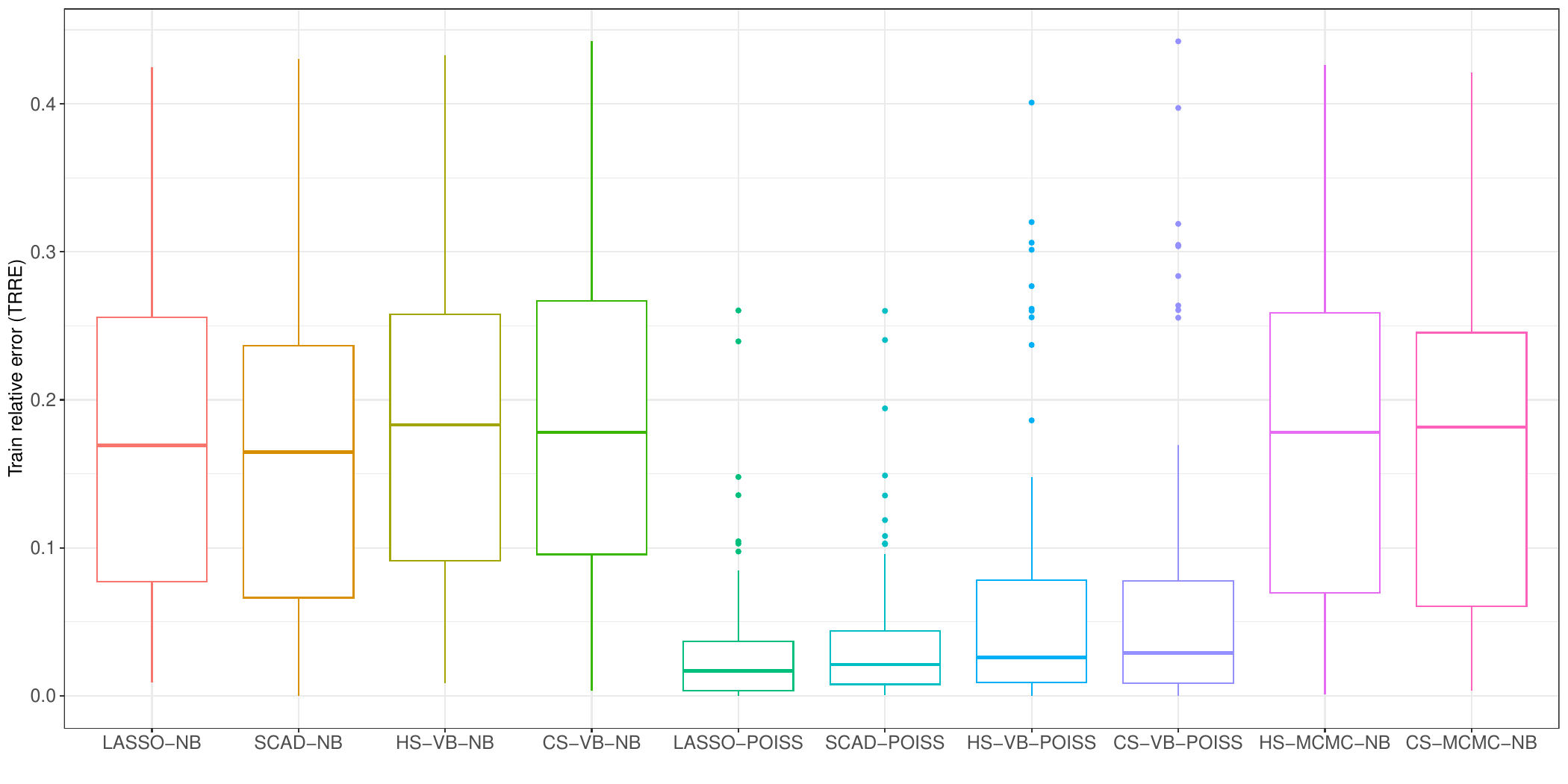}}
\centerline{\includegraphics[scale=0.4]{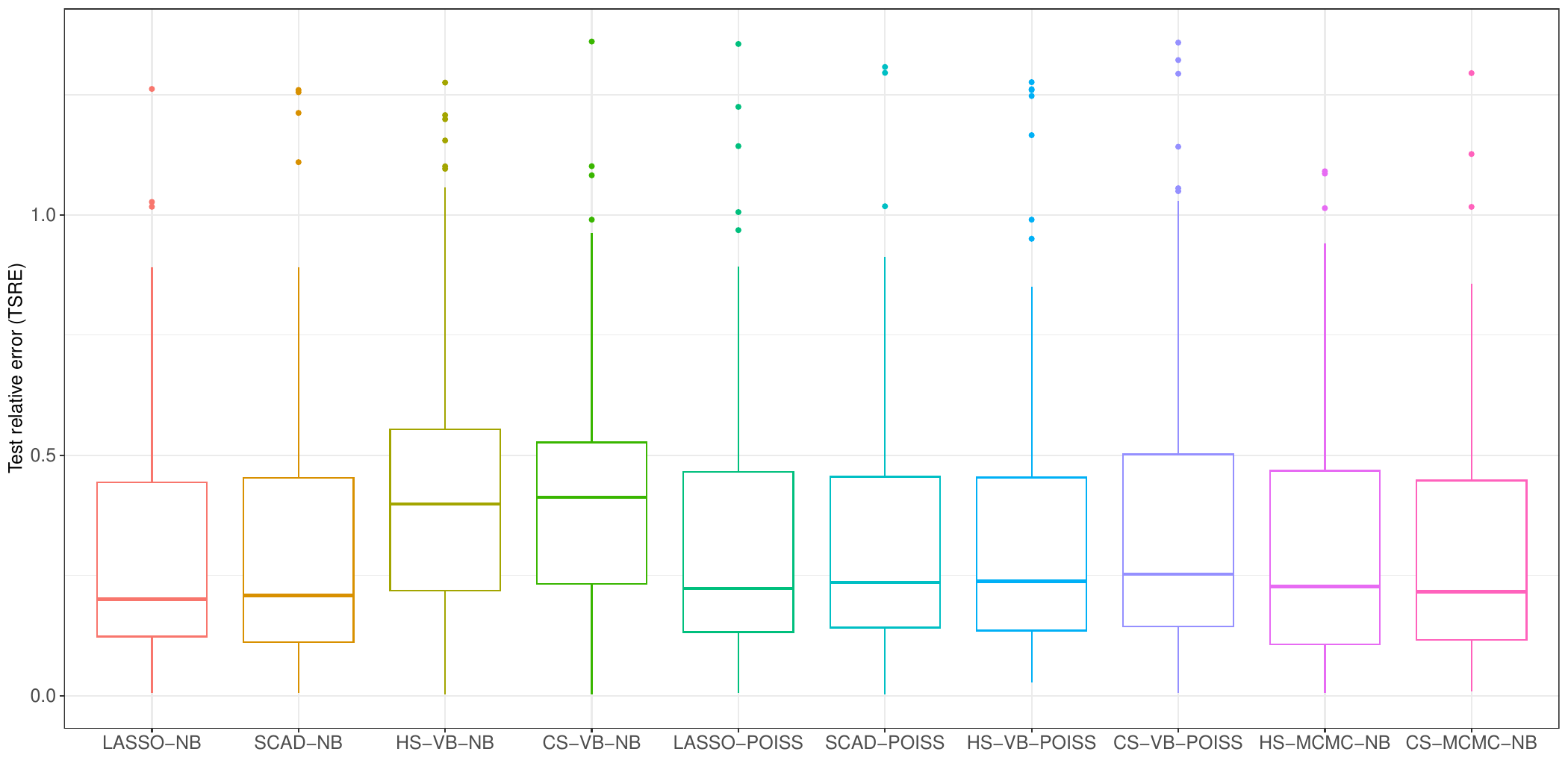}}
\caption{The low dimensional scenario simulation results: the coefficient relative error (top), the train relative error (middle), and the test relative error (bottom) for different methods.}\label{box1}
\end{figure}

\begin{figure}
\centerline{\includegraphics[scale=0.4]{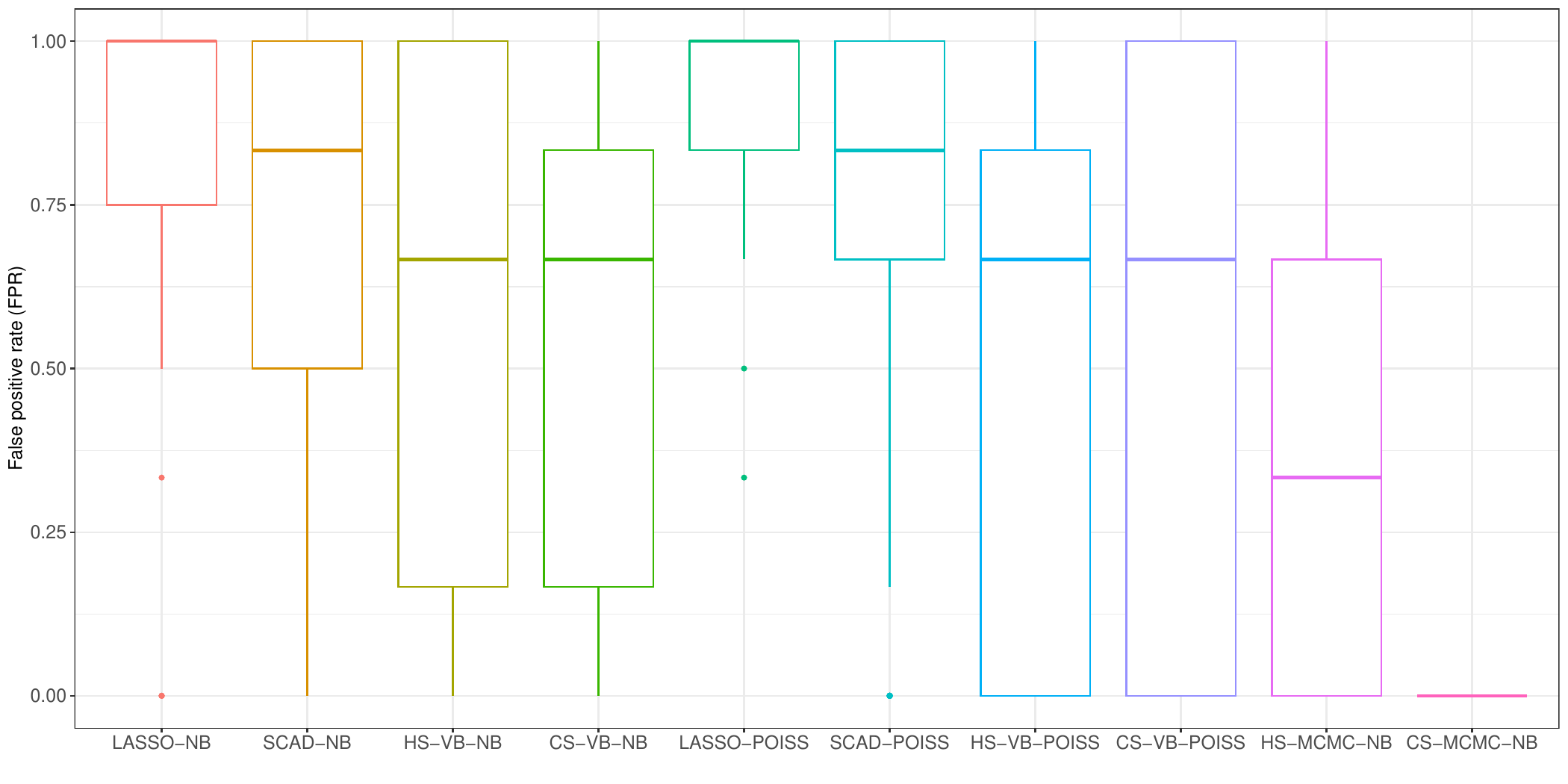}}
\centerline{\includegraphics[scale=0.4]{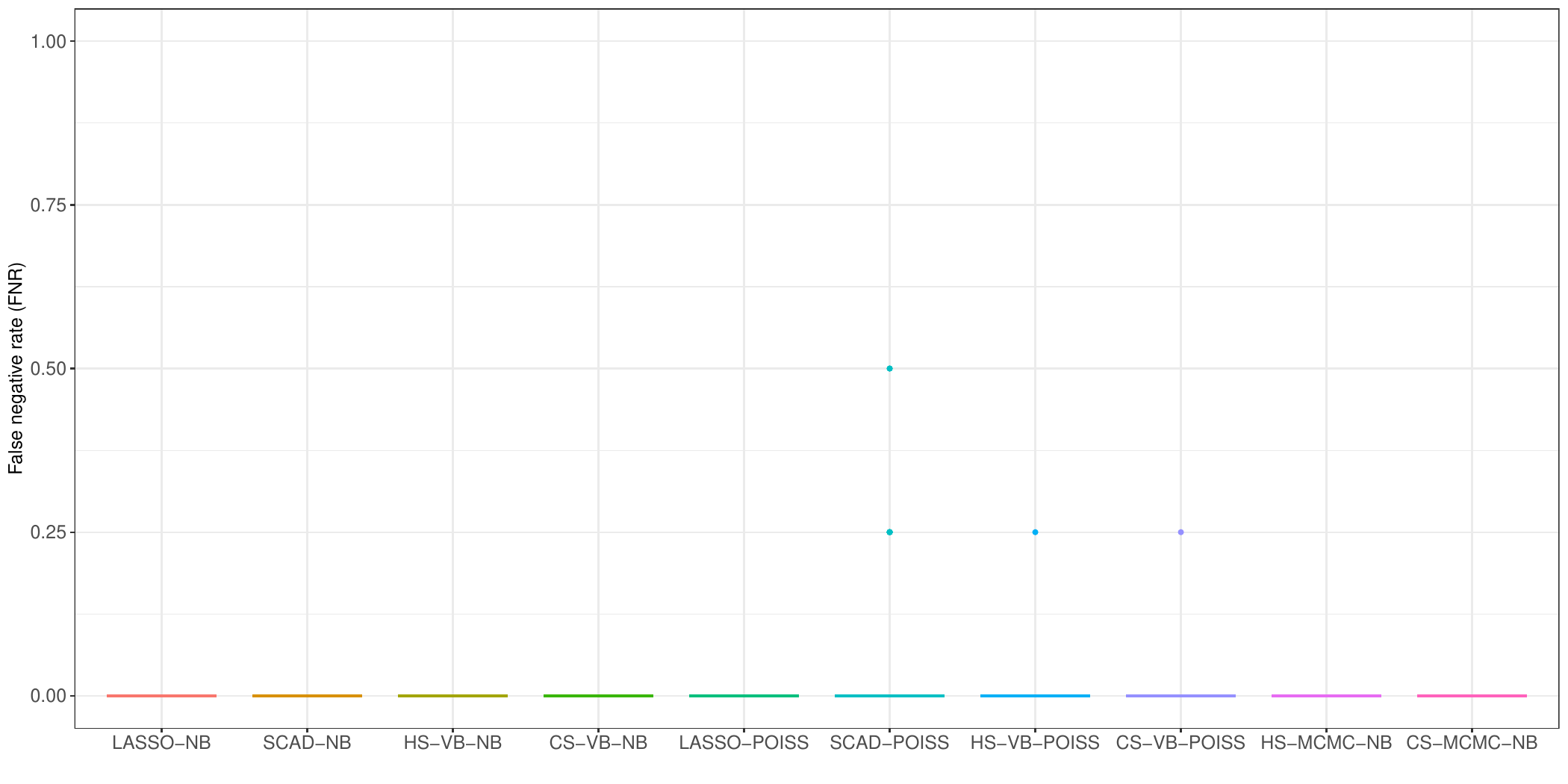}}
\centerline{\includegraphics[scale=0.4]{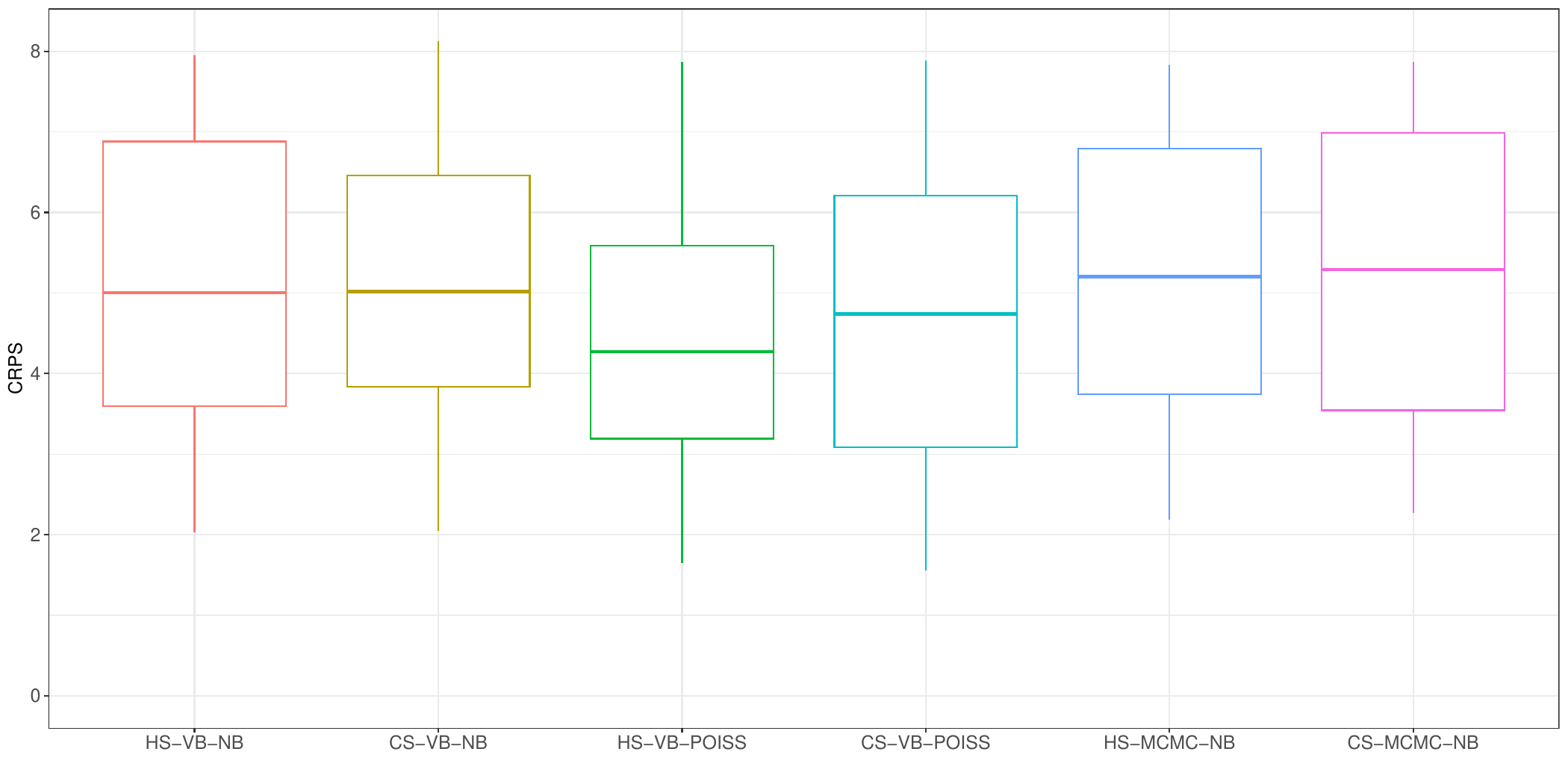}}
\caption{The low dimensional scenario simulation results: the FPR (top), the {\color{black} FNR} (middle), and the CRPS (bottom) for different methods.}\label{box2}
\end{figure}

\begin{figure}
\centerline{\includegraphics[scale=0.4]{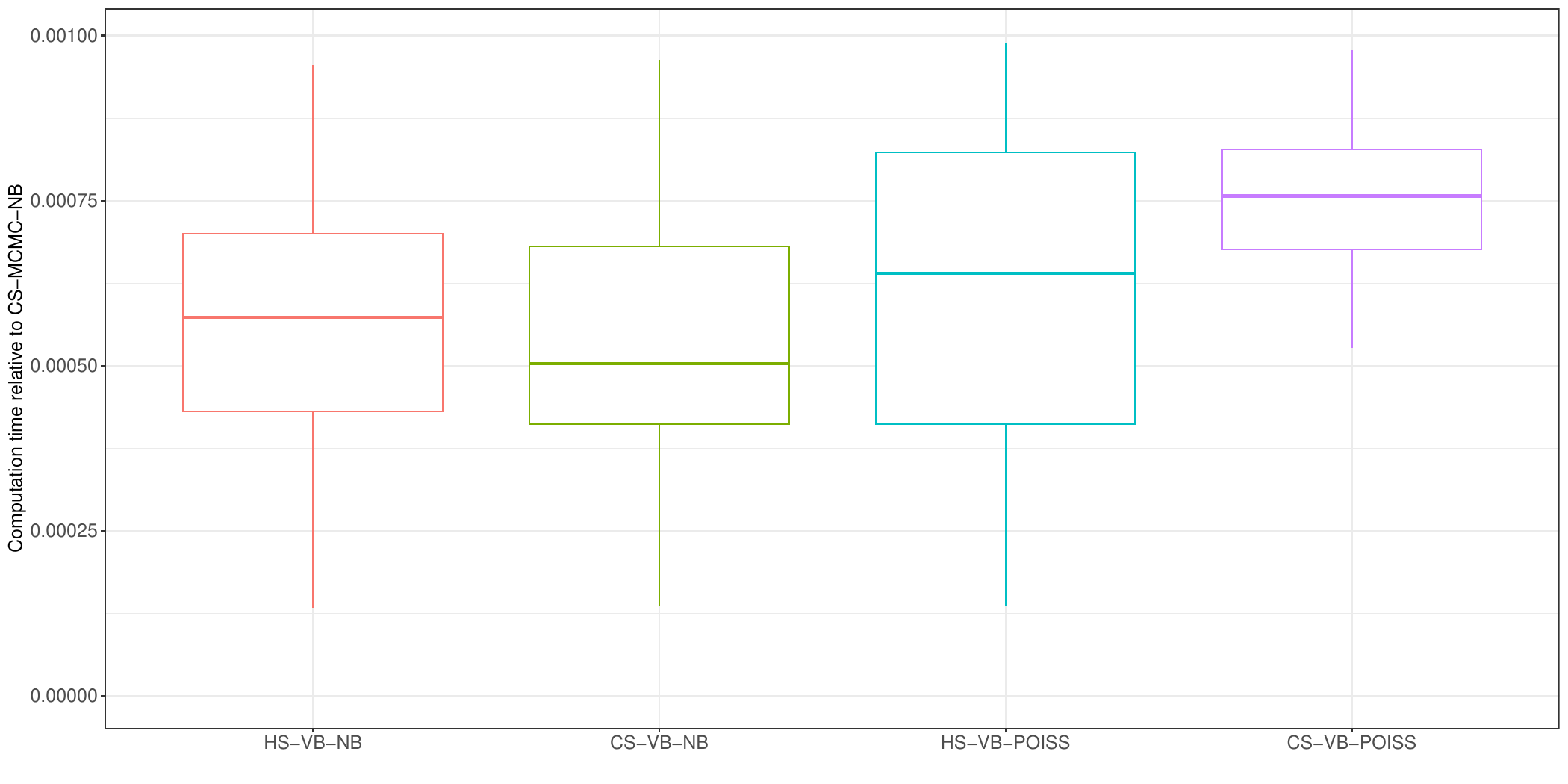}}
\caption{The low dimensional scenario relative computation time for different methods.}\label{box3}
\end{figure}

\begin{figure}
\centerline{\includegraphics[scale=0.4]{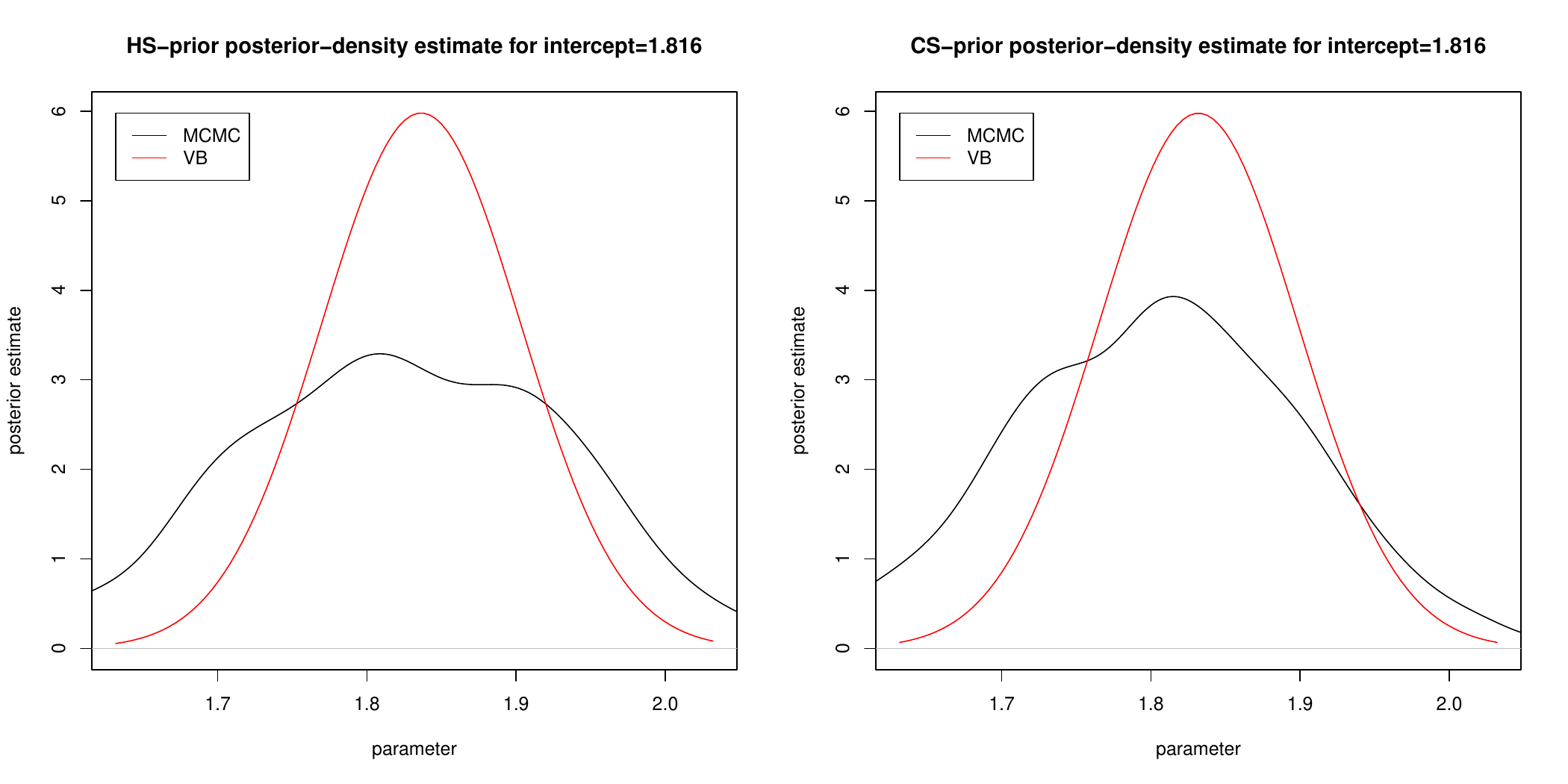}}
\centerline{\includegraphics[scale=0.4]{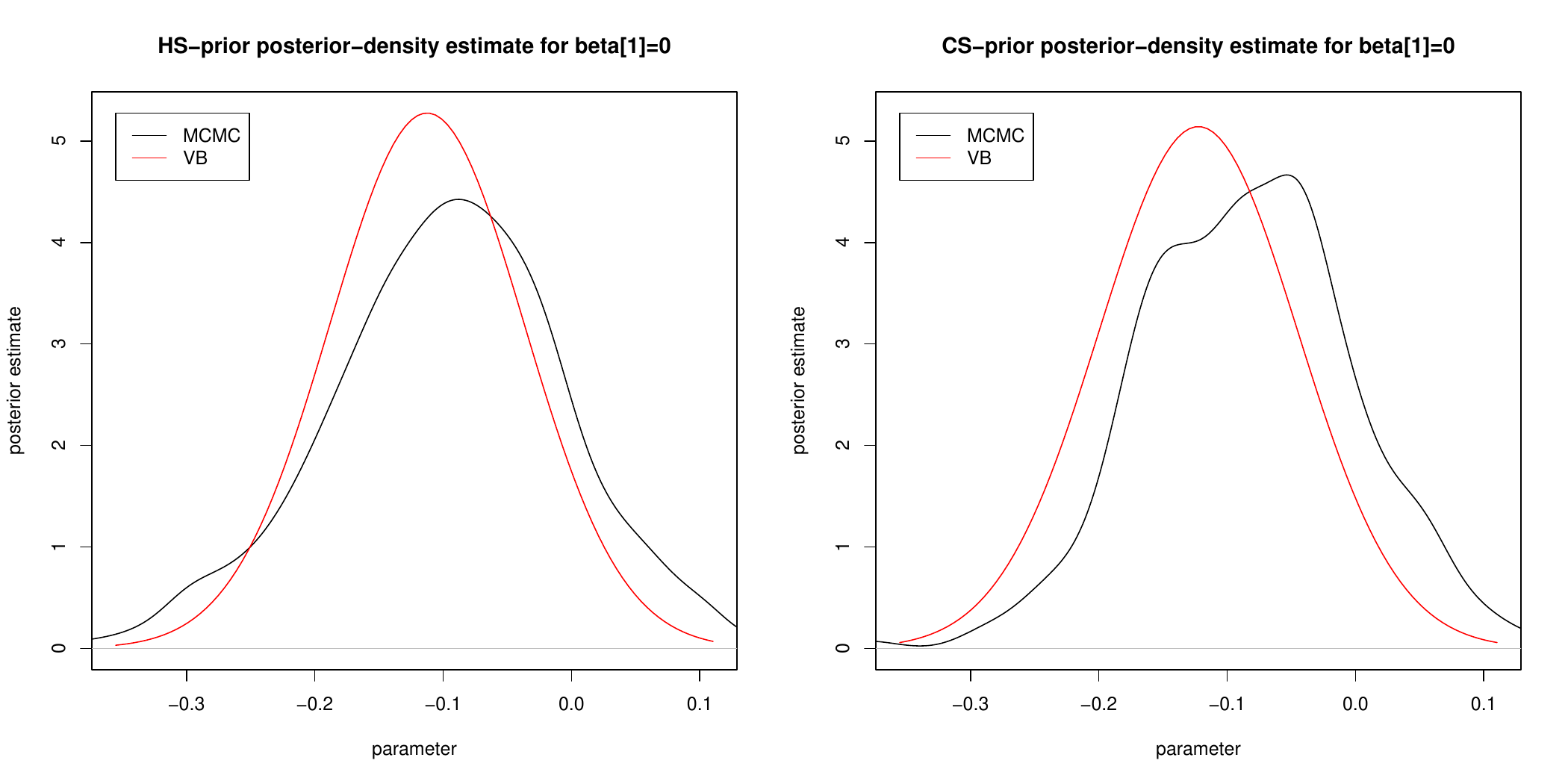}}
\centerline{\includegraphics[scale=0.4]{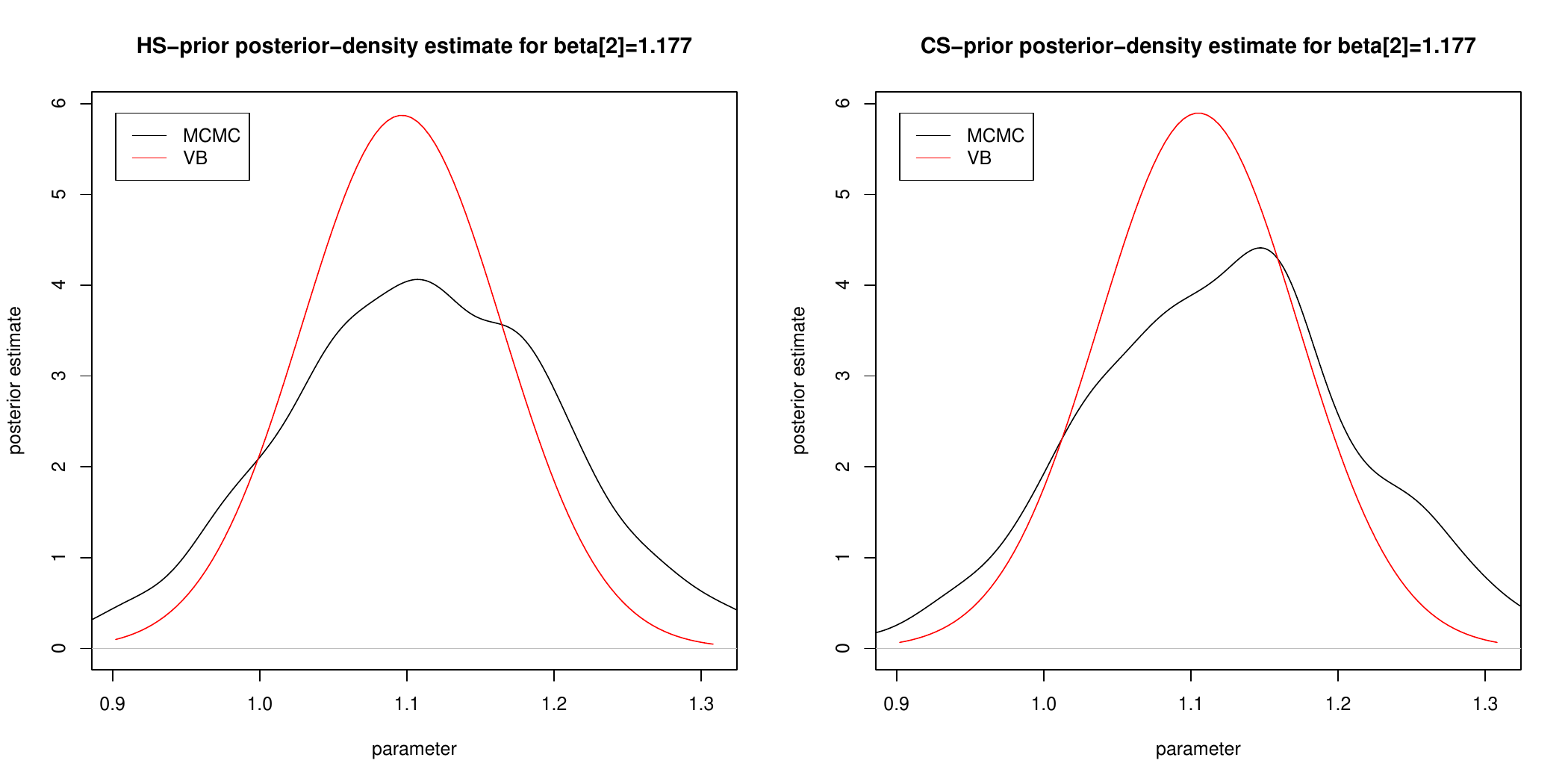}}
\caption{ Approximate posterior density functions of the first three regression coefficients for different models.}\label{dens1}
\end{figure}

\begin{figure}
\centerline{\includegraphics[scale=0.4]{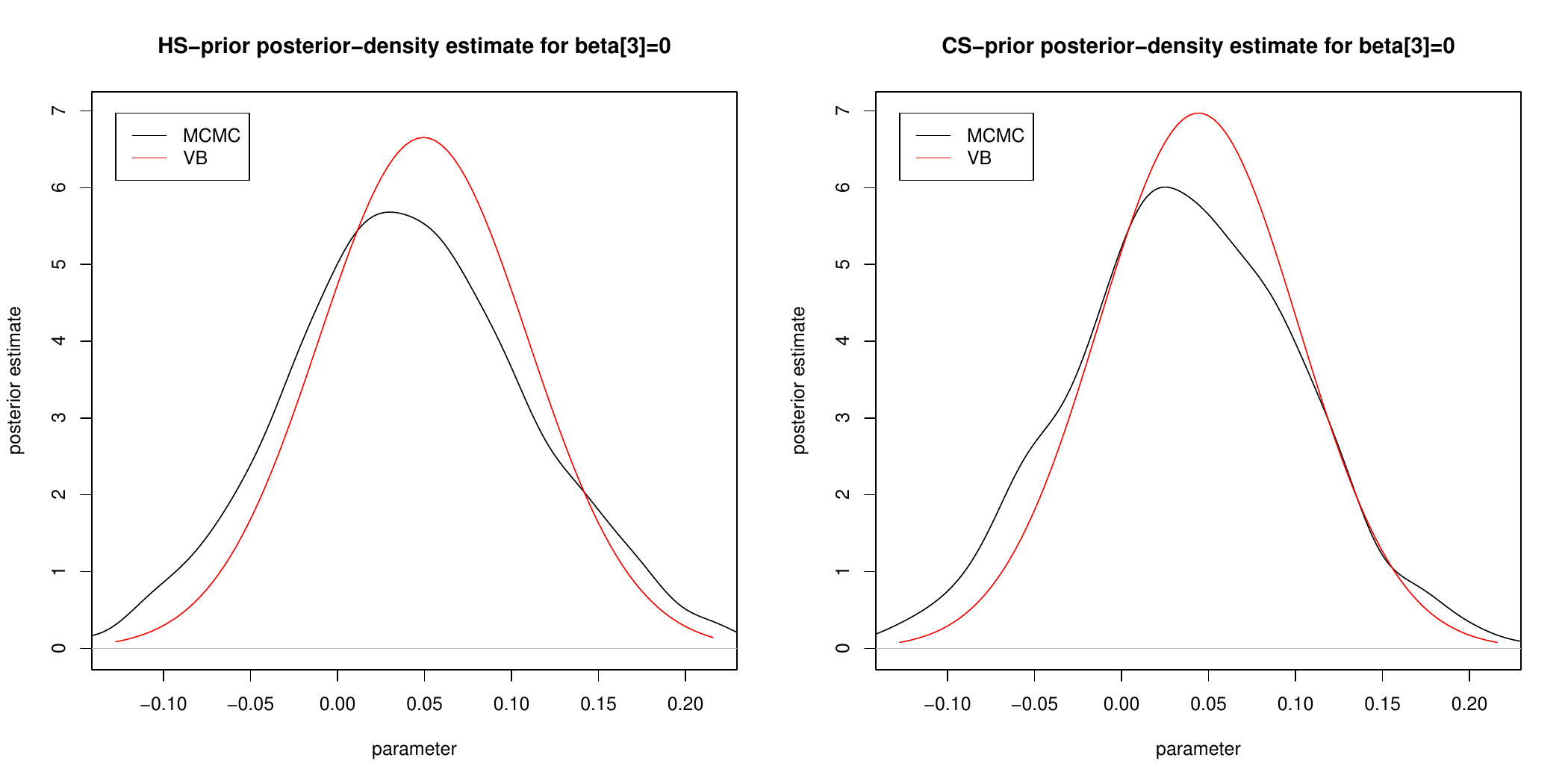}}
\centerline{\includegraphics[scale=0.4]{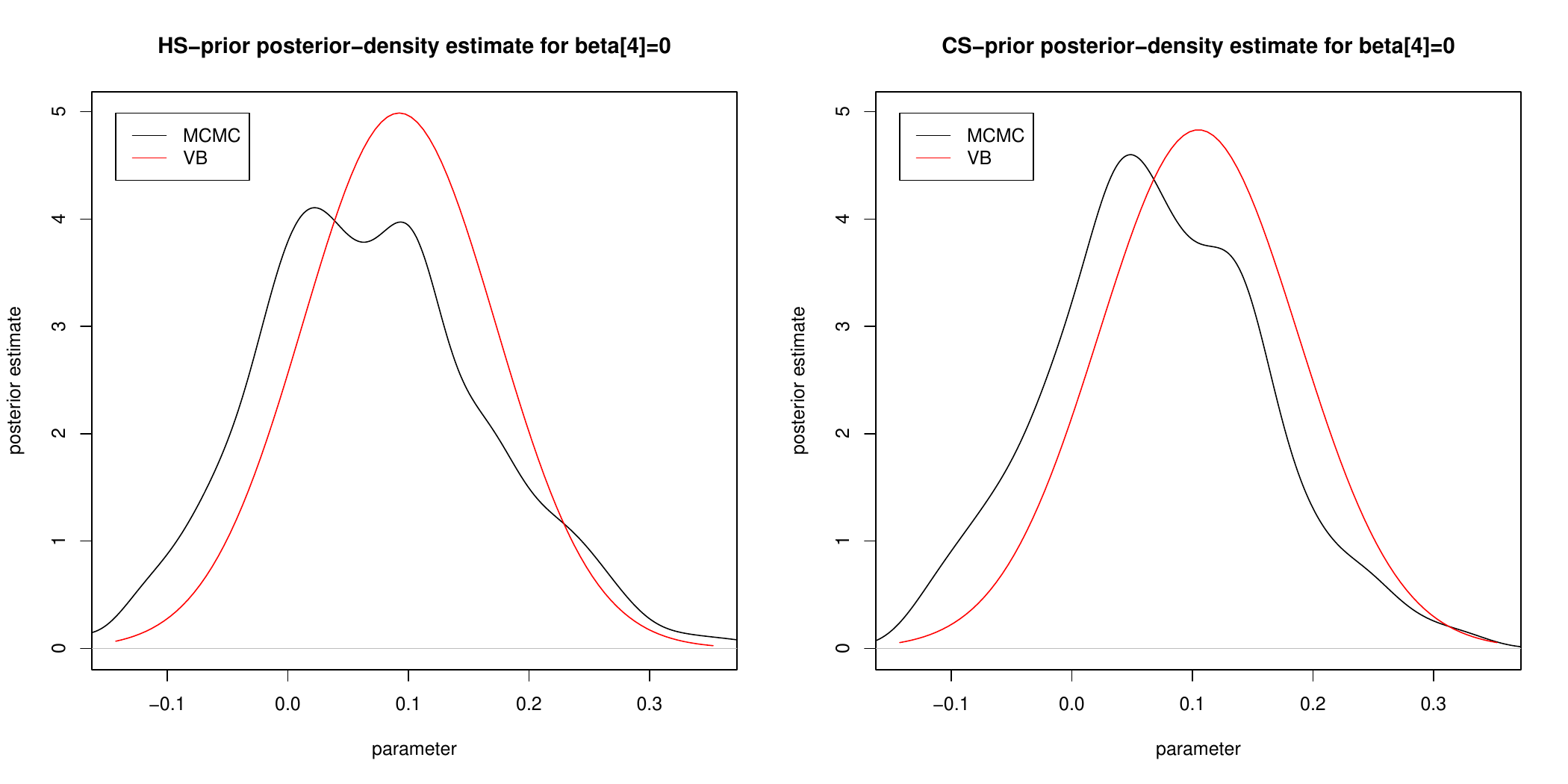}}
\centerline{\includegraphics[scale=0.4]{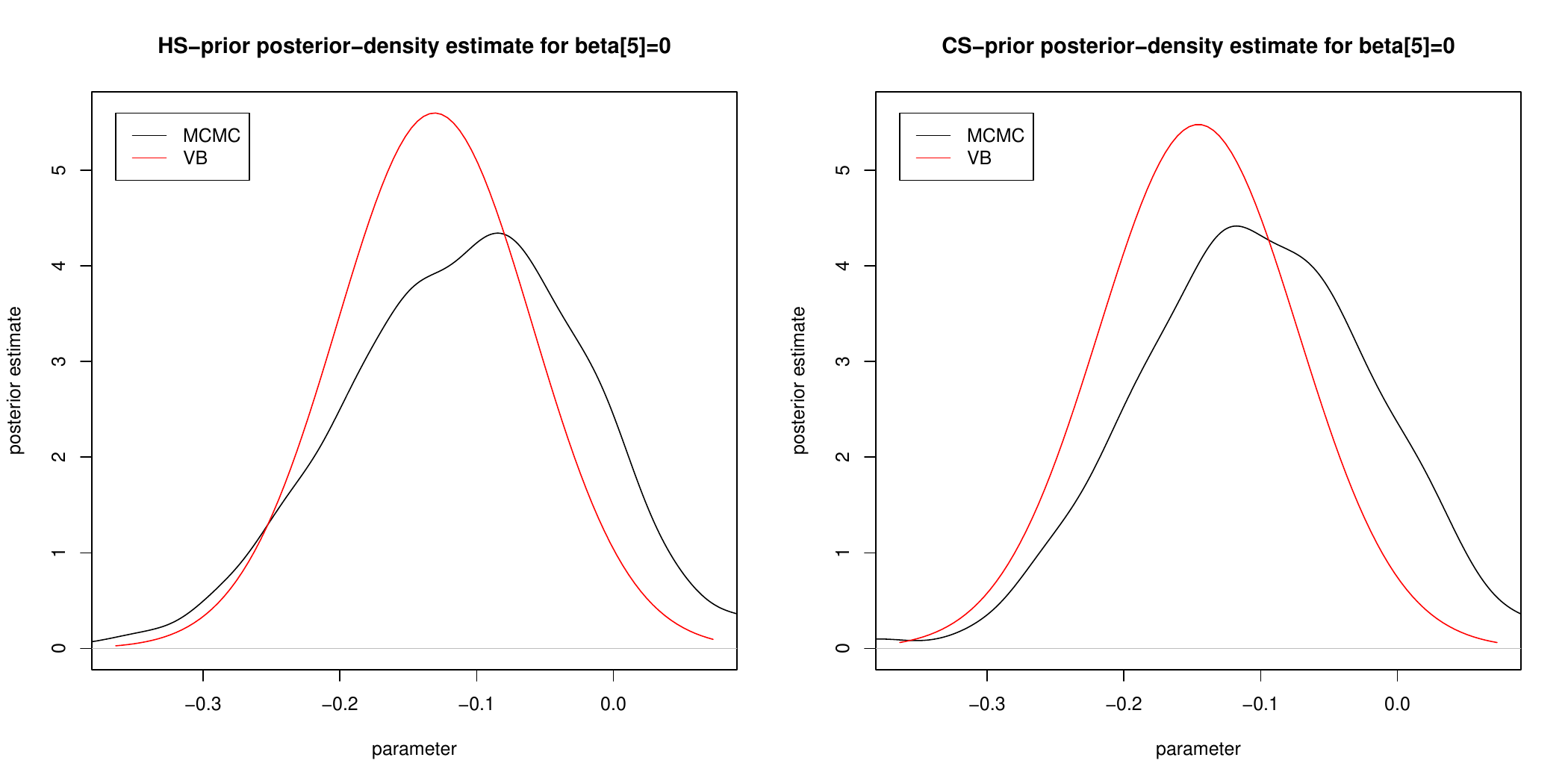}}
\caption{ Approximate posterior density functions of the second three regression coefficients for different models.}\label{dens2}
\end{figure}
\begin{figure}
\centerline{\includegraphics[scale=0.3]{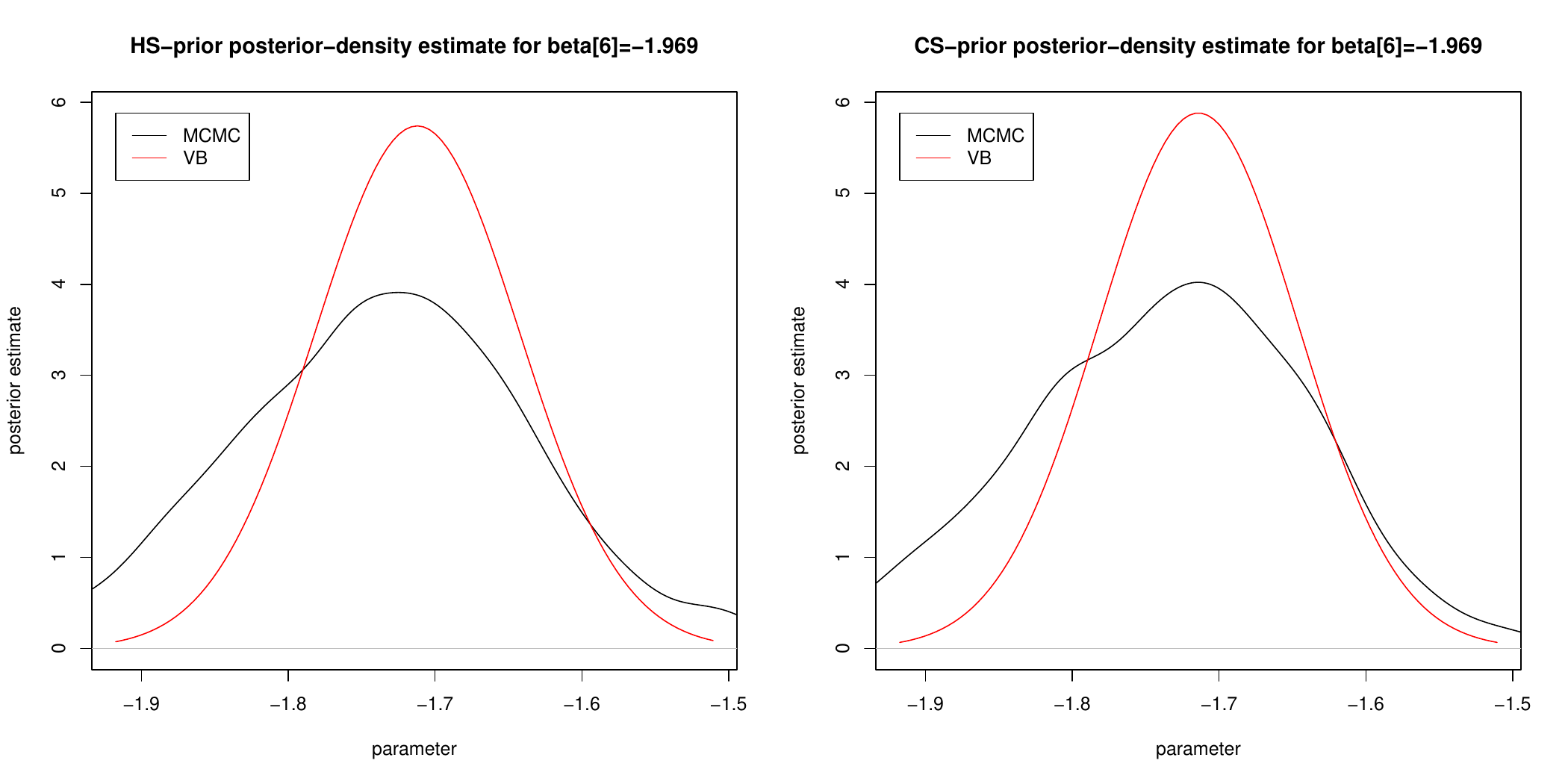}}
\centerline{\includegraphics[scale=0.3]{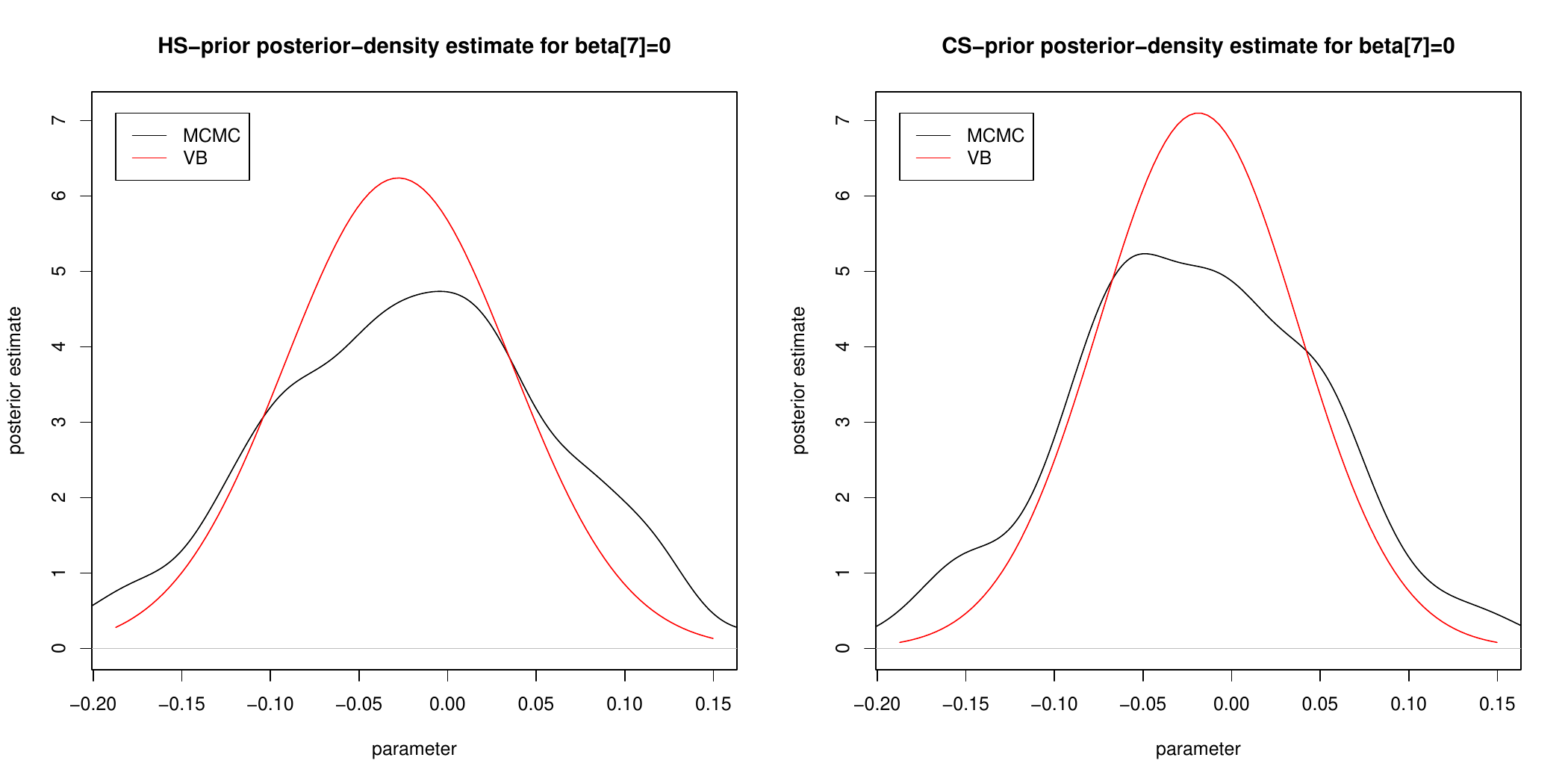}}
\centerline{\includegraphics[scale=0.3]{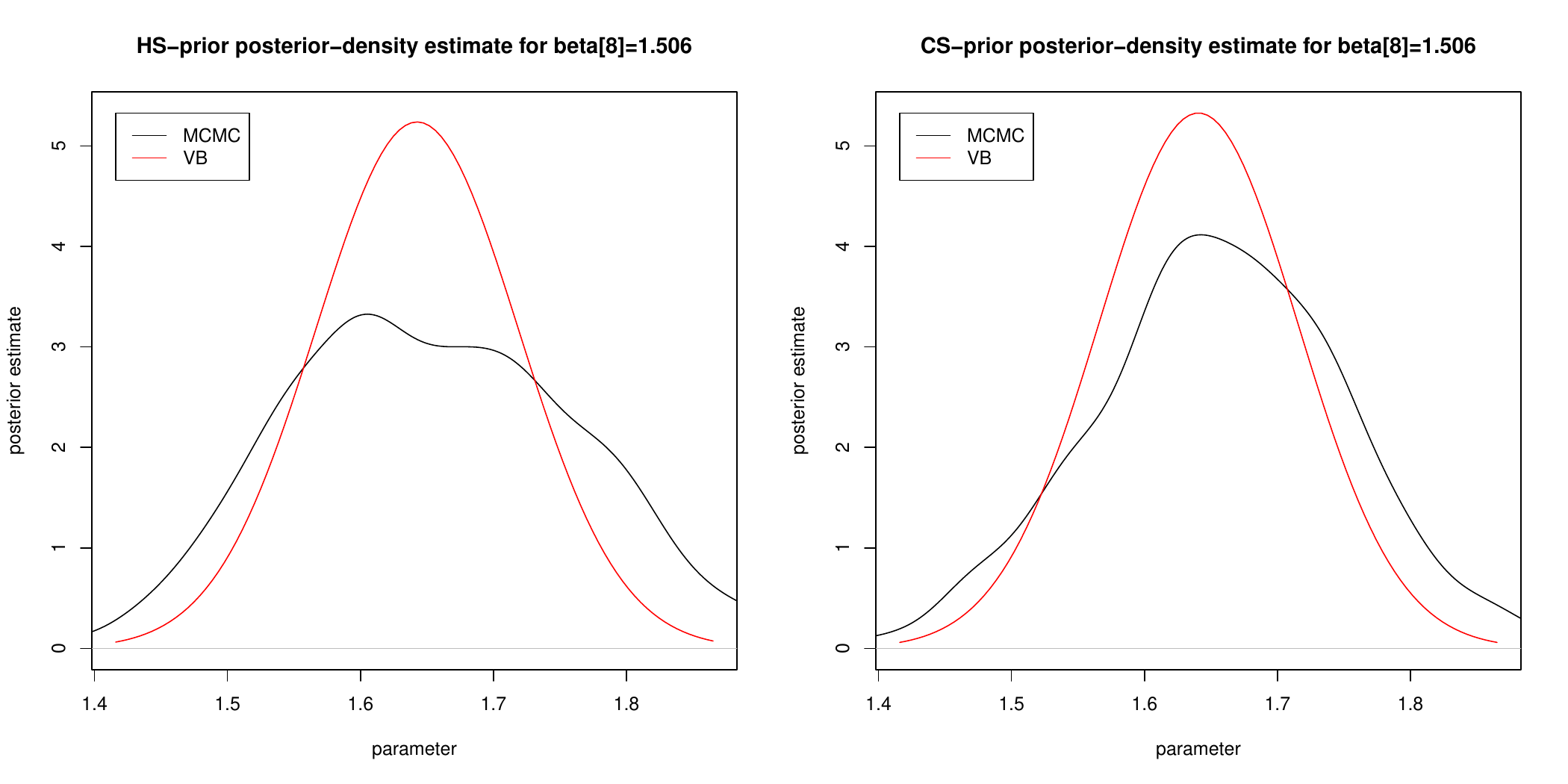}}
\centerline{\includegraphics[scale=0.3]{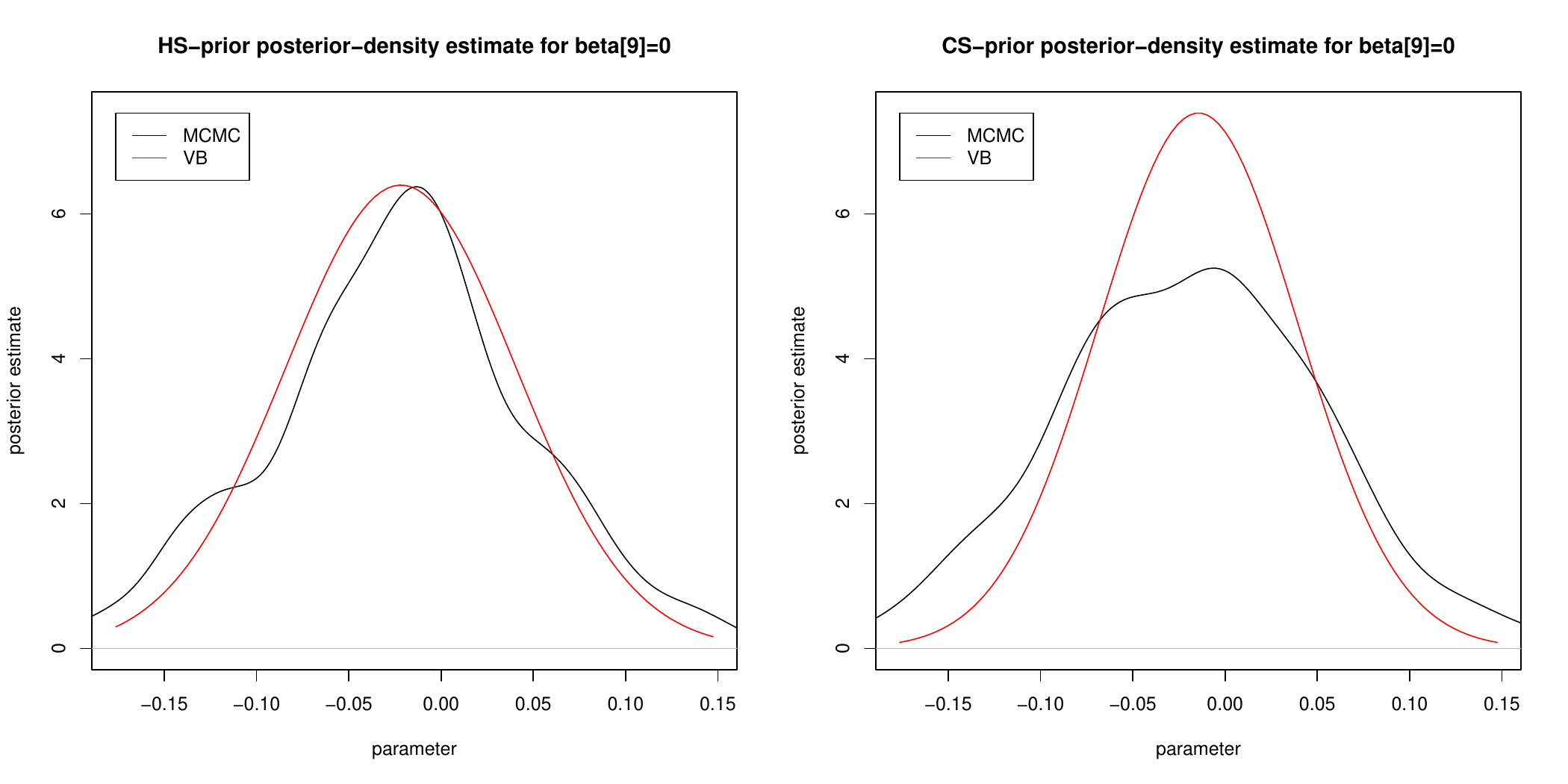}}
\caption{ Approximate posterior density functions of the last four regression coefficients for different models.}\label{dens3}
\end{figure}

{To examine} the accuracy of the approximation of the posterior distribution by VB methods, the accuracy measure \citep[see][]{luts15} is computed for each parameter component $\theta_j,\; j=1,\ldots,k,$ as follows
\begin{equation}\label{acc}
{\rm accuracy}(q(\theta_j)) = 100\left(1-\frac{1}{2}\int_{-\infty}^\infty |q(\theta_j) - p(\theta_j|y)| d\theta\right)\%,
\end{equation}
where $p(\theta_j|y)$ is the kernel density estimator using the MCMC sample.
{If $q(\theta) = p(\theta|y)$ for all values of $\theta$, we have ${\rm accuracy}(q(\theta_j)) = 100\%,$ while since both $q(\cdot)$ and $p(\cdot|y)$ are probability density functions, the minimum value of ${\rm accuracy}(q(\theta_j))$ is zero, which occurs when $q(\cdot)$ and $p(\cdot|y)$ have separate supports.}

For the Bernoulli components $q(Z_j)$ for CS-VB-NB method, the accuracy is computed as follows
$${\rm accuracy}(q(\theta_j)) =100\left(1-\frac{1}{2}\left(\left|q(\theta_j = 1) - \frac{1}{T}\sum_{t=1}^T I(\theta_{jt} = 1)\right| +\left|q(\theta_j = 0) - \frac{1}{T}\sum_{t=1}^T I(\theta_{jt} = 0)\right|\right)\right)\%,$$
where $\theta_{j1},\ldots,\theta_{jT}$ are MCMC samples. { If $q({ \theta}_j = k) = \frac{1}{T}\sum_{t=1}^T I({ \theta}_{jt} = k)$ for $k = 0,1$, we have ${\rm accuracy}(q(\theta_j)) = 100\%,$ and when $q({ \theta}_j = k) = 1$ and $\frac{1}{T}\sum_{t=1}^T I({ \theta}_{jt} = k) = 0$ for some $k \in \{0,1\}$, we have ${\rm accuracy}(q(\theta_j)) = 0\%$.}

The average (standard error) of accuracy values are given in Table \ref{tab-acc-nb}. The table presents the accuracy of the posterior approximation for the regression coefficients, auxiliary variables $\mathbf{Z}$, hyperparameters $\boldsymbol\pi$, local shrinkage parameters $\boldsymbol\lambda$, and global shrinkage parameter $\sigma$ for the CS-VB-NB and HS-VB-NB methods against an MCMC benchmark. {The boxplots for the accuracy of the regression coefficients are given in Figure \ref{boxac1}.} Furthermore, the $q$-functions as well as the kernel density estimator of the marginal posteriors of the regression coefficients, for a single iteration of the simulation study, are given in Figures \ref{dens1}, \ref{dens2}, and the corresponding plots for the other parameters are given in {the supplementary material}. 

As shown in Table \ref{tab-acc-nb}, the accuracy of the posterior approximation for the regression coefficients is high for both HS-VB-NB and CS-VB-NB methods, with average accuracies ranging from 76.41\% to 88.18\% for HS-VB-NB and 80.58\% to 84.53\% for CS-VB-NB. The HS-VB-NB method generally achieves slightly higher accuracies than CS-VB-NB across most coefficients, with notably smaller standard errors, indicating more stable approximations. For the auxiliary variables $\mathbf{Z}$, CS-VB-NB achieves excellent accuracy for the active components ($\mathbf{Z}_2$, $\mathbf{Z}_6$, $\mathbf{Z}_8$) with accuracies above 97\%, while the accuracies for the inactive components are moderate (around 14-17\%). The hyperparameter $\boldsymbol\pi$ shows similarly high accuracy for active components (approximately 94-95\%) and moderate accuracy for inactive components (approximately 60-61\%). The local shrinkage parameters $\boldsymbol\lambda$ exhibit consistent accuracy around 65\% across all components with very small standard errors, while the global shrinkage parameter $\sigma$ achieves an accuracy of 79.23\% with a standard error of 6.69. Overall, these results demonstrate that the variational approximations reliably recover the true parameter values and that the accuracy patterns reflect the underlying sparsity structure of the model.

\begin{figure}
\centerline{\includegraphics[scale=0.4]{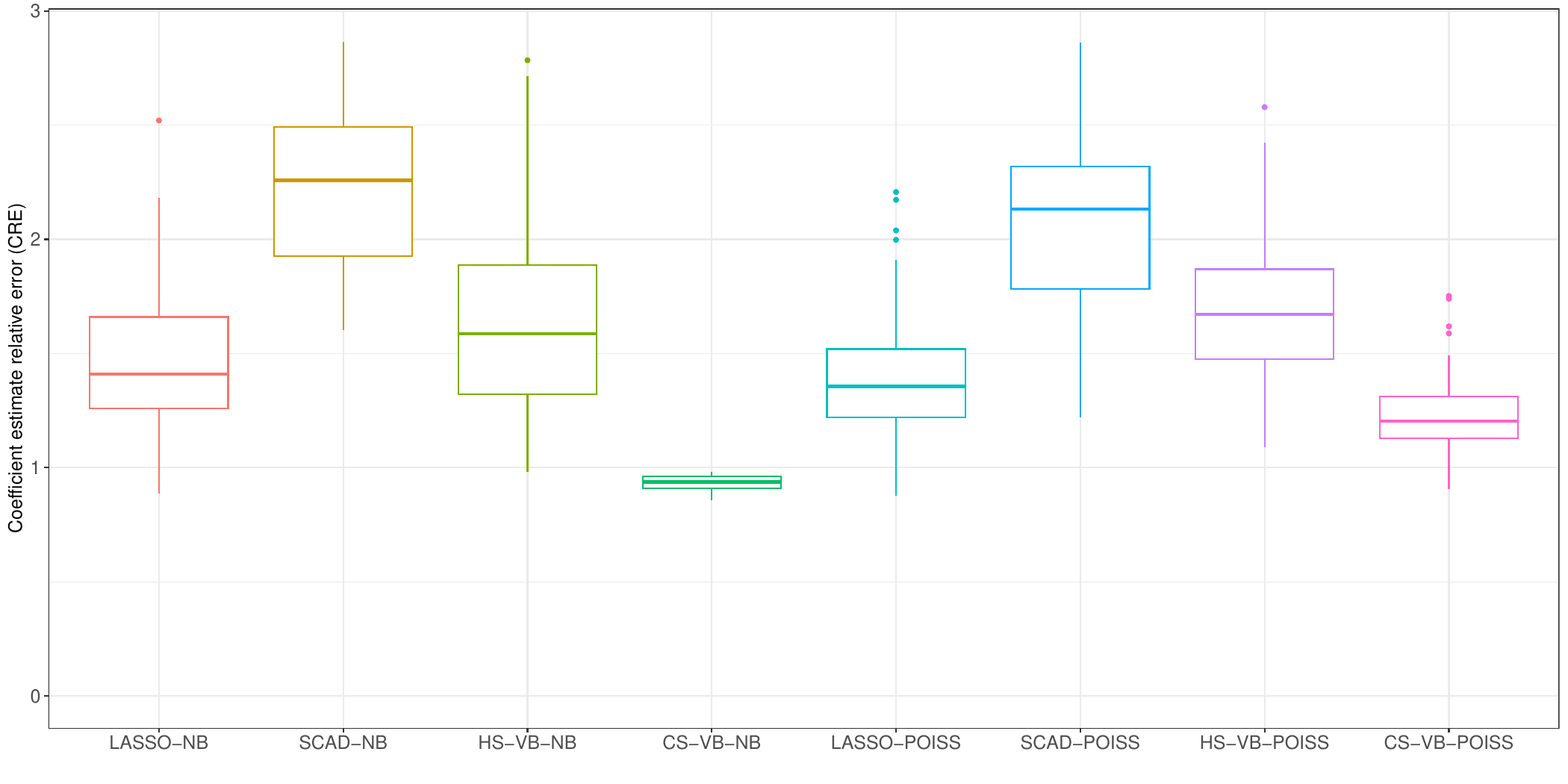}}
\centerline{\includegraphics[scale=0.4]{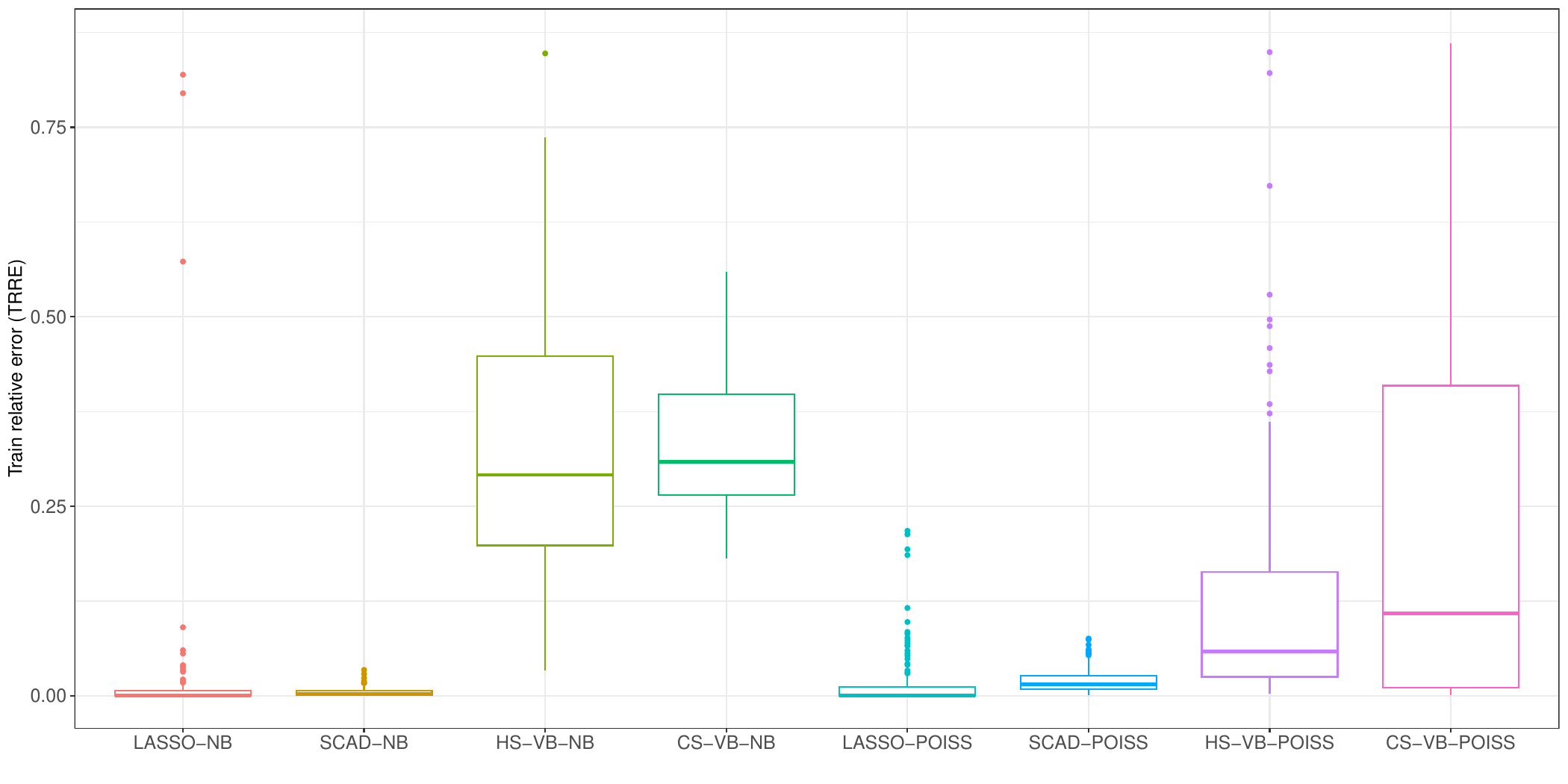}}
\centerline{\includegraphics[scale=0.4]{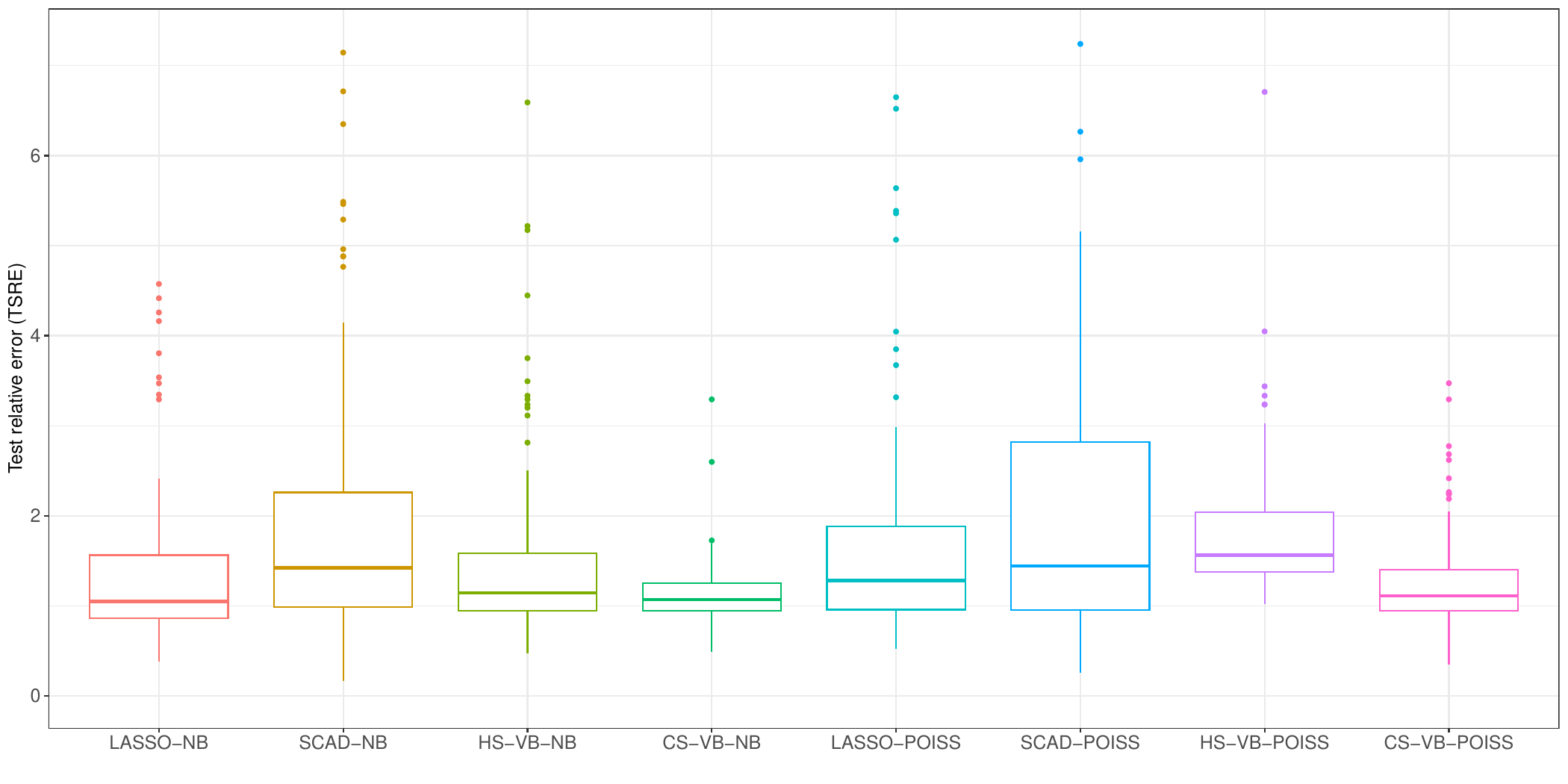}}
\caption{The high-dimensional scenario simulation results: the coefficient relative error (top), the train relative error (middle), and the test relative error (bottom) for different methods.}\label{box4}
\end{figure}

\begin{figure}
\centerline{\includegraphics[scale=0.4]{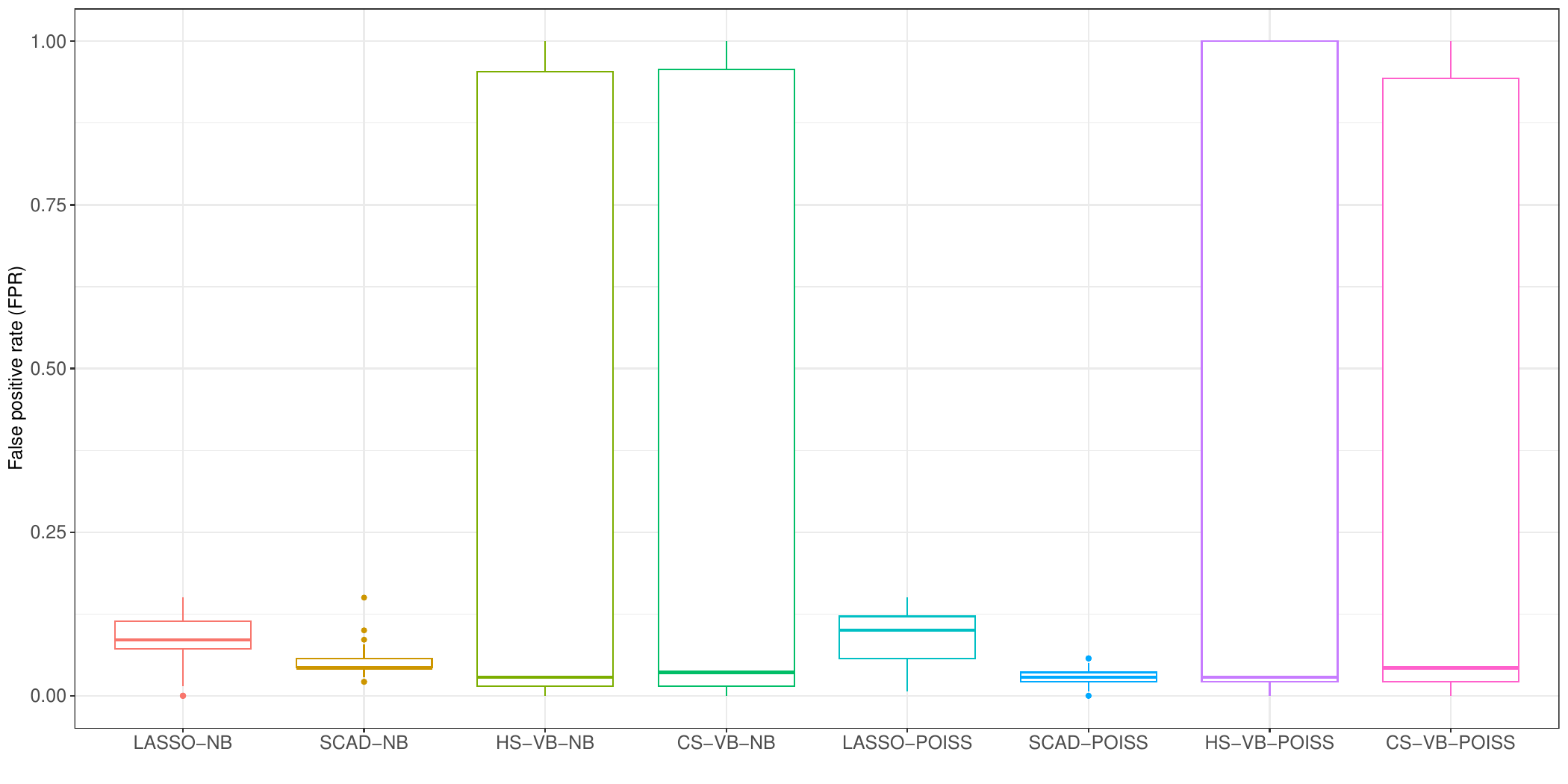}}
\centerline{\includegraphics[scale=0.4]{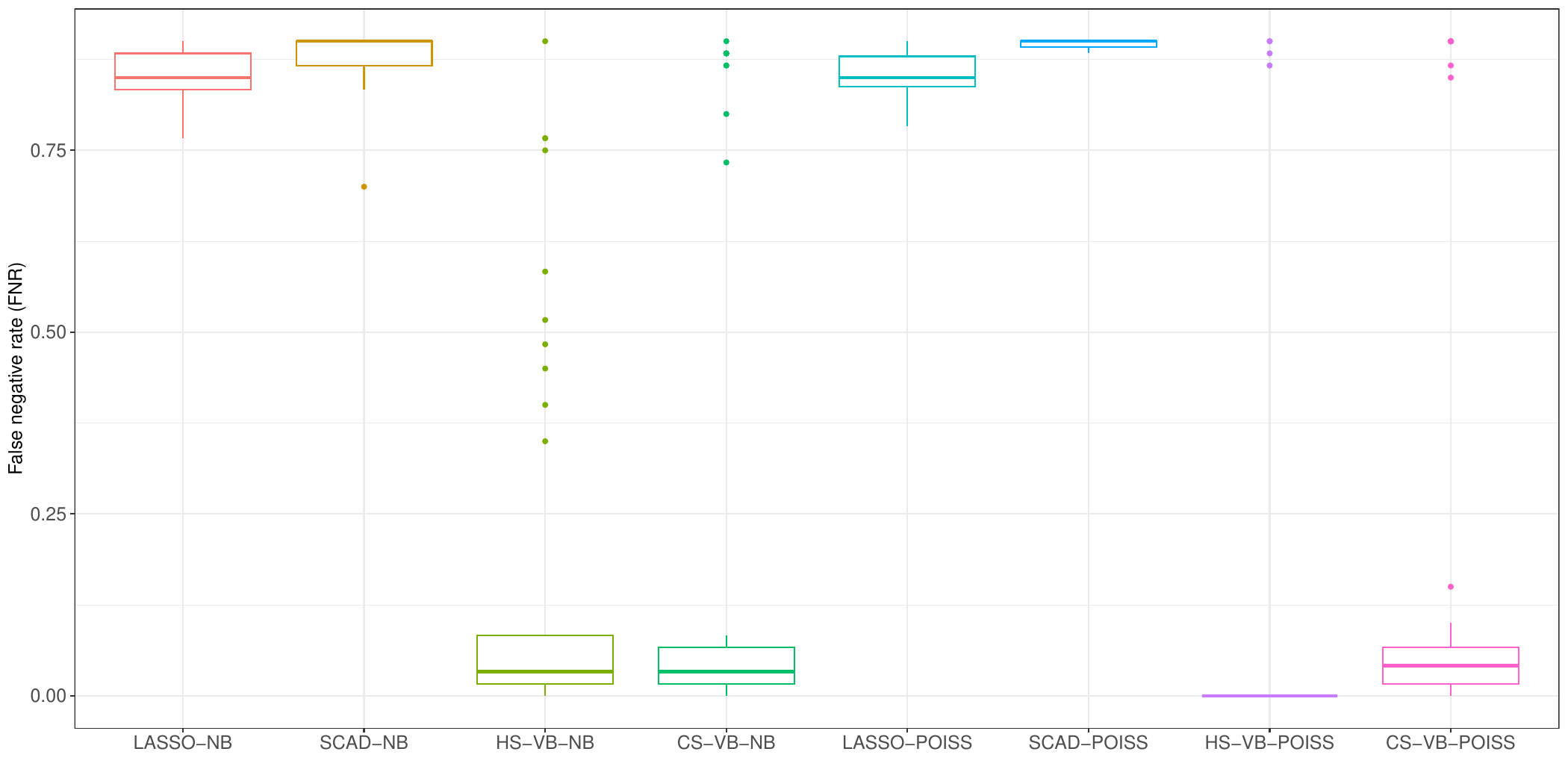}}
\centerline{\includegraphics[scale=0.4]{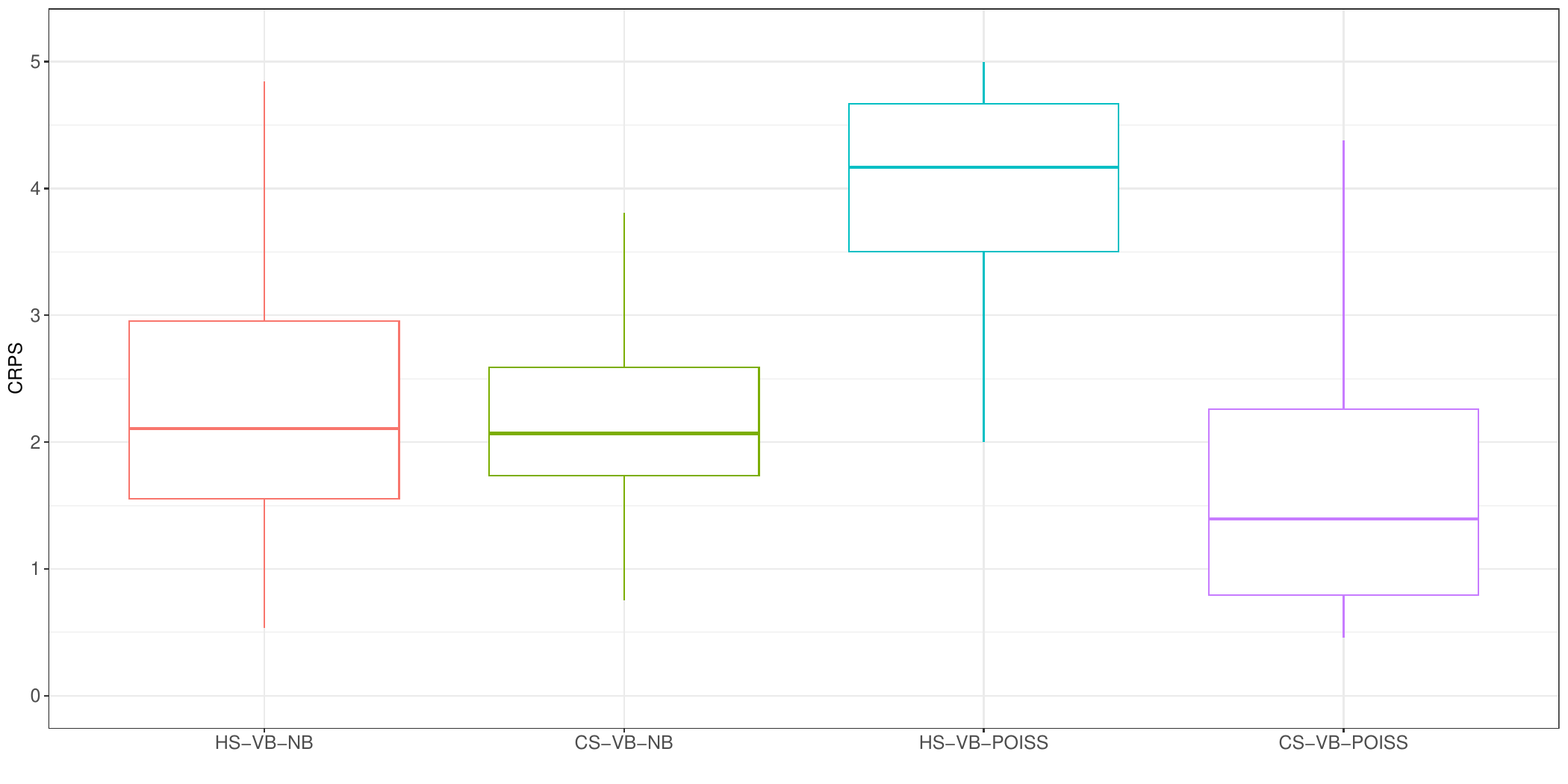}}
\caption{The high-dimensional scenario simulation results: the FPR (top), the {\color{black} FNR} (middle), and CRPS (bottom) for different methods.}\label{box5}
\end{figure}

\begin{figure}
\centerline{\includegraphics[scale=0.4]{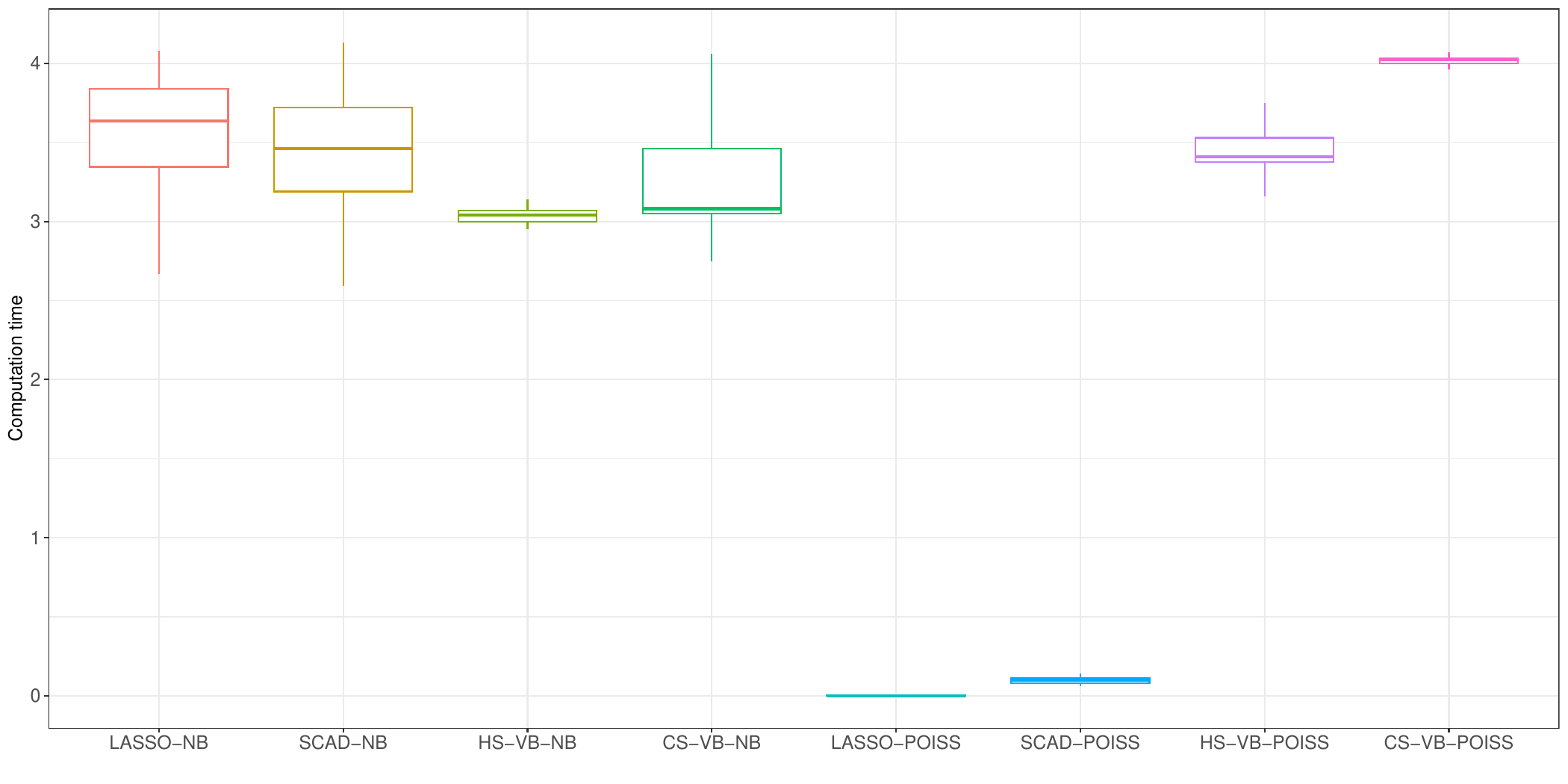}}
\caption{The high-dimensional scenario computation time (bottom) for different methods.}\label{box6}
\end{figure}

\subsection{High dimensional scenario}

To evaluate the performance of the proposed methods in a high-dimensional setting, we consider the case with
$p = 200$, $n = 30$, and {We randomly selected 60 values of 
$\mathbf{z}$ (including the first) to be set to 1, with the remaining values assigned 0.}. Figures \ref{box4} -- \ref{box6} show the boxplots of all aforementioned criteria for all competitive methods. Notably, MCMC methods are not included in these comparisons due to their prohibitive computational cost in high-dimensional settings, which would render them impractical for routine use.

Figure \ref{box4} displays the values of CRE, TRRE, and test TSRE, for the high dimensional scenario. 
The CRE of the CS-VB-NB method is consistently lower than that of the other methods. The negative binomial VB methods (HS-VB-NB and CS-VB-NB) also yield slightly lower CRE values compared to their Poisson counterparts. As in the low-dimensional scenario, the TRRE of the Poisson VB methods is lower than that of the NB methods, indicating overfitting by the Poisson models. However, the TSRE of the VB-NB methods is consistently lower than that of the VB-POISS methods, suggesting better predictive performance for the negative binomial approaches.

Figure \ref{box5} presents the FPR, FNR, and CRPS, for the high dimensional scenario. 
All VB methods exhibit higher FPR values than the frequentist LASSO and SCAD methods, while their FNR values are lower. This indicates that the VB methods tend to select more variables overall, including some false positives, but they are more successful at recovering the truly active predictors, thereby avoiding missing important signals. This trade-off reflects the greater flexibility of the Bayesian shrinkage priors in accommodating uncertainty about which variables should be included, at the cost of a slightly higher false discovery rate. With the exception of the HS-VB-POISS method, which exhibits notably higher CRPS values than the other VB approaches, all remaining VB methods achieve broadly similar CRPS values. Although the CS-VB-POISS method shows slightly lower CRPS compared to the VB-NB methods, this may be attributable to the narrower predictive distributions induced by the Poisson model when the data are not heavily overdispersed, rather than indicating superior predictive performance.

Figure \ref{box6} shows the computation time for each method. 
The variational Bayes methods (HS-VB-NB and CS-VB-NB) are slightly faster than the frequentist LASSO and SCAD methods. The Poisson VB methods are marginally faster than their NB counterparts due to the simpler model structure, but the difference is minimal. 

\section{Benchmark real data analysis}

\begin{figure}
\centerline{\includegraphics[scale=0.33]{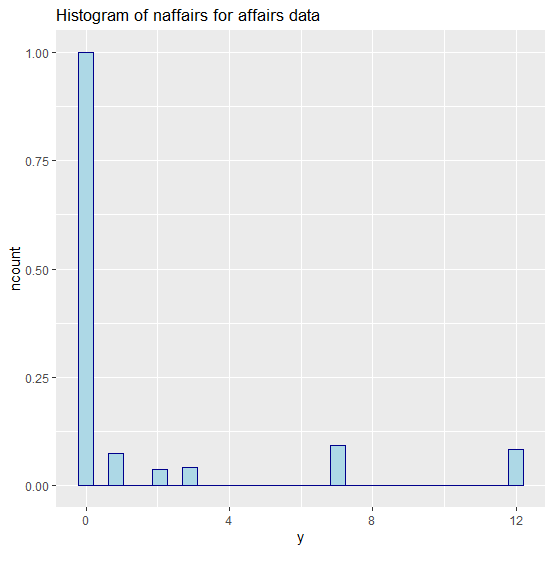}\includegraphics[scale=0.33]{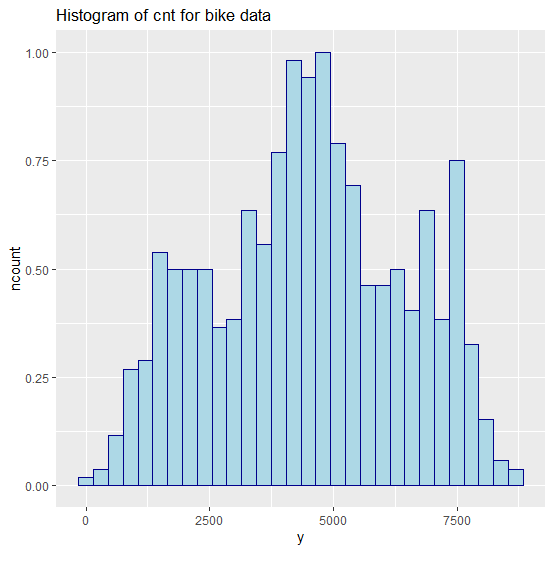}\includegraphics[scale=0.33]{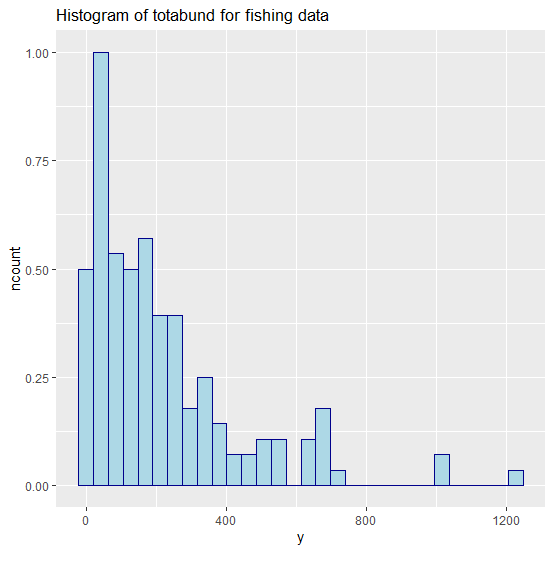}}
\centerline{\includegraphics[scale=0.33]{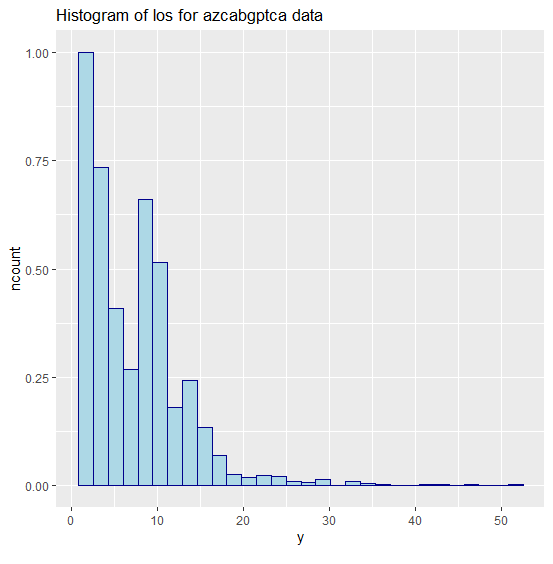}\includegraphics[scale=0.33]{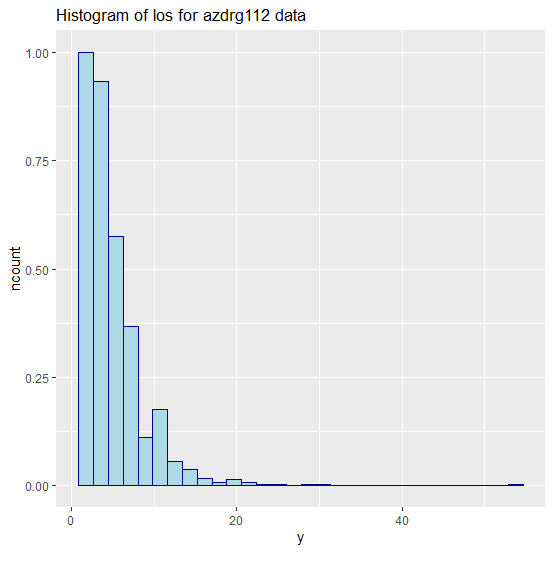}\includegraphics[scale=0.33]{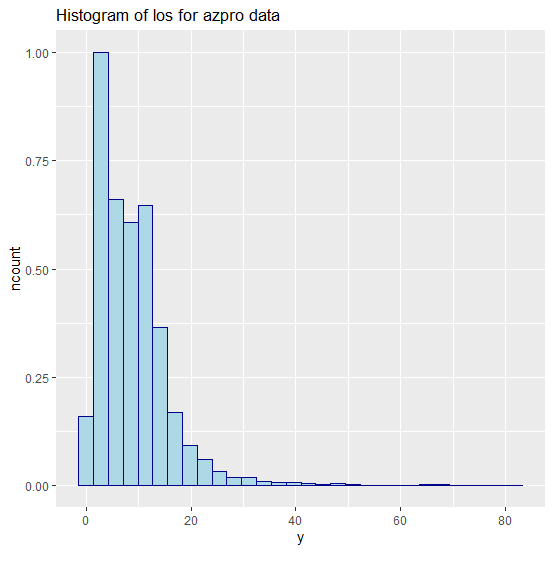}}
\caption{Histograms of the responses for the benchmark real data sets. Top left; affairs, top middle; bike sharing, top right; fishing, bottom left; azcabgptca, bottom middle; azdrg112, bottom right; azpro.}\label{hists}
\end{figure}

To examine the prediction performance of the proposed VB methods, we have considered 5 benchmark real data sets with a count response variable. 

 {The affairs data set \citep{f79}, available in the R package COUNT, contains 601 observations and 18 variables, including naffairs (the number of affairs in the past year) and 17 covariates related to children, marital happiness, religiosity, and age.} \cite{gr03}, modeled this data using Poisson regression, although given the amount of over-dispersion in the data, employing a negative binomial model is an appropriate strategy. The variable naffairs is considered the response variable. The top left graph of Figure \ref{hists} shows the histogram of the response variable naffairs. {The bike-sharing dataset records hourly and daily rental counts from 2011 to 2012 in the Capital Bikeshare system, along with corresponding weather and seasonal information.} This data set includes 731 observations and 14 variables (except date and instant number), including cnt (the response variable), and 13 covariates, including season, year, month, holiday, weekday, working day, 5 weather variables, casual, and registered. The top middle graph in Figure \ref{hists} presents the histogram of the response variable for the bike-sharing data set. {The three remaining data sets, azcabgptca, azdrg112, and azpro, from the R package COUNT pertain to samples from Arizona hospital cardiovascular patient files, collected in 1991 (azcabgptca, azpro) and 1995 (azdrg112), involving patients who received one of two standard procedures: CABG or PTCA.} The data set azcabgptca has 1959 observations on 6 variables, azdrg112 has 1,798 observations on 4 variables, and azpro has 3589 observations on 6 variables. { In all three data sets, the response variable is los (length of hospital stay), with covariates including procedure type, sex, age, and additional factors.} The bottom plots of Figure \ref{hists} show the histograms of the los response variable in azcabgptca, azdrg112, and azpro, respectively, from left to right. 

To formally assess the presence of overdispersion in the benchmark datasets, we conducted the overdispersion test of \citet{zeileis2008} for the Poisson regression models. Table \ref{tab:overdispersion} reports the Akaike Information Criterion (AIC), the sum of squared Pearson residuals, and the dispersion test results for both Poisson and negative binomial models across all five datasets. The results reveal strong evidence of overdispersion in all datasets, with dispersion parameter estimates ranging from 2.23 (azcabgptca) to 53.51 (bike sharing), and all p-values being highly significant (p-value $< 0.001$). The negative binomial models consistently yield substantially lower AIC values compared to their Poisson counterparts, with the largest improvements observed for the bike sharing dataset (AIC reduction of 39,943.77) and the affairs dataset (AIC reduction of 1,347.07). The sum of squared residuals is also considerably smaller for the negative binomial models across all datasets, further confirming the inadequacy of the Poisson assumption. These findings justify our use of the negative binomial model as the primary framework for the proposed variational Bayesian methods, while also motivating the inclusion of Poisson-based VB methods as a benchmark to assess the impact of model misspecification.

\begin{table}[htbp]
\centering
\caption{Overdispersion diagnostics and model comparison for the benchmark datasets.}
\label{tab:overdispersion}
\begin{tabular}{lcccccc}
\toprule
\multirow{2}{*}{Dataset} & \multicolumn{2}{c}{AIC} & \multicolumn{2}{c}{Sum of Squared Residuals} & \multirow{2}{*}{Dispersion} & \multirow{2}{*}{p-value} \\
\cmidrule(lr){2-3} \cmidrule(lr){4-5}
 & Poisson & NB & Poisson & NB & & \\
\midrule
affairs & 2827.15 & 1480.09 & 2303.16 & 339.86 & 6.74 & $1.07\times10^{-8}$ \\
azcabgptca & 10430.37 & 9916.14 & 3442.21 & 1887.81 & 2.23 & $1.19\times10^{-11}$ \\
azdrg112 & 9178.55 & 8568.92 & 3364.05 & 1674.99 & 2.31 & $2.27\times10^{-8}$ \\
azpro & 22391.79 & 19961.07 & 8874.15 & 3525.63 & 3.21 & $<2.2\times10^{-16}$ \\
azprocedure & 22391.79 & 19961.07 & 8874.15 & 3525.63 & 3.21 & $<2.2\times10^{-16}$ \\
bike sharing & 51680.01 & 11736.24 & 44263.18 & 800.75 & 53.51 & $<2.2\times10^{-16}$ \\
\bottomrule
\end{tabular}
\end{table}

\begin{table}[htbp]
\centering
\caption{Test relative error (TSRE) and continuous ranked probability score (CRPS) means (standard deviations) for 10 random partitions of 6 benchmark real data sets.}
\label{tab4-nb}
\vspace{2mm}
\begin{tabular}{l l c c c c}
\hline\hline
Data set & Criterion & LASSO-NB & SCAD-NB & HS-VB-NB & CS-VB-NB \\
\hline
\multirow{2}{*}{affairs} & TSRE & 0.933 (0.110) & 0.944 (0.142) & 0.963 (0.033) & 0.947 (0.042) \\
 & CRPS & -- & -- & 3.968 (1.704) & 4.642 (1.742) \\
\cline{2-6}
 &  & LASSO-POISS & SCAD-POISS & HS-VB-POISS & CS-VB-POISS \\
\cline{2-6}
 & TSRE & 0.901 (0.116) & 0.904 (0.116) & 1.038 (0.194) & 0.898 (0.088) \\
 & CRPS & -- & -- & 0.852 (0.030) & 1.900 (0.589) \\
\hline
\multirow{2}{*}{bike sharing} & TSRE & 0.132 (0.025) & 0.134 (0.026) & 0.117 (0.020) & 0.118 (0.020) \\
 & CRPS & -- & -- & 218.103 (84.131) & 215.180 (84.632) \\
\cline{2-6}
 &  & LASSO-POISS & SCAD-POISS & HS-VB-POISS & CS-VB-POISS \\
\cline{2-6}
 & TSRE & 0.054 (0.005) & 0.059 (0.010) & 0.061 (0.012) & 0.061 (0.012) \\
 & CRPS & -- & -- & 15.892 (7.935) & 20.399 (6.347) \\
\hline
\multirow{2}{*}{azcabgptca} & TSRE & 0.539 (0.020) & 0.540 (0.020) & 0.724 (0.023) & 0.764 (0.020) \\
 & CRPS & -- & -- & 8.231 (1.710) & 8.166 (2.161) \\
\cline{2-6}
 &  & LASSO-POISS & SCAD-POISS & HS-VB-POISS & CS-VB-POISS \\
\cline{2-6}
 & TSRE & 0.537 (0.019) & 0.538 (0.019) & 0.548 (0.018) & 0.549 (0.019) \\
 & CRPS & -- & -- & 2.405 (0.221) & 2.439 (0.253) \\
\hline
\multirow{2}{*}{azdrg112} & TSRE & 0.867 (0.041) & 0.868 (0.041) & 0.945 (0.033) & 0.945 (0.033) \\
 & CRPS & -- & -- & 4.787 (0.213) & 4.787 (0.213) \\
\cline{2-6}
 &  & LASSO-POISS & SCAD-POISS & HS-VB-POISS & CS-VB-POISS \\
\cline{2-6}
 & TSRE & 0.867 (0.041) & 0.868 (0.041) & 0.875 (0.027) & 0.866 (0.031) \\
 & CRPS & -- & -- & 1.660 (0.222) & 1.684 (0.220) \\
\hline
\multirow{2}{*}{azprocedure} & TSRE & 0.600 (0.049) & 0.602 (0.049) & 0.737 (0.045) & 0.738 (0.045) \\
 & CRPS & -- & -- & 10.545 (1.166) & 10.801 (1.229) \\
\cline{2-6}
 &  & LASSO-POISS & SCAD-POISS & HS-VB-POISS & CS-VB-POISS \\
\cline{2-6}
 & TSRE & 0.598 (0.049) & 0.598 (0.050) & 0.600 (0.049) & 0.602 (0.049) \\
 & CRPS & -- & -- & 2.559 (0.198) & 2.552 (0.204) \\
\hline
\multirow{2}{*}{azpro} & TSRE & 0.634 (0.039) & 0.636 (0.040) & 0.774 (0.034) & 0.773 (0.034) \\
 & CRPS & -- & -- & 10.450 (1.002) & 10.619 (1.154) \\
\cline{2-6}
 &  & LASSO-POISS & SCAD-POISS & HS-VB-POISS & CS-VB-POISS \\
\cline{2-6}
 & TSRE & 0.632 (0.038) & 0.632 (0.038) & 0.634 (0.038) & 0.636 (0.038) \\
 & CRPS & -- & -- & 2.514 (0.187) & 2.563 (0.174) \\
\hline\hline
\end{tabular}
\end{table}

We randomly partition the observations of each dataset into a training set (80 \%) and a test set (20 \%), and replicate the random partitioning 10 times to compute the test set relative prediction error and compare it for all considered competitors. Table \ref{tab4-nb} presents the TSRE and CRPS for all methods across the six benchmark data sets. The results demonstrate that the proposed VB methods perform comparably to frequentist sparse regression approaches. The HS-VB-NB and CS-VB-NB methods show competitive predictive performance relative to LASSO-NB and SCAD-NB, with slightly higher test errors on some datasets but smaller standard deviations, indicating more stable predictions. For the Poisson-based methods, the LASSO-POISS and SCAD-POISS approaches generally achieve the lowest test errors on the bike sharing dataset, while the VB-Poiss methods perform similarly to their frequentist counterparts. Notably, on the affairs dataset, HS-VB-POISS exhibits a substantially higher test error (1.038) with a large standard deviation (0.194), suggesting potential instability in the variational approximation for this particular dataset under the Poisson model. Regarding the CRPS, which evaluates the full predictive distribution, the Poisson-based VB methods produce substantially lower scores than their negative binomial counterparts across all datasets, reflecting the narrower predictive distributions of the Poisson model. Among the negative binomial VB methods, HS-VB-NB and CS-VB-NB yield comparable CRPS values, with HS-VB-NB performing slightly better on most datasets.

\begin{figure}
\centerline{\includegraphics[scale=0.37]{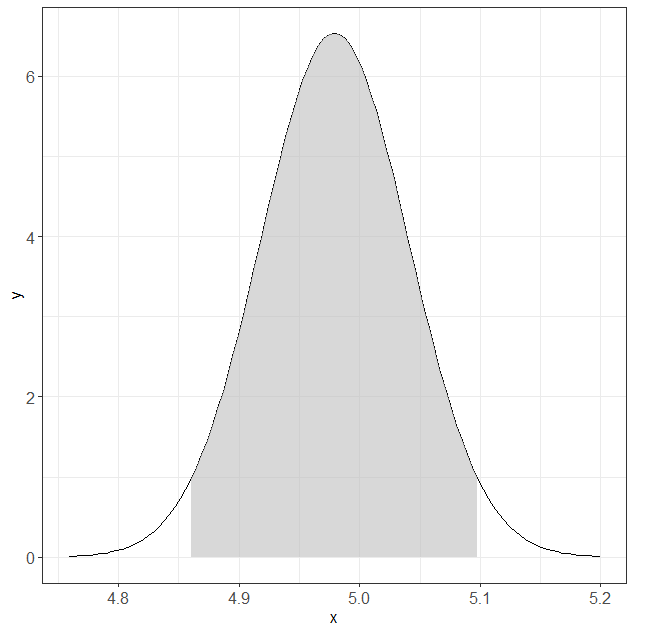}\includegraphics[scale=0.37]{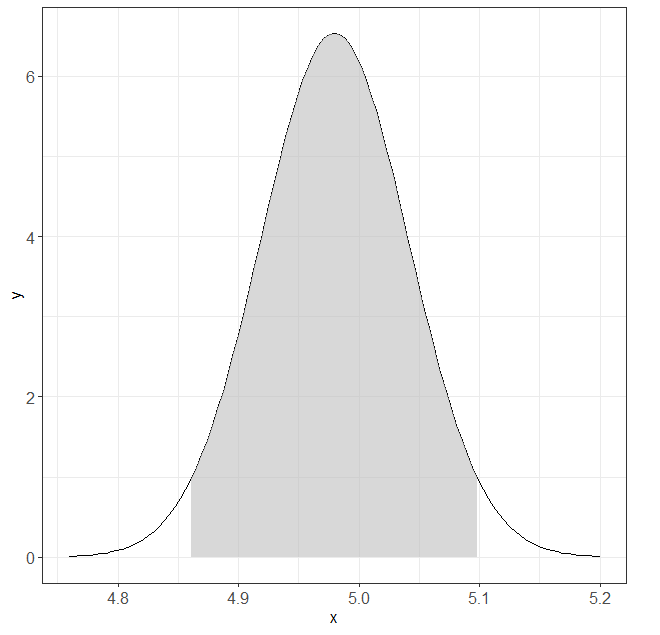}}
\centerline{\includegraphics[scale=0.37]{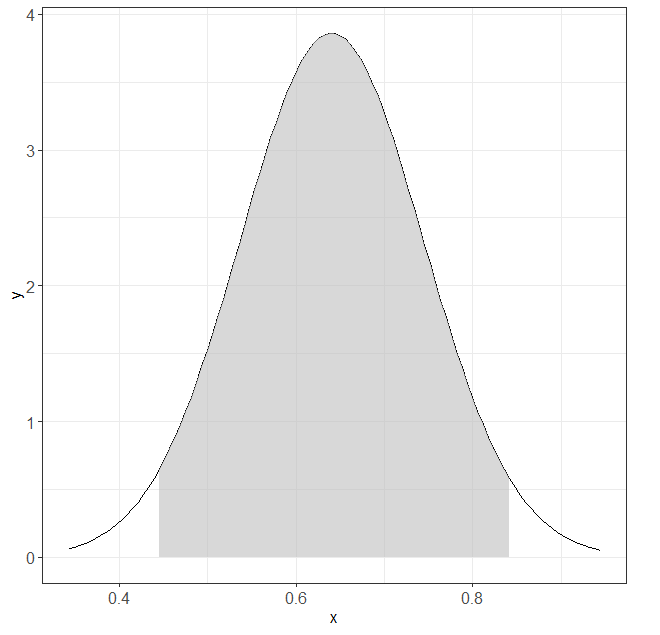}\includegraphics[scale=0.37]{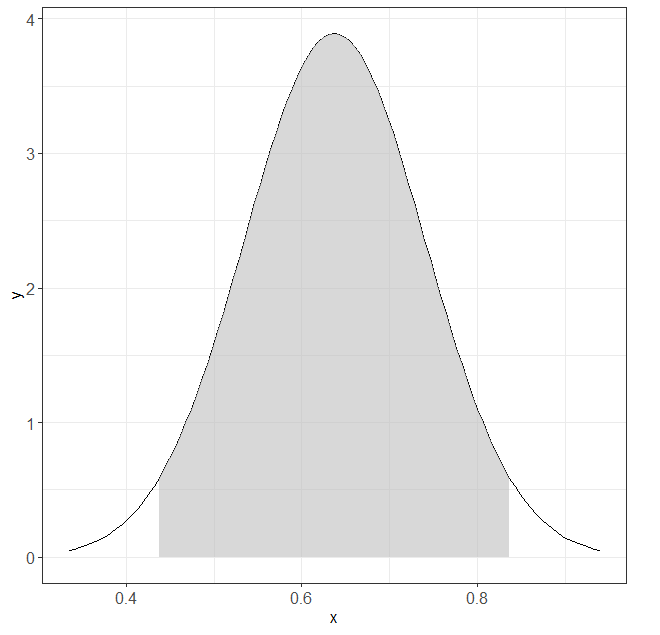}}
\centerline{\includegraphics[scale=0.37]{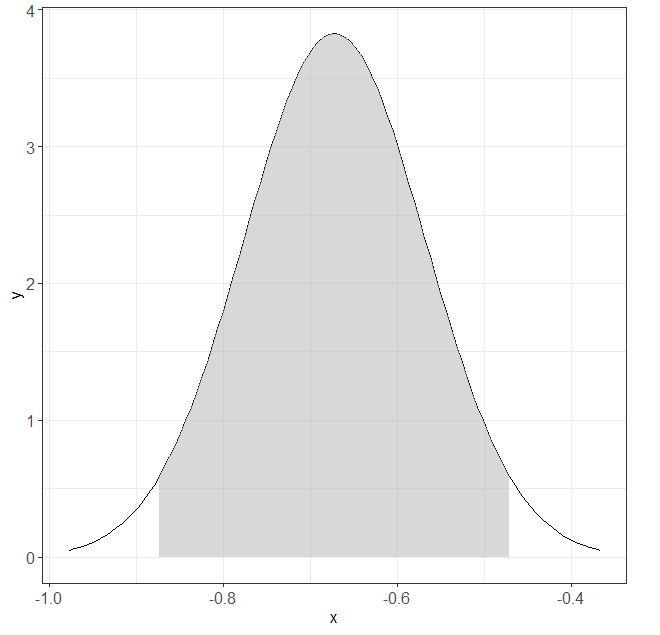}\includegraphics[scale=0.37]{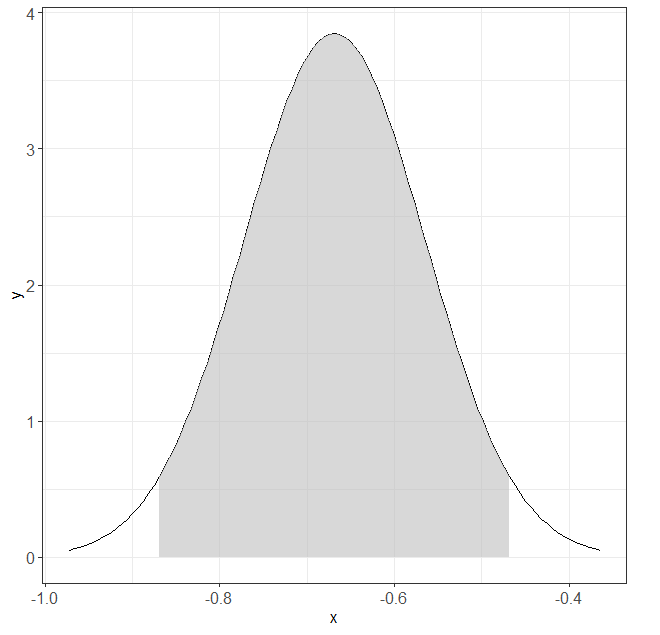}}
\centerline{\includegraphics[scale=0.37]{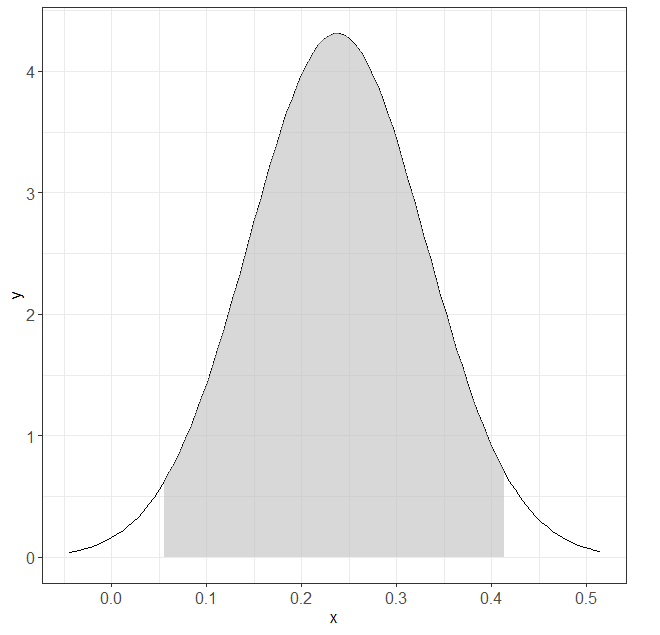}\includegraphics[scale=0.37]{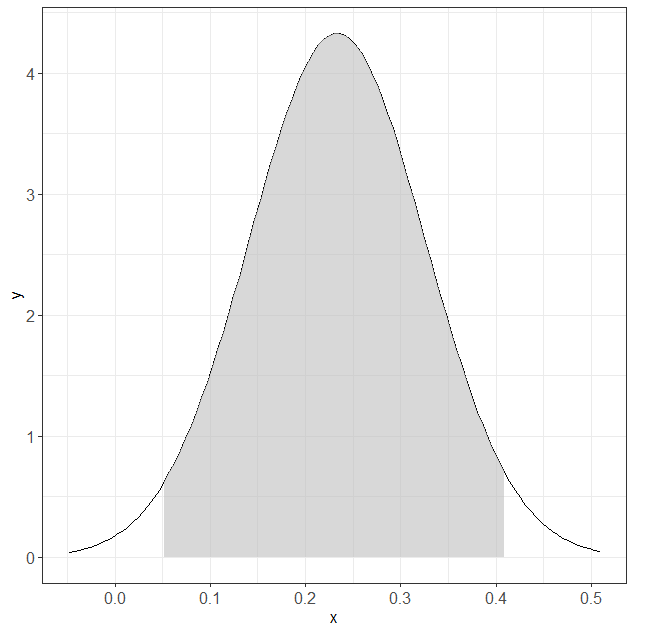}}
\caption{The posterior approximated densities of the regression coefficients, and the corresponding HPD intervals, for the fishing data set. Columns from left to right are associated with HS-VB-NB, and CS-VB-NB, and rows from top to bottom are associated with the intercept and 3 regression coefficients: density, mean-depth, and swept-area.}\label{fishing}
\end{figure}

\subsection{Fishing data set}

\begin{sidewaysfigure}
\centerline{\includegraphics[scale=0.45]{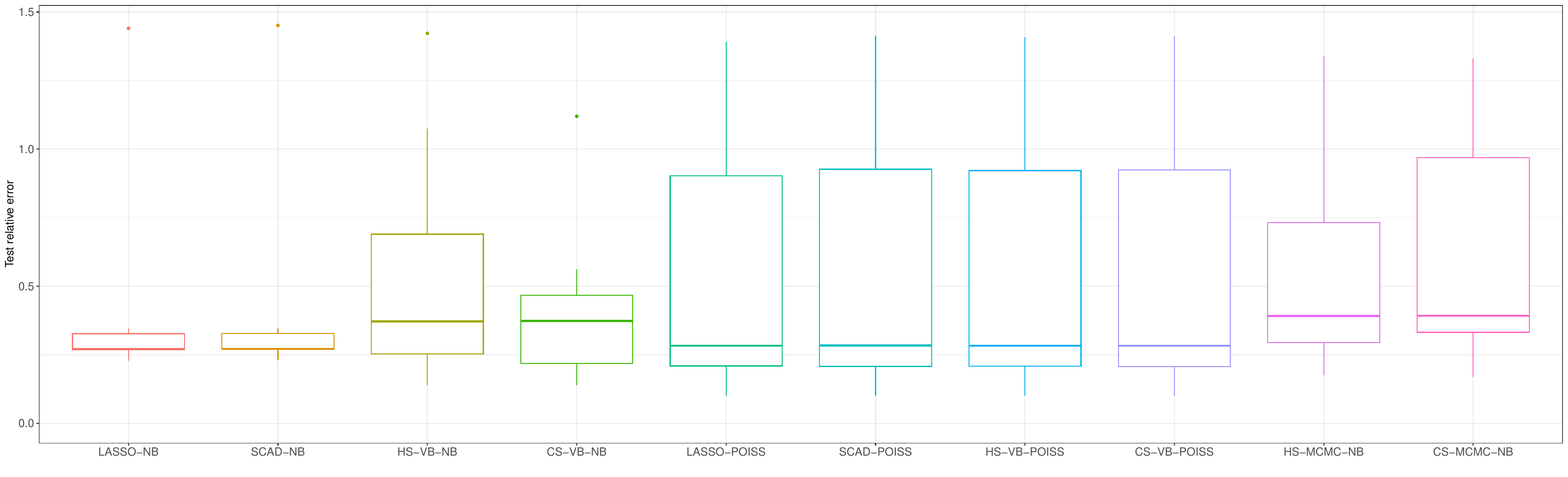}}
\caption{Test relative error boxplots for 10 random partitions of fishing data set.}\label{boxf}
\end{sidewaysfigure}

We have chosen one of the benchmark data sets, that is the fishing data \citep{zea13}, for a more extensive numerical study. {For this dataset, we further compare the predictive performance of the VB methods against their MCMC counterparts, and present the HPD intervals for the model coefficients.

The fishing data, adapted from \cite{bea09}, investigate the effects of commercial fishing on certain deep-sea fish populations when operations expanded into deeper waters than in prior years. Observations from 147 sites include totabund (total fish per site) and six covariates: depth, area, foliage density, catch site, year, and period. In this study, only three covariates, density, mean depth, and swept area, are used.  The top right plot in Figure \ref{hists} shows the histogram for this data set. This dataset, available from the COUNT package in R, contains information on recreational fishing trips and is commonly used for count regression modeling. The response variable is the number of fish caught, with covariates including various demographic and trip-related factors.

Figure \ref{boxf} presents the test relative error boxplots for the fishing dataset, which was not included in the main real data analysis.  The boxplots show the predictive performance of all methods across 10 random data partitions. 
Figure \ref{boxf} shows that, the negative binomial VB methods (HS-VB-NB and CS-VB-NB) achieve median test relative errors comparable to their MCMC counterparts (HS-MCMC-NB and CS-MCMC-NB) which exhibit similar performance. The MCMC benchmarks confirm that the VB approximations closely replicate the predictive performance of full MCMC inference, while offering substantial computational savings. The Poisson-based VB methods (HS-VB-POISS and CS-VB-POISS) show slightly higher test errors, confirming the presence of overdispersion in the fishing data. The frequentist LASSO-NB and SCAD-NB methods achieve median errors slightly lower than the VB-NB methods. Notably, the CS-VB-NB method shows the lowest median test error and the smallest interquartile range among VB methods, indicating both high predictive accuracy and stability.

\section{Concluding remarks}

This paper has developed a comprehensive variational Bayesian framework for sparse negative binomial regression, addressing the critical need for computationally efficient inference in high-dimensional count data settings. The proposed HS-VB-NB and CS-VB-NB methods simultaneously handle overdispersion, variable selection, and parameter estimation, providing a unified approach that overcomes the limitations of both Poisson-based models and computationally intensive MCMC methods. Our extensive simulation studies demonstrate that the variational approximations achieve estimation accuracy, variable selection performance, and uncertainty quantification comparable to MCMC benchmarks in low-dimensional settings, while requiring less computational time.

In high-dimensional scenarios where MCMC becomes infeasible, our methods maintain excellent performance, with the negative binomial specification proving essential when overdispersion is present. The Poisson-based VB methods exhibit substantial degradation in all performance metrics under overdispersion, while the proposed NB methods remain robust even when the data are generated from a Poisson model, making them a safer and more reliable default choice for practitioners.

While the HS-VB-NB and CS-VB-NB methods perform similarly in the low-dimensional setting, the CS-VB-NB method appears to outperform its HS counterpart in the high-dimensional case, achieving lower estimation and prediction errors along with more stable variable selection performance.

The results further reveal that the primary limitation of the VB-POISS methods on overdispersed datasets is the under-coverage of the HPD intervals, indicating that the Poisson-based approximations fail to adequately capture the uncertainty associated with the dispersion in the data. This shortcoming underscores the importance of adopting a negative binomial framework when overdispersion is present.

Several limitations of the current work should be acknowledged. First, while the variational Bayes approximation substantially reduces computational cost, the mean-field assumption may underestimate posterior uncertainty, particularly for parameters with strong posterior dependence. Second, the Gaussian approximation used for the non-conjugate negative binomial model, based on the variational Bayes framework of \citep{wand11}, may not capture highly non-Gaussian posterior shapes. Future research directions include extending the proposed framework to zero-inflated Poisson and negative binomial models, or considering other sparsity enforcing priors.  Additionally, exploring structured sparsity patterns, such as group sparsity or spatial dependence, would broaden the applicability of the approach. The codes and data for this paper are available online on GitHub at \url{https://github.com/mortamini/VBSparseNegBin}.

\section*{Appendix (A): Derivation of $q(\cdot)$ density functions}
\setcounter{equation}{0}
 \renewcommand{\theequation}{A\thesection.\arabic{equation}}
 Here, we propose the details of the derivation of the VB  components,  as well as, the computation of the ELBO, for the two proposed VB models.
For model \eqref{model22}, we can see that 

\begin{align*}
{\rm E}_q =  {\rm E}_q \left[ \log P \left( g , \beta , \tau^2 , a , \kappa , b , Z , \zeta \right)  \right] = & {\rm E}_q \left[ \log P (g \vert \kappa , \beta) + \log P (\beta \vert Z , \sigma^2) \right. \\
 &  + \log P (\sigma^2 \vert a) + \log P (a) + \log P (\kappa \vert b)  \\
& \left. + \log P (b)  + \log P (Z \vert \zeta)  + \log P (\zeta) \right] \\
= & {\rm E}_q \left[ \log P (g \vert \kappa , \beta) + \log P (\beta \vert Z , \sigma^2)  \right] + \text{Const.} \\
= & \sum_{i=1}^{n} {\rm E}_q \left[ - \kappa \mathbf{X}_i \beta -\kappa \exp (- \mathbf{X}_i \beta ) g_i \right] \\
& +  {\rm E}_q \left[ -\dfrac{1}{2} \sigma^{-2} \beta^{\top} (\text{diag}(Z) + c^{-1}\text{diag}(1-Z))\beta \right] + \text{Const.} \\
= & \sum_{i=1}^{n} \left[ -\mu_{\kappa} \mathbf{X}_i { \boldsymbol\mu }_{{ \boldsymbol\beta}(C)} - \mu_{\kappa} \mu_{g_i} \omega_i^{(C)} \right] \allowdisplaybreaks\\
&  -\frac{1}{2} (1- \frac{1}{c}) {\rm E}({ \sigma^{-2}}) \text{tr}\left(\text{diag}(P^{(C)})D_{\boldsymbol\beta}^{(C)}\right)\\
& - \frac{1}{2c} {\rm E}({ \sigma^{-2}}) \text{tr}\left(D_{\boldsymbol\beta}^{(C)}\right)  + \text{Const.}
\end{align*}
Therefore, 
\begin{align*}
d_{{ \boldsymbol\mu }_{{ \boldsymbol\beta}(C)}} {\rm E}_q  = & -\mu_{\kappa}\mathbf{X}^\top (\mathbf{1}_n - \mu_{\boldsymbol g} \odot \boldsymbol\omega^{(C)}) -  {\rm E}(\tau^{-2}) ((1-1/c)P^{(C)}+1/c\mathbf{1}_p)\odot \mu_{\beta(C)}\\
d_{{ \boldsymbol\Sigma}_{{ \boldsymbol\beta}(C)}} {\rm E}_q = & -\frac{1}{2} \mu_{\kappa}\mathbf{X}^\top \text{diag}(\mu_{\boldsymbol g} \odot \boldsymbol\omega^{(C)}) \mathbf{X} - \frac{1}{2} {\rm E}(\tau^{-2}) \text{diag}((1-1/c)P^{(C)}+1/c\mathbf{1}_p).
\end{align*}
where ${ P^{(C)}} = (1,P_1^{(C)},\ldots,P_{p-1}^{(C)})^{\top}$. 
For model \eqref{model32}, we can see that 

\begin{align*}
{\rm E}_q = &  {\rm E}_q \left[ \log P \left( g , \mathbf{\beta} , \kappa , b , \mathbf{\Xi} , \xi , \mathbf{\Lambda} , \lambda , \beta_0 , \tau_0 , a)  \right)  \right] \\
= & {\rm E}_q \left[ \log P (g \vert \kappa , \beta) + \log P (\kappa \vert b) + \log P (b) + \log P( \beta \vert \Lambda , \Xi) \right. \\
 &  + \log P (\Lambda \vert \lambda) + \log P (\Xi \vert \xi) + \log P (\lambda) + \log P(\xi)  \\
& \left. + \log P (\beta_0 \vert \tau_0)  + \log P (\tau_0 \vert a)  + \log P (a) \right] \\
= & {\rm E}_q \left[ \log P (g \vert \kappa , \beta) + \log P (\beta \vert \Lambda , \Xi)  \right] + \text{Const.} \\
= & \sum_{i=1}^{n} {\rm E}_q \left[ - \kappa \mathbf{X}_i \beta - \kappa \exp (- \mathbf{X}_i \beta ) g_i \right] -\dfrac{1}{2}  {\rm E}_q \left[ \Xi^{-2} \beta^{\top} \text{diag}( \Lambda^{-2}) \beta \right] + \text{Const.} \\
= & \sum_{i=1}^{n} \left[ -\mu_{\kappa} \mathbf{X}_i { \boldsymbol\mu }_{{ \boldsymbol\beta}(H)} -\mu_{\kappa} \mu_{g_i} \omega_i^{(H)} \right] \\
&  - \dfrac{1}{2}   \left[  {\text{tr}}\left( {\rm E}({\Xi^{-2} }) \text{diag} \left( {\rm E} ({\Lambda^{-2}}) \right)  D^{(H)}_{{ \boldsymbol\beta}} \right) \right]  + \text{Const.}
\end{align*}
Therefore, 
\begin{align*}
d_{{ \boldsymbol\mu }_{{ \boldsymbol\beta}(H)}} {\rm E}_q  = & -\mu_{\kappa}\mathbf{X}^\top (\mathbf{1}_n - \mu_{\boldsymbol g} \odot \boldsymbol\omega^{(H)})  - {\rm E}({\Xi^{-2} }) \text{diag} \left( {\rm E} ({\Lambda^{-2}}) \right) { \boldsymbol\mu }_{{ \boldsymbol\beta}(H)}\\
d_{{ \boldsymbol\Sigma}_{{\boldsymbol\beta}(H)}} {\rm E}_q = & -\frac{1}{2} \mu_{\kappa}\mathbf{X}^\top \text{diag}(\mu_{\boldsymbol g} \odot \boldsymbol\omega^{(H)}) \mathbf{X}  -\frac{1}{2} {\rm E}({\Xi^{-2} }) \text{diag} \left( {\rm E} ({\Lambda^{-2}}) \right).
\end{align*}

Furthermore, for $M = C, H$, 
\begin{align*}
\log q(g) = &  \sum_{i=1}^{n} {\rm E}_{-g_i} \left[ \log P(y_i \vert g_i) + \log P(g_i \vert \kappa , \beta) \right] +  \text{Const.} \\
= & \sum_{i=1}^{n} {\rm E}_{-g_i}  \left[ y_i \log g_i - g_i + (\kappa -1) \log g_i - \kappa g_i \exp (- \mathbf{X}_i \beta) \right] + \text{Const.} \\
= &  \sum_{i=1}^{n} [ ( y_i + \mu_{\kappa} - 1 )\log g_i - ( \mu_{\kappa}\omega_i^{(M)} + 1 ) g_i ] + \text{Const.}
\end{align*}

The variational distribution for the parameter $ \kappa $ is obtained, for $M = C, H$, as follows
\begin{equation}\label{logka}
\begin{aligned}
\log q(\kappa) = & E_{-\kappa} \left[ \sum_{i=1}^{n} \log P (g_i | \kappa , \beta ) + \log P (\kappa | b) \right] + \text{Const.} \\
= & -\kappa 1^T X \mu_{\beta} + (\kappa - 1) \sum_{i=1}^{n} {\rm E}( \log g_i ) - n \log \Gamma ( \kappa ) \\
& - \dfrac{1}{2} \log \kappa + n \kappa  \log \kappa - \kappa \sum_{i=1}^{n} \mu_{g_i} \omega_i^{(M)} - \kappa {\rm E}(b^{-1}).
\end{aligned}
\end{equation}

Also,
\begin{equation}\label{logb}
\begin{aligned}
\log q(b) 
= &\; E \left[- \frac{\kappa}{b} - \frac{3}{2} \log b - \frac{\mathcal{M}_{\kappa}}{b} \right] + \text{Const.} \\
= &\; - \frac{3}{2} \log b  - \frac{1}{b} \left( \mathcal{M}_{\kappa} + \mu_{\kappa} \right)+ \text{Const.}
\end{aligned}
\end{equation}

For the CS-VB-NB model, we have 
\begin{align*}
\log q({ \sigma^2} ) = & {\rm E}_{- { \sigma^2}} \left[  \log p({ \boldsymbol\beta} \vert Z,{ \sigma^2})+ \log p({ \sigma^2} \vert a) \right] + \text{Const.} \\
= & -\frac{p}{2} \log ({ \sigma^2}) - \frac{1}{2 { \sigma^2}} E \left[ \text{tr(diag}(Z) { \boldsymbol\beta} { \boldsymbol\beta}^{\top}) \right] \\
& - \dfrac{1}{2 c { \sigma^2}} E \left[ \text{tr(diag}(1-Z) { \boldsymbol\beta} { \boldsymbol\beta}^{\top}) \right] - \frac{1}{{ \sigma^2}} {\rm E}(a^{-1})  \\
& - \dfrac{1}{2} \log { \sigma^2} + \text{Const.} \\
= & - \dfrac{p+1}{2} \log { \sigma^2} - \dfrac{1}{{ \sigma^2}} \left[ \frac{1}{2} \text{tr(diag}({ P^{(C)}}) D_{{ \boldsymbol\beta}}^{(C)}) \right. \\
& + \left.  \dfrac{1}{2c}\text{tr(diag}(1-{ P^{(C)}}) D_{{ \boldsymbol\beta}}^{(C)})  + {\rm E}(a^{-1}) \right] + \text{Const.}, 
\end{align*}
which is the logarithm of an inverse gamma density, and 
\begin{align*}
\log q(a) = & {\rm E}_{-a} \left[  \log p({ \sigma^2} \vert a) + \log p(a) \right] + \text{Const.}  \\
 = &  {\rm E}_{-a} \left[ \dfrac{1}{2} \log a - \dfrac{1}{a { \sigma^2}} + \dfrac{3}{2} \log a - \dfrac{1}{\mathcal{M}_{\sigma^2}a} \right] + \text{Const.} \\
 = & 2 \log a - \dfrac{1}{a} ({\rm E}({ \sigma^{-2}} )+ \mathcal{M}_{\sigma^2}^{-1})+ \text{Const.}, 
\end{align*}
which corresponds with an inverse gamma distribution. Furthermore, 
\begin{align*}
\log q(Z) = & \sum_{j=1}^{p-1} {\rm E}_{-Z_j} \left[  \log p(\beta_j \vert Z_j,{ \sigma^2}) + \log p(Z_j \vert \zeta_j) \right] + \text{Const.} \\
= &  \sum_{j=1}^{p-1}{\rm E}_{-Z_j} \Big [Z_j (- \dfrac{1}{2 { \sigma^2}} \beta_j^2 ) + (1 - Z_j) (-  \dfrac{1}{2 c { \sigma^2}} \beta_j^2 )   \\
& + Z_j \log  \zeta_j + (1- Z_j ) \log (1-  \zeta_j) \Big ] + \text{Const.}\\
& = \sum_{j=1}^{p-1} Z_j \left(- {\rm E}({ \sigma^{-2}}){ d_{jj}^{(C)}}(1-1/c)/{2 }+ \sum_{j=1}^{p-1}{\rm E}(\log  \zeta_j )\right. \\
&- \left.\sum_{j=1}^{p-1} {\rm E}(\log(1-  \zeta_j ))\right)+\text{Const.},
\end{align*}
which is the kernel of a Bernoulli probability mass function,  and 
\begin{align*}
\log q( \zeta) = & \sum_{j=1}^{p-1} {\rm E}_{- \zeta_j} \left[ \log p(Z_j \vert  \zeta_j) +  \log p( \zeta_j )  \right] + \text{Const.} \\
= & \sum_{j=1}^{p-1} E \Big [Z_j \log  \zeta_j + (1 -Z_j) (1-  \zeta_j ) + ({ \alpha}_1 - 1) \log  \zeta_j \\
& + (  {\alpha}_2 - 1) \log (1-\pi_j) \Big ] + \text{Const.} \\
= & \sum_{j=1}^{p-1} ({ P_j^{(C)}} + { \alpha}_1 - 1) \log \zeta_j+ \sum_{j=1}^{p-1} ( { \alpha}_2 - { P_j^{(C)}}) \log(1-  \zeta_j)+ \text{Const.}, 
\end{align*}
which corresponds with the beta distribution. 

To obtain the ELBO for the CS-VB-NB model, we compute the following terms

\begin{align*}
{\rm E}_q ( \log P(y \vert g) ) = &  \sum_{i=1}^{n}{\rm E}_q \left[ -g_i+ y_i \log g_i \right] +\text{Const.} \\
= & - 1^\top\mu_{\boldsymbol g} + y^\top {\rm E}(\log g) +\text{Const.}
\end{align*}
\begin{align*}
{\rm E}_q (\log P (g \vert \kappa, \beta) ) = & - \mu_{\kappa} 1^T X \mu_{\beta} + (\mu_{\kappa} -1) 1^\top {\rm E}(\log g) \\
& - n {\rm E}\left[ \log \Gamma (\kappa) \right]+ n E\left[ \kappa \log \kappa \right] - \mu_{\kappa} 1^\top (\mu_{\boldsymbol g}\odot \boldsymbol\omega^{(C)}),
\end{align*}
\begin{align*}
{\rm E}_q (\log P(\kappa | b)) = -\dfrac{1}{2} {\rm E} (\log \kappa) - \mu_{\kappa}{\rm E} (b^{-1}) + \text{Const.}
\end{align*}
\begin{align*}
{\rm E}_q (\log P(b)) = -\dfrac{3}{2} {\rm E} (\log b) - \mathcal{M}_{\kappa} {\rm E} (b^{-1}) + \text{Const.}
\end{align*}
\begin{align*}
{\rm E}_q (\log p({ \boldsymbol\beta} \vert \mathbf{Z},{ \sigma^2})) = & {\rm E}_q \left[ {\sum_{j=1}^{p-1}} Z_j  \big(- \frac{1}{2} \log { \sigma^2} - \dfrac{1}{2 { \sigma^2}} \beta_j^2 \big)\right. \\
& \left.- \sum_{j=1}^{p-1}(1-Z_j) \big( - \frac{1}{2} \log c { \sigma^2} - \dfrac{1}{2 c { \sigma^2}} \beta_j^2 \big) \right] + \text{Const.}\\
= &- \frac{1}{2} {\sum_{j=1}^{p-1}} { P_j^{(C)}} \big( {\rm E}(\log{ \sigma^2})+{\rm E}({ \sigma^{-2}}) { d_{jj}^{(C)}} \big) \\
&  - \frac{1}{2} {\sum_{j=1}^{p-1}} (1- { P_j^{(C)}}) \big({\rm E}(\log{ \sigma^2})+\frac{1}{c} {\rm E}({ \sigma^{-2}}) { d_{jj}^{(C)}} \big) \\
= & -\frac{1}{2} (1- \frac{1}{c}) {\rm E}({ \sigma^{-2}}) {\sum_{j=1}^{p-1}} { P_j^{(C)}} { d_{jj}^{(C)}} - \frac{1}{2c} {\rm E}({ \sigma^{-2}}) {\sum_{j=1}^{p-1}} { d_{jj}^{(C)}} \\
&- \frac{p-1}{2}{\rm E}(\log{ \sigma^2})+ \text{Const.},
\end{align*}
\begin{align*}
{\rm E}_q(\log p(\mathbf{Z} \vert \zeta)) = & {\sum_{j=1}^{p-1}} E \left[Z_j \log \zeta_j + (1- Z_j) \log (1 - \zeta_j) \right] + \text{Const.}\\
= & {\sum_{j=1}^{p-1}} [{ P_j^{(C)}} E\left(\log{\zeta_j}\right)+(1-{ P_j^{(C)}}){\rm E}(\log(1-\zeta_j))] +\text{Const.},
\end{align*}
\begin{align*}
{\rm E}_q(\log p(\zeta)) = &  ({ \alpha}_1 -1 )  {\sum_{j=1}^{p-1}}  {\rm E}(\log\zeta_j) +  ({ \alpha}_2 -1 )    {\sum_{j=1}^{p-1}} {\rm E}(\log (1-\zeta_j))+ \text{Const.},
\end{align*}
\begin{align*}
{\rm E}_q(\log p({ \sigma^2}\vert a)) = &   -\frac{1}{2} {\rm E}(\log a) - \frac{3}{2} {\rm E}(\log { \sigma^2}) - \frac{1}{2} {\rm E}({ \sigma^{-2}}) {\rm E}(a^{-1}) + \text{Const.},
\end{align*} 
\begin{align*}
{\rm E}_q (\log p(a)) = &\; {\rm E}_q \left(-\frac{3}{2} \log a - \dfrac{1}{\mathcal{M}_{\sigma^2}a} \right) + \text{Const.}\\
= & -\frac{3}{2} {\rm E}(\log a )- \mathcal{M}_{\sigma^2}^{-1}{\rm E}(a^{-1})+ \text{Const.},
\end{align*}
Furthermore, 
\begin{align*}
-{\rm E}_q (\log q(g)) = & -\sum_{i=1}^{n} \left( y_i + \mu_{\kappa} -1 \right) {\rm E}(\log g_i) \\
&  + \sum_{i=1}^{n} (\mu_{\kappa}\omega_i^{(C)} + 1)\mu_{g_i} + \sum_{i=1}^{n} \log \Gamma(y_i + \mu_{\kappa}) \\
& - \sum_{i=1}^{n} (y_i + \mu_{\kappa}) \log (\mu_{\kappa}\omega_i^{(C)}+ 1) + \text{Const.}
\end{align*}
\begin{align*}
-{\rm E}_q (\log q(\kappa)) = & \mu_{\kappa}1^\top X\mu_{\beta}-(\mu_{\kappa}-1)1^\top {\rm E}(\log g) + \mu_{\kappa}\mathbf{1}_n^\top (\mu_{\boldsymbol g}\odot\boldsymbol\omega^{(C)}) + \mu_{\kappa}{\rm E}(b^{-1})-n {\rm E} (\kappa \log \kappa)\\
& + n {\rm E} (\log \Gamma(\kappa)) + \dfrac{1}{2} {\rm E}(\log \kappa) + \mathcal{H} (-\frac{1}{2} , 0, 1, n, C_1) + \text{Const.},
\end{align*}
\begin{align*}
-{\rm E}_q (\log q(b) ) = \frac{3}{2} {\rm E}(\log b) + {\rm E} (b^{-1}) (\mathcal{M}_{\kappa} + \mu_{\kappa}) -\frac{1}{2}\log (\mathcal{M}_{\kappa} + \mu_{\kappa}) + \text{Const.}
\end{align*}
\begin{align*}
-{\rm E}_q(\log q({ \boldsymbol\beta})) = \frac{1}{2} \log \vert { \boldsymbol\Sigma}_{{ \boldsymbol\beta}(C)} \vert+ \text{Const.},
\end{align*}
\begin{align*}
-{\rm E}_q(\log q(\mathbf{Z})) = & -E \left[ {\sum_{j=1}^{p-1}} Z_j \log ({ P_j^{(C)}}) + (1-Z_j) \log(1-{ P_j^{(C)}}) \right] + \text{Const.}\\
= & - {\sum_{j=1}^{p-1}} \left[ { P_j^{(C)}} \log ({ P_j^{(C)}}) + (1-{ P_j^{(C)}}) \log(1-{ P_j^{(C)}}) \right]+ \text{Const.},
\end{align*}
\begin{align*}
-{\rm E}_q(\log q(\zeta)) = & -\sum_{j=1}^{p-1}{\rm E} \left[ \log \Gamma (\alpha_1 + \alpha_2 + 1) - \log \Gamma ({ P_j^{(C)}} + \alpha_1)  - \log \Gamma (\alpha_2 - { P_j^{(C)}} + 1) \right. \\
& \left. + (\alpha_ 1 + { P_j^{(C)}}) \log \zeta_j + (\alpha_2 - { P_j^{(C)}} ) \log (1 - \zeta_j) \right] + \text{Const.}\\
= & \sum_{j=1}^{p-1} \left[  \log \Gamma ({ P_j^{(C)}} + \alpha_1)  + \log \Gamma (\alpha_2 - { P_j^{(C)}} + 1) \right. \\
& \left. - \left(\alpha_ 1 + { P_j^{(C)}} -1\right) {\rm E} (\log \zeta_j) - \left(\alpha_2 - { P_j^{(C)}} \right) {\rm E} (\log (1- \zeta_j)) \right] + \text{Const.},
\end{align*}
\begin{align*}
-{\rm E}_q(\log q ({ \sigma^2})) = & - \alpha_{\sigma^2} \log \beta_{\sigma^2} + \log \Gamma (\alpha_{\sigma^2}) \\
& + (\alpha_{\sigma^2} -1 ) {\rm E}(\log { \sigma^2}) + \beta_{\sigma^2} E ({ \sigma^{-2}}) + \text{Const.},
\end{align*}
and 
\begin{align*}
-{\rm E}_q (\log q (a)) = & -\log \Big(\mathcal{M}_{\sigma^2}^{-1}+ {\rm E}({ \tau^{-2}}) \Big) + 2 {\rm E}(\log a) +  \Big(\mathcal{M}_{\sigma^2}^{-1} + {\rm E}({ \tau^{-2}}) \Big)  {\rm E}(a^{-1}) + \text{Const.}
\end{align*}

Summation of the above terms would result in \eqref{elbo22}. For the HS-VB-NB model, and for $j=0,1,\ldots,p-1$,
\begin{align*}
\log q(\Lambda_j) & = E_{-\Lambda_j} \left[ \log p(\beta_j \vert \Lambda_j, \Xi) + \log p(\Lambda_j \vert \lambda)   \right] + \text{Const.}\\
& = -\frac{1}{2} \log \Lambda_j - \frac{1}{2\Lambda_j} {\rm E}(\Xi^{-1}) {\rm E}(\beta_j^2) - \frac{3}{2} \log \Lambda_j - {\rm E}(\lambda^{-1}) \Lambda_j^{-1} + \text{Const.}
\end{align*}

Also,
\begin{align*}
\log q(\Xi) & = E_{-\Xi} \left[\sum_{j=0}^{p-1} \log p(\beta_j \vert \Lambda_j, \Xi) + \log p(\Xi \vert \xi)   \right] + \text{Const.}\\
& = -\frac{p}{2} \log \Xi - \frac{1}{2\Xi} \sum_{j=0}^{p} {\rm E}(\Lambda_j^{-1}) {\rm E}(\beta_j^2) - \frac{3}{2} \log \Xi - {\rm E}(\xi^{-1}) \Xi^{-1} + \text{Const.},
\end{align*}
\begin{align*}
\log q(\lambda) & = E_{-\lambda} \left[\sum_{j=0}^{p-1}  \log p(\Lambda_j \vert \lambda) + \log p(\lambda)  \right] + \text{Const.}\\
& = -\frac{1}{\lambda} \sum_{j=0}^{p-1}  {\rm E}(\Lambda_j^{-1}) - \frac{3}{2} \log \lambda - \mathcal{M}_{\lambda}^{-1} \lambda^{-1} + \text{Const.},
\end{align*}
and
\begin{align*}
\log q(\xi) & = E_{-\xi} \left[\log p(\Xi \vert \xi) + \log p(\xi)  \right] + \text{Const.}\\
& = -\frac{1}{\xi} {\rm E}(\Xi^{-1}) - \frac{3}{2} \log \xi - \mathcal{M}_{\xi}^{-1} \xi^{-1} + \text{Const.}
\end{align*}

The computation of the ELBO for the HS-VB-NB model is done through computation of the following terms
\begin{align*}
{\rm E}_q\left[\sum_{j=0}^{p-1} \log p(\beta_j \vert \Lambda_j, \Xi) \right] & =
-\frac{p}{2} E[\log \Xi] -\frac{1}{2} \sum_{j=0}^{p-1} E[\log \Lambda_j] - \frac{1}{2} {\rm E}(\Xi^{-1}) \sum_{j=0}^{p-1} {\rm E}(\Lambda_j^{-1}) {\rm E}(\beta_j^2) + \text{Const.},
\end{align*}
\begin{align*}
{\rm E}_q\left[\sum_{j=0}^{p-1}  \log p(\Lambda_j \vert \lambda) \right] & =
-{\rm E}(\lambda^{-1}) \sum_{j=0}^{p-1}  {\rm E}(\Lambda_j^{-1}) - \frac{3}{2} \sum_{j=0}^{p-1} {\rm E}(\log \Lambda_j)  + \text{Const.},
\end{align*}
\begin{align*}
{\rm E}_q\left[\log p(\Xi \vert \xi) \right] & =
- \frac{3}{2} {\rm E}(\log \Xi) - {\rm E}(\xi^{-1}) {\rm E}(\Xi^{-1}) + \text{Const.},
\end{align*}
\begin{align*}
{\rm E}_q\left[\log p(\lambda) \right] & =
- \frac{3}{2} {\rm E}(\log \lambda) - M_{\lambda}^{-1} {\rm E}(\lambda^{-1}) + \text{Const.},
\end{align*}
\begin{align*}
{\rm E}_q\left[\log p(\xi) \right] & =
- \frac{3}{2} {\rm E}(\log \xi) - M_{\xi}^{-1} {\rm E}(\xi^{-1}) + \text{Const.},
\end{align*}
\begin{align*}
-{\rm E}_q[\sum_{j=0}^{p-1}\log q(\Lambda_j)] & = 2 \sum_{j=0}^{p-1} {\rm E}(\log \Lambda_j) + \frac{1}{2} {\rm E}(\Xi^{-1}) \sum_{j=0}^{p-1} {\rm E}(\Lambda_j^{-1}) {\rm E}(\beta_j^2) + {\rm E}(\lambda^{-1}) \sum_{j=0}^{p-1} {\rm E}(\Lambda_j^{-1}) + \text{Const.},
\end{align*}
\begin{align*}
-{\rm E}_q[\log q(\Xi)] & = \frac{p+3}{2} {\rm E}(\log \Xi) + \frac{1}{2}{\rm E}(\Xi^{-1}) \sum_{j=0}^{p-1} {\rm E}(\Lambda_j^{-1}) {\rm E}(\beta_j^2) + {\rm E}(\xi^{-1}) {\rm E}(\Xi^{-1}) + \text{Const.},
\end{align*}
\begin{align*}
-{\rm E}_q[\log q(\lambda)]  & =\left(\sum_{j=0}^{p-1}  {\rm E}(\Lambda_j^{-1}) +\mathcal{M}_{\lambda}^{-1}\right) {\rm E}(\lambda^{-1}) + \frac{3}{2} {\rm E}(\log \lambda) + \text{Const.},
\end{align*}
and
\begin{align*}
-{\rm E}_q[\log q(\xi)] & = \left({\rm E}(\Xi^{-1}) +\mathcal{M}_{\xi}^{-1}\right) {\rm E}(\xi^{-1}) + \frac{3}{2} {\rm E}(\log \xi) + \text{Const.}
\end{align*}


\end{document}